\documentclass[twocolumn,english,showpacs,preprintnumbers,amsmath,amssymb,floatfix,nofootinbib]{revtex4-1}

\RequirePackage{graphicx}
\usepackage[T1]{fontenc}
\usepackage[latin9]{inputenc}
\usepackage{color}
\usepackage{array}
\usepackage{amstext}
\usepackage{amsmath}
\usepackage{graphicx}
\usepackage{esint}
\usepackage{xspace}
\usepackage{adjustbox}
\usepackage{hhline,multirow,tabularx} 
\usepackage{balance}
\usepackage{subfigure}

\usepackage{lineno}

\begin{document}

\begin{sloppypar}

\title{Parton distribution functions and constraints on the intrinsic charm content of the proton using BHPS approach} 
\author{H. Abdolmaleki}
\email{Abdolmalki@semnan.ac.ir}

\author{A. Khorramian}
\email{Khorramiana@semnan.ac.ir}

\affiliation{Faculty of Physics, Semnan University, 35131-19111, Semnan, Iran}

\begin{abstract}
In this work, a new set of parton distribution functions taking into account the intrinsic charm (IC) contribution is presented. We focus on the impact of the EMC measurements on the large $x$ charm structure function as the strongest evidence for the intrinsic charm when combined with the HERA, SLAC and BCDMS data.  
The main goal of this paper is the simultaneous determination of the intrinsic charm probability ${P}_{c{\bar c/p}}$ and strong coupling $\alpha_s$. This allows us to study the interaction of these two quantities as well as the influence on the PDFs in the presence of IC contributions. By considering  $\alpha_s$  which can be fix or free parameter from our QCD analysis, we find that although there is not a significant change in the extracted central value of PDFs and their uncertainties, the obtained value of ${P}_{c{\bar c/p}}$ change by factor 7.8\%. The extracted value of ${P}_{c{\bar c/p}}$ in the present QCD analysis is consistent with the recent reported upper limit of $1.93\%$, which is obtained for the first time from LHC measurements. We show the intrinsic charm probability is sensitive to the strong coupling constant and also the charm mass. The extracted value of the strong coupling constant $\alpha_s(M_Z^2) = 0.1191 \pm 0.0008$ at NLO is in good agreement with world average value and available theoretical models.
\end{abstract} 

\maketitle
\tableofcontents{}

\section{Introduction}
\label{intro}
The distribution of heavy quarks in the proton can provide a comprehensive overview of the nucleon structure, and it is essential for processes in the perturbative QCD (pQCD) calculations  at the LHC physics.

According to the Brodsky, Hoyer, Peterson, and Sakai (BHPS) model \cite{BHPS1,BHPS2}, the charm quark distribution of the nucleon consists of two separate contributions. The first one is generated perturbatively by gluon splitting to $c\bar{c}$ ($g\rightarrow c\bar{c}$) from DGLAP evaluation equations \cite{Altarelli:1977zs}, which is generally referred to ``extrinsic charm'', see \cite{Brodsky:2001yt,Vogt:1994zf,Vogt:1995tf}.  This extrinsic charm distribution depends logarithmically on the charm mass $m_c$ and is most important at low $x$. Next is the ``intrinsic charm'' (IC) which has a non-perturbative origin, in contrast to extrinsic charm, and is associated with bound state hadron dynamics. This distribution is described by Fock state in the nucleon structure which is dominant at large $x$ arises from multiply connected charm quark pairs by the valence quarks. With same rapidities of the quarks which constituent the nucleons, this contribution is maximal in the minimal off-shellness and depends on the non-perturbative structure of the nucleon. While the intrinsic charm contribution is dominant at high $x$ and depends on $1/{m_c^2}$, the extrinsic charm distribution is dominant at small $x$ and depends logarithmically on charm mass.

 A review of the intrinsic heavy quark content of the nucleon
has been reported in Ref.~\cite{Brodsky:2015fna}. Several theoretical and phenomenological studies have been performed taking into account the IC component in the proton 
\cite{Brodsky:2018zdh,Brodsky:2016tew,Dulat:2013hea,Pumplin:2005yf,Blumlein:2015qcn,Brodsky:2016fyh,Koshkarev:2016rci,Anjos:2001jr,Montano:1996nj,Bednyakov:2013zta}. Also, according to BHPS approach,  the intrinsic light-quark sea in the proton are studied in Refs.~\cite{Chang:2014lea,An:2017flb}.

Other studies have been performed using the intrinsic quark components to find the effects on charm quark production  \cite{Freeman:2012ry,Breidenbach:2008ua,Brodsky:1997fj,Vogt:1995fsa,Gutierrez:1998bc,Vogt:1992ki,Brodsky:2006wb,Ball:2016neh,Duan:2016rkr,Aubert:1981ix,Martynenko:1994kk,Vafaee:2016jxl,Ermolaev:2017qei,Carvalho:2017zge,Bi:2016vbt,Zhou:2017bhq,Hsiao:2015nna,Polyakov:2015foa}. Recently, QCD analyses of parton distribution functions (PDFs) taking into account IC contribution are reported by NNPDF3IC \cite{Ball:2017nwa} and CT14 \cite{Hou:2017khm} collaborations. Also, there are explanations of how one can probe the intrinsic charm in different ways, {\it{i.e.}},  directly \cite{Boettcher:2015sqn} and indirectly \cite{Halzen:2013bqa}. 

The intrinsic charm contribution is important for estimating  the flux of high energy neutrino which observed in the IceCube measurement. In Ref.~\cite{Brodsky:2016fyh},  the prompt neutrino spectrum using the intrinsic charm contribution is estimated. It is shown that the prompt atmospheric neutrino flux taking into account the intrinsic charm contribution is comparable with the extracted results of QCD calculations without taking to account the intrinsic charm. Undoubtedly,  the IceCube measurements will constrain the IC component in the proton. This kind of measurements will also contribute to the main questions in high energy physics phenomenology in the future \cite{Laha:2016dri}.
 
Not only intrinsic charm quark but also intrinsic strange and bottom quarks are  principle property of the wave functions of hadronic bound states.
The application of the operator product expansion shows that the probability for the analogous intrinsic bottom contribution from $|uud b \bar b>$ Fock states is suppressed relative to intrinsic charm by a factor of $m^2_c / m^2_b$. 

 There are some processes which are sensitive to the charm quark distribution in the large $ x $ region. 
For example the European Muon Collaboration (EMC) \cite{Aubert:1982tt} provided the charm structure function data $F_2^{c} $ as  strong evidence for
intrinsic component at large $x$ \cite{Aubert:1982tt}. The measurement at $x$=0.42, $Q^2$= 75 GeV$^2$ is approximately 30 times higher
than predicted from gluon splitting. It should be noted that the EMC data is the only DIS evidence for intrinsic charm at high $x$ and it is worthwhile to include these data in the study of intrinsic charm.

A description of the EMC data with intrinsic charm is presented in Refs.~\cite{Ball:2017nwa,Hou:2017khm,Hoffmann:1983ah,Harris:1995jx}. In  Ref.~\cite{Ball:2017nwa}, the NNPDF included the EMC data with using the standard  $W^2 > 12.5$ GeV$^2$ cut to avoid the presence of dynamical higher twists. Although the EMC data are included without nuclear corrections in NNPDF collaboration \cite{Ball:2017nwa}, several fit variants have been performed to assess the impact of nuclear corrections on the EMC data and find a moderate impact.

In Ref. \cite{Hoffmann:1983ah}, Hoffmann and Moore performed the NLO calculation of the BHPS model using EMC data. The first NLO analysis of both extrinsic and intrinsic contributions with EMC data was done by Harris, Smith, and Vogt \cite{Harris:1995jx}.

Since the first experimental evidence of IC originated from the EMC measurement at large $x$, a variety of charm hadrons measurements are consistent with the existence of IC. In the case of hadron-hadron collisions, the intrinsic charm leads to the production of charm hadrons  such as the $\Lambda_c(cud)$ as observed in ISRF experiments \cite{Chauvat:1987kb} and more recently by SELEX \cite{Garcia:2001xj} at high $x_F = x_c + x_u + x_d$  from the coalescence of the charm quark with its coming valence quarks, as well as quarkonium productions at high $x_F= x_c + x_{\bar c}$.   
Double intrinsic charm Fock states such as $|uud c \bar c c \bar c>$ in the proton lead to the production of double-charm baryons as well as double quarkonia at high $x_F$, see Ref. \cite{Badier:1983dg}. Previous fixed target $J/\psi$ measurements also show signs of significant IC contribution,  taking into account the nuclear mass dependence, as measured at CERN and Fermilab, see {\it e.g.}  \cite{Vogt:1991qd,Gabbiani:2002ti}.

On the other hand, the production of the prompt photons at hadron colliders
in $ pp(\bar{p}) \rightarrow \gamma +c $-jet is sensitive to the charm distribution and may provide good evidence for IC at high $p_T$. 
In Refs.~\cite{Lipatov:2016feu,Bailas:2015jlc,Beauchemin:2014rya} the results of $ c $-jet
production accompanied by vector bosons $ Z,W^{\pm} $ is studied by using the intrinsic charm quark component.
The impact of IC contributions for the prediction of $ \gamma+c $-jet production in $pp$ collisions at the LHC is reported in Ref.~\cite{Rostami:2015iva}, where this impact for large $p_T^\gamma $ can be distinguished
from the case in which the IC contribution is not considered. Without considering IC contribution, the D0 data for $ \gamma+c $-jet differential cross section \cite{D0:2012gw} can not be described by solely using an extrinsic charm PDF based on the DGLAP evolution equation. In contrast, the D0 data for $ \gamma+b $-jet differential cross section is explained by the extrinsic standard PDFs. This is consistent with the fact that the ratio of the intrinsic bottom distribution to the intrinsic charm distribution in the proton is suppressed by $m^2_c /m^2_b$.

In addition to the typical observables for IC, the intrinsic bottom (IB) quarks also contribute
to diffractive Higgs production in which the Higgs boson carries a remarkable fraction of the proton momentum. 
Measuring the high $x_F$ process $ p p \to H X$ via intrinsic quarks (IQ) at the LHC would give new constraints on the Higgs couplings to quark pairs, including $H \to b \bar b$ \cite{Brodsky:2007yz}. 

Additionally, the new ATLAS measurements at the LHC \cite{Aaboud:2017skj} have shown that the production of prompt
photons accompanied by a charm-quark jet in $pp$ collisions are sensitive to the intrinsic charm content of the nucleon. By having these new ATLAS measurements at 8 TeV, the upper limit of the intrinsic charm probability is found to be about ${P}_{c{\bar c/p}}$=1.93\% \cite{Bednyakov:2017vck}. They proposed a method that reduced the uncertainty on the determination of intrinsic charm probability. Undoubtedly, demonstrating the compatibility of the DIS data and the new LHC data for the ${P}_{c{\bar c/p}}$ extraction \cite{Bednyakov:2017vck}, would be worth. Regarding the importance of intrinsic heavy quarks, now we have sufficient motivation to incorporate the IC contributions in our QCD analysis.

In this paper, we perform our QCD analysis based on the $\mathsf{xFitter}$ open source framework \cite{xFitter,Alekhin:2014irh}, which was previously known as HERAfitter. Recently, the low $x$ resummation of QCD analysis is performed by $\mathsf{xFitter}$ \cite{Abdolmaleki:2018jln}, which leads to the better description of the data at low $x$ and low $Q^2$. Other QCD analysis based on $\mathsf{xFitter}$ framework are performed in Ref.~\cite{vafaee:2017nze,Abdolmaleki:2017wlg}. In Refs. \cite{vafaee:2017nze, GHAFARI-PRC} we extracted the strong coupling constant using HERA I and II combined data and Neutrino-nucleon structure function data using  $\mathsf{xFitter}$. More recently, we also determined the strong coupling constant with polarized data \cite{Salimi-Amiri:2018had}. Note that one of the main purposes of this paper is to determine $\alpha_s$ concurrently with the IC probability.

The paper is organized as follows. In the Sec.\ref{Intrinsic Heavy}, we briefly outline the basic formalism and provide a theoretical overview of the intrinsic heavy quark distributions. In this section, we also introduce the $Q^2$ evolution of the intrinsic heavy quark distribution functions and the intrinsic heavy quark component of the structure functions. The experimental data which we use in the present analysis is presented in Sec.\ref{data sets}.
 The PDF parametrizations are discussed in Sec.\ref{PDF}. In Sec.\ref{fit results}, we present the fit results for the PDFs including the intrinsic charm at NLO, and we make comparisons with other different theoretical analyses from literature. Finally, our discussion and conclusion are given in Sec.\ref{Discussion}.

\section{BHPS model and evolution of intrinsic heavy quark structure function}
\label{Intrinsic Heavy}
According to the light-cone formalism, the proton wave function can be represented as a sum over the complete basis of free
quark and gluon states based on a superposition of the proton wave function and $n$-particle Fock fluctuation components, $\vert q_1q_2q_3 \rangle$,  $\vert q_1q_2q_3g \rangle$, $\vert q_1q_2q_3q\bar q \rangle$, $etc$, \cite{BHPS1} as 
\begin{eqnarray}
\vert \rm \Psi_p \rangle &=& \psi _{3q/p}(x_i, {k}_{\perp i})  \vert uud  \rangle
 +\psi _{3qg/p}(x_i, {k}_{\perp i})  \vert uudg \rangle \nonumber \\
 &+&\psi _{5q/p}(x_i, {k}_{\perp i}) \vert uudq\bar{q} \rangle+\ldots~, 
\label{F1}
\end{eqnarray}
where the light-front wave functions $\psi _{j/p}(x_i, {k}_{\perp i})$ depends on the relative momentum coordinates $ k_{\perp i}$ and $x_i={k_i^+}/{P^+}$ in which 
$k_i$ explains the parton momenta
and $P$ denotes the hadron momentum.
For a proton with mass of $m_p$, the general form of the Fock state wave function is
\begin{equation}
\psi_{n/p}(x_i,{\vec k}_{\perp i})= \frac{\Gamma(x_i,\vec{k}_{\perp i})}
{m_p^2-M^2}= \frac{\Gamma(x_i,\vec{k}_{\perp i})}
{m_p^2-\sum_i^n \widehat{m}_i^2/
x_i}~,\label{F2}
\end{equation}
where,  $\widehat{m}_i^2 = m_i^2 + \langle
\vec{k}_{\perp i}^2 \rangle$ is the square of the average transverse mass
of parton $i$. Note that by decreasing $m_p^2-M^2$, the  $\Gamma$ as a vertex function, expected to be a slowly varying. This form can be applied to the higher Fock components for an arbitrary number of light and heavy quarks. 
The momentum conservation for $n$ number of partons in state  $\vert j \rangle$ demands $\sum _{i=1}^n \vec{k}_{\perp i}={0}$  and $\sum _{i=1}^n x_i=1$.
 
In the BHPS model \cite{BHPS1}, the probability distribution in the $5$-particle intrinsic Fock state is
\begin{equation}
\label{icprobtot}
\frac{dP_{\rm IC}}{dx_1 \cdots dx_5}=N_5 \frac{\delta (1-\sum_{i=1}^{5} x_i)}{[m_p^2-\sum_{i=1}^{5} {\widehat{m}_i^2}/{x_i}]^{2}}~, 
\end{equation}
where $N_5$ normalizes the $5$-particle Fock state probability that is determined by ${P}_{c{\bar c/p}}= \int_0^1 dx_1...dx_5 \frac{dP_{\rm IC}}{dx_1,...,dx_5}$, where ${P}_{c{\bar c/p}}$ is the intrinsic charm probability in the proton. In the heavy quark limit,  $\widehat{m}_c$, $\widehat{m}_{\overline c} \gg m_p$,
$\widehat{m}_q$,  where the mass of light quark and proton are negligible compared to the heavy quarks mass, the Fock state probability distribution is written as
 \begin{equation}\label{E311}
\frac{dP_{\rm IC}}{dx_1 \cdots  dx_{\bar c} dx_c} = {N}_5\delta (1-\sum _{i=1}^5 x_i)\frac{x_c^2x_{\bar c}^2}{(x_c+x_{\bar c})^2}~,
\end{equation}
where $ {N}_5= { N}/m_{c,\bar{c}}^4$ is determined from Eq.~(\ref{E311}) by integrating over $dx_1...dx_{\bar c}$, so the intrinsic heavy quark probability distribution is given by 
\begin{eqnarray}
\label{heavyBHPS}
c_{int}(x_c) &=& \int dx_1...dx_{\bar c} \frac{dP_{\rm IC}}{dx_1,...,dx_{\bar c}dx_c} \nonumber \\ 
& =&P_{c{\bar c}/p} 1800~x_c^2 \Big[ \frac{(1-x_c)}{3} \left( 1 + 10x_c + x_c^2 \right) \nonumber \\ 
		  && + 2 x_c (1+x_c) \ln(x_c) \Big]~.	
 \end{eqnarray}

 We can simplify the above equation by considering $P_{c{\bar c}/p}=1\%$ IC probability. For more details see Refs. \cite{Vogt:1995fsa,Gutierrez:1998bc,Vogt:1992ki}. 
 
According to Eq.~(\ref{heavyBHPS}), as the BHPS model predicts \cite{BHPS1}, the existence of the intrinsic bottom (IB) distribution is very similar to the IC, but differs in the normalization factor by a coefficient $m_c^2/m_b^2$. Therefore, ${P}_{b{\bar b/p}} = {P}_{c{\bar c/p}}(m_c^2/m_b^2)\sim 0.001$ taking into account the $1\%$ IC probability distribution.

As mentioned above, the heavy quark distribution in the standard approach,  which is used by almost all global PDFs analyses, is generated by quark-gluon fusion in the DGLAP equations at the starting scale on the order of the heavy quark mass. As the BHPS model predicts, the purely perturbative treatment can not give a good description of the proton structure. Therefore the full charm parton distribution must be expressed by the sum $xc(x,Q^2)=xc_{ext}(x,Q^2)+xc_{int}(x,Q^2)$, where $xc_{ext}(x,Q^2)$ and $xc_{int}(x,Q^2)$ indicate the extrinsic charm that is radiatively generate by the DGLAP equation (perturbative) and the intrinsic charm (non-perturbative) at an initial scale $Q_0 \simeq m_c$, respectively.
This decomposition is a good approximation at any scale since the intrinsic charm is controlled by non-singlet evaluation equations \cite{Lyonnet:2015dca}. Therefore the evaluation of heavy quarks must be divided into independent parts, singlet and non-singlet. The compact form of the DGLAP equation used by the standard global analyses, is given by
\begin{equation}
\dot f_i = \sum_{j=q,g,Q}P_{ij}\otimes f_j~, 
\end{equation}
where $\dot f_i$ denotes the light quarks, heavy quarks and gluon and $P_{ij}$ is the splitting function which is known up to three-loop order in the massless $\overline{MS}$ scheme \cite{Vogt:2004mw,Moch:2004pa}. By considering the intrinsic heavy quark distribution $Q_{int}$ in the proton, one can find the non-singlet equation as 
 \cite{Lyonnet:2015dca}
\begin{equation}
\dot Q_{int} = P_{QQ}\otimes Q_{int}~. 
\end{equation}

In perturbative QCD, the structure functions can be written as a convolution between the hard scattering coefficient function and parton distribution functions which are parametrised and determined from experimental data. 

The deep-inelastic structure functions $F(x,Q^2)$  consists of the  heavy and light flavor component 
\begin{eqnarray}
\label{eq:F3}
F(x , Q^2) &=& F^{h}(x , Q^2)+F^{l}(x , Q^2)~,
\end{eqnarray}
where $F^h$ is the heavy contribution of DIS structure function, which is valid if only the heavy quark electric charge is non-zero.

 As we plan to study the intrinsic charm component of the proton structure function, then we need to add the intrinsic structure function $F_{(int)}(x, Q^2)$ to the extrinsic heavy structure function $F_{(ext)}^{h,RT}(x , Q^2)$ using the Thorne- Roberts (RT) scheme \cite{Thorne:1997ga} as 
\begin{eqnarray}
\label{eq:F4}
F^h(x , Q^2) &=& F_{(ext)}^{h,RT}(x , Q^2)+F_{(int)}^{c\bar{c}/p}(x,Q^2)~.
\end{eqnarray}

The definition of the heavy quark probability distribution in Eq.~(\ref{heavyBHPS}), using ${P}_{c{\bar c/p}} = 1$\%, for the $\vert uudc\bar{c} \rangle$ Fock state, leads to a simple form for the IC component in the proton structure function, as $F_{2 (int)}^{c\bar{c}/p}(x , Q^2) = \frac{8}{9}x\int dx_1 \cdots dx_{\bar c} \frac{dP_{IC}}{dx_i \cdots dx_{\bar c}dx_c}=8/9 xc_{int}(x,Q^2) $ with $c_{int}$ from Eq.~(\ref{heavyBHPS}), when the charmed mass is negligible in the leading order  \cite{Brodsky:2015fna}. By considering mass effects, the IC component in the proton structure $F_{2 (int)}^{c\bar{c}/p}(x , Q^2,m_c)$ using the mass variable $\xi = 2ax[1+(1+4m^2_p x^2/Q^2)^{1/2}]^{-1}$ where $a = [(1+4m^2_c/Q^2)^{1/2} +1]/2$, is as follows \cite{Hoffmann:1983ah}
\begin{eqnarray}
\label{eq:F4}
F_{2 (int)}^{c\bar{c}/p}(x , Q^2,m_c) &=& \frac{8x^2}{9(1 + 4{m^2_p x^2/Q^2}  )^{3/2}} \nonumber \\ && [\frac{1+4m^2_c/Q^2}{\xi}c(\xi , \gamma)
+ 3\hat{g}(\xi , \gamma)] ~,\nonumber \\
\end{eqnarray}
with $c(z,\gamma ) = c_{int}(z)-zc_{int}(\gamma )/\gamma$ for $z \leq \gamma$. Also,  $\gamma$ as another mass scaling variable which depends on $Q^2$ and is defined by $\gamma = 2a \hat{x}[1+(1+4m^2_p \hat{x}^2/Q^2)^{1/2}]^{-1} $,
where  $\hat{x}=[1+4 m^2_c/Q^2-m^2_p/Q^2]^{-1}$. In the above,   $\hat{g}(\xi , \gamma)$ is

\begin{eqnarray}
\label{eq:F5}
\hat{g}(\xi , \gamma) &=& \frac{2m^2_p x/Q^2}{(1 + 4m^2_p x^2/Q^2  )}\int_{\xi }^{\gamma }dt \frac{c(t,\gamma )}{t}\nonumber \\  
&&(1-\frac{m^2_c}{m^2_p t^2})[ 1+2m^2_p/Q^2 xt + \frac{2m^2_c x/Q^2}{t}] ~.\nonumber \\
\end{eqnarray}
For more detail see Refs. \cite{Hoffmann:1983ah,Kretzer:1998ju}.

\section{\label{data sets}Data sets}
For the QCD analysis of PDFs including intrinsic charm, different DIS processes which are generally important in the presence of intrinsic charm are used to determine the unknown parameters in the PDFs parametrization,  strong coupling constant $\alpha_s$ and ${P}_{c{\bar c/p}}$ as the extra fit parameters.

The experiments contributing to this analysis are the combined HERA inclusive proton DIS cross sections  \cite{Abramowicz:2015mha}, fixed target inclusive DIS $F_2^p$ BCDMS  \cite{Benvenuti:1989rh}, DIS $F_2^p$ SLAC  \cite{Whitlow:1991uw}, H1-ZEUS combined charm cross section  \cite{Abramowicz:1900rp}, and DIS $F_2^c$ EMC data \cite{Aubert:1982tt}.

The neutral current (NC) and charged current (CC) cross section data at HERA have explored a wide kinematic region of the Bjorken variable $x$ and negative boson transverse momentum squared $Q^2$. In this paper we used the NC and CC inclusive DIS experimental data collected by H1 ans ZEUS \cite{Aaron:2009bp,Aaron:2009kv,Adloff:1999ah,Adloff:2000qj,Adloff:2003uh,Aaron:2012qi,Andreev:2013vha,Collaboration:2010ry} from both HERAI and HERAII with the proton beam energy $E_p = 460, 575, 820$ and $920$ GeV$^2$ related to center of mass energy $\sqrt{s}= 225, 251, 300$ and $320$ GeV, respectively.

High $Q^2$ CC data, together with difference between the NC $e^- p$ and $e^+ p$ at high $Q^2$, constrain the $u$ and $d$-valence PDFs  which dominate at large $x$ \cite{Abramowicz:2015mha}. The wide kinematic region of the precise NC and CC cross-sections data allow us to extract PDF sets. 

We also choose the proton structure function data which are reported by BCDMS \cite{Benvenuti:1989rh} in deep-inelastic scattering of muons on the hydrogen target at the beam energy of 100, 120, 200 and 280 GeV.

The SLAC data on $ep$ scattering used in our analysis, as well. It is obvious that our motivation to choose this data set of experimental data is due to the kinematic range of large $x$, where the intrinsic charm is dominant.

Also, the combine charm reduced cross sections, $\sigma_{red}^{c\bar{c}}$ by H1 and ZEUS are used in the present QCD analysis. The charm quark predominantly produce by boson gluon fusion, $\gamma g \to c \bar{c}$, which is sensitive to the gluon distribution in the proton \cite{Aaron:2009bp,Aaron:2009kv,Adloff:1999ah,Adloff:2000qj,Adloff:2003uh,Aaron:2012qi,Andreev:2013vha,Adloff:1996xq,Breitweg:1997mj,Aaron:2009af}. 

In the HERA kinematic domain, where the virtuality $Q^2$ of the exchanged boson is small, $Q^2 \ll M_Z^2$, 
charm production is dominated by virtual photon exchange, where the $xF_3^{c \bar{c}}(x,Q^2)$ is negligible.  Therefore the neutral current deep-inelastic $ep$ cross section by considering the running electromagnetic coupling, $\alpha(Q^2)$, without QED and electro-weak radiative corrections, may be written in terms of the structure functions $F^{c\bar {c}}_2(x,Q^2)$ and $F^{c\bar c}_L(x,Q^2)$. Since the $F_L^{c\bar{c}}$ contribution is suppressed only at low rapidity $y$ \cite{Daum:1996ec} at HERA, it is possible to neglect this contribution. 
So the important contribution to the reduced charm cross section is the charm structure function, $F_2^{c\bar{c}}$. Therefore additional correction must be done in this contribution in the presence of IC contribution.
The importance of this kind of data set is due to the existence of charm PDFs, where the IC contribution can be included.

Since the main goal of this paper is to determine the intrinsic charm probability in the proton, we need to include the charm structure function data in the high $x$ region, where the intrinsic charm is dominant. Therefore, in this analysis, we choose the EMC measurement of the large $x$ charm structure function as a first experimental evidence of IC.
These data are produced in inclusive dimuon and trimuon iron target. Since the EMC $F_2^c$ data are extracted using the nuclear target, we need to consider the nuclear corrections in the present QCD analysis, as we will explain in the next section. 

Although the fixed-target Drell-Yan data \cite{Towell:2001nh} are sensitive to the distributions of anti-quark in large $x$ and can be used in the QCD global analysis,  we did not find a significant impact on PDF parametrization by including this kind of dataset in the presence of IC contribution.  On the other hand, in our previous analysis~\cite{Abdolmaleki:2017wlg}, we have also investigated that including the jet and $W$, $Z$ boson data at the LHC \cite{Aad:2013lpa,Aad:2013iua,Aad:2014qja,Aad:2011dm} do not any impacts on the PDFs in the presence of IC contribution. So we can exclude these  data sets in the present analysis.

In this analysis, we apply the  $(W^2-W_{th}^2)/W^2$ with $W_{th}^2 = 16$ on heavy quark structure function to suppress charm production near threshold as suggested in Ref. \cite{bt}.  Note that,  we exclude the data using  $Q^2 < 4 $ GeV$^2$ and $W < 15$ GeV$^2$ cuts,  where the higher twist correction might become relevant. So the total number of data points in our QCD analysis with considering these cuts are reduced from 1939 to 1535.

\section{\label{PDF}PDF parametrization}
In this article, we perform our QCD analysis using the $\mathsf{xFitter}$ open source framework \cite{xFitter,Alekhin:2014irh}, which previously was known as HERAfitter \cite{Sapronov}. Very recent analyses using $\mathsf{xFitter}$ package are reported in Refs.~\cite{Abdolmaleki:2018jln,vafaee:2017nze}. 

The numerical solutions of the DGLAP evolution equations for PDFs in pQCD framework at NLO are implemented in the QCDNUM package in $x$-space \cite{Botje:2010ay}. To compute the perturbative part of heavy quark contributions of the DIS structure-function, we used the optimal Thorne-Roberts (RT-opt) \cite{Thorne:1997ga}, General-Mass Variable Flavor Number (GM-VFN) scheme. This scheme assumes that the charm distribution is generated perturbatively by gluon and light quark splittings, and its value depends strongly on the charm mass. If the PDFs are parameterized as a function of $x$ at initial scale $Q^2_0$, the QCD evolution will help us to achieve it at any value of $Q^2$. In the present analysis, the initial QCD scale is chosen to be $Q_0^2 = 1.9$  GeV$^2$, which is below the charm threshold.  In this approach, we choose the heavy quark masses $m_c = 1.50$ GeV and $m_b = 4.5$ GeV, and we take $\mu_r = \mu_f = Q$ for the QCD re-normalization and factorization scale. 

We choose a standard parametrization for the PDFs at the input scale  $Q_0^2=1.9$  GeV$^2$ \cite{Abramowicz:2015mha} to be:
\begin{eqnarray}
xf_i(x,Q^2_0) &=  & A_i x^{B_i}  (1-x)^{C_i}\left(1+Dx + Ex^2\right)~. 
\label{eq:xpar}
\end{eqnarray}
The parametrised PDFs are the valence distributions, $xu_v$, $xd_v$, the u-type and d-type anti-quarks distributions, $x\bar{U}$, $x\bar{D}$, and the gluon distribution, $xg$. We assume the relations $x\bar{U} = x\bar{u}$ and $x\bar{D} = x\bar{d} + x\bar{s}$ at the starting scale.

In summary, our parametrisation is:
\begin{eqnarray}
xu_v(x) &=  & A_{u_v} x^{B_{u_v}}  (1-x)^{C_{u_v}}\left(1+E_{u_v}x^2 \right)~, \nonumber\\
xd_v(x) &=  & A_{d_v} x^{B_{d_v}}  (1-x)^{C_{d_v}}~,\nonumber\\
x\bar{U}(x) &=  & A_{\bar{U}} x^{B_{\bar{U}}} (1-x)^{C_{\bar{U}}}\left(1+D_{\bar{U}}x\right)~, \nonumber\\
x\bar{D}(x)&=&A_{\bar{D}} x^{B_{\bar{D}}} (1-x)^{C_{\bar{D}}}~, \nonumber\\
xg(x) &=   & A_g x^{B_g} (1-x)^{C_g} - A_g' x^{B_g'} (1-x)^{C_g'}~.
\label{eq:xpar-general}
\end{eqnarray}
In the above equations, to ensure the same behavior of the $x\bar{u}$ and $x\bar{d}$  as $ x\rightarrow 0 $, one can impose the additional constraints $B_{\bar{U}}=B_{\bar{D}}$  and $ A_{\bar{U}}=A_{\bar{D}}(1-f_s)$ \cite{Abramowicz:2015mha}. In this analysis, we fixed the $D_{\bar{U}}$ parameter after the first minimization because, in the presence of IC contribution, the selected DIS data set does not constrain the $D_{\bar{U}}$ well enough. The strange-quark distribution is expressed as an $x$-independent fraction, $f_s$, of the $d$-type sea, $x\bar{s}= f_sx\bar{D}$, at $Q_0^2$. The value $f_s=0.31$ was chosen in the strange quark density as suggested in Ref.~\cite{Mason:2007zz}.  For the gluon PDF,  $C'_g$ is fixed to $C'_g=25$ to ensure a positive gluon density at large $x$, as suggested in Ref.~\cite{Martin:2009iq}. 

The normalization parameters for gluon and valence distributions, $A_g$, $A_{u_v}$ and $A_{d_v}$ are constrained by the fermion number and momentum sum rules,
\begin{eqnarray}
\int_0^1u_{v}dx=2~, \int_0^1d_{v}dx=1~,
\end{eqnarray}
\begin{eqnarray}
\int_0^1 x[g &+& \sum_i (q_i+\bar{q}_i) + c_{ext} + \bar{c}_{ext} +c_{int} + \bar{c}_{int}] dx = 1~.\nonumber \\
\end{eqnarray}

In fact, in the above sum rule the total intrinsic charm quark momentum fraction is    
\begin{equation}
\int_0^1 x(
c_{int} + \bar{c}_{int})dx\equiv <x>_{c+\bar{c}}~.
\label{eq:m fraction}
\end{equation}
Notice that in Eq.~(\ref{heavyBHPS}) we can fix the $P_{c\bar{c}/p}$ value with 1\%, which is only included in the intrinsic part of above equation, or can be considered as a free parameter. It should be noted that the extrinsic charm generated perturbatively  using the DGLAP evaluation equation and the intrinsic charm evolve by a non-singlet evaluation equation \cite{Rostami:2015iva}. 

In this analysis, we use the nuclear correction on the EMC data because these data come from a nuclear target.
This correction creates a connection between the parton distributions in the nucleus $A$ and parton distribution in the proton which we model as:
\begin{equation}
f_i^A(x,Q^2)= R_i(x,A,Z)f_i(x,Q^2)~,
\label{eq:nuclear}
\end{equation} 
where $f_i^A(x,Q^2)$ is the parton distribution with type $i$ in the nucleus and $f_i(x,Q^2)$ is the corresponding parton distribution in the proton. $A$ and $Z$ are the mass number and atomic number, respectively. To study the impact of the EMC data on the PDFs behavior considering IC contributions, we apply the nuclear correction factor from Ref.~\cite{deFlorian:2011fp} on EMC data.

By comparing the theoretical and experimental measurements of various physical observables, we determine the unknown PDFs parameters by minimizing the $\chi^2$ function taking into account the correlated and uncorrelated measurement uncertainties. 
The $ \chi^2$ function is minimized using the MINUIT package \cite{James:1975dr}, and is defined as \cite{Abdolmaleki:2018jln,Aaron:2009aa}
\begin{equation}
         \chi^2 = \sum_{i,j} (t_i -  d_i)(C^{-1})_{i,j}(t_j - d_j)~,
 \label{eq:chi2-cov}
\end{equation}
where $C^{-1}_{ij}$ is the covariance matrix. If the full correlated uncertainties of the experimental data are available, the $\chi^2$ function is  as follows
\begin{equation}
         \chi^2 = \sum_i \frac{ [d_i -  t_i(1-\sum_j \beta_{j}^{i} s_j )]^2 }{\delta^2_{i,unc} t^2_i +\delta^2_{i,stat} d_i t_i } + \sum_j s^2_j~,
 \label{eq:chi2}
\end{equation} 
where  $t_i$ is the theoretical prediction, $d_i$ is the measured value of the $i$-$th$ data point, $(Q^2,x,s)$, and  $\delta_{i,stat}$ and $\delta_{i,unc}$ are the relative statistical and uncorrelated systematic
uncertainties. In the above, $\beta^i_j$ are the corresponding systematic uncertainties, in which case  $s_j$ are the nuisance parameters associated with the correlated systematic error.  Here $j$ labels the sources of correlated systematic uncertainties, and in the Hessian method the $s_j$ are not fixed. If the $s_j$ are fixed to zero, the correlated systematic errors are ignored.

The Hessian method for the PDF uncertainties are obtained from $\Delta \chi^2 = T^2$. A tolerance parameter $T$ is selected such that the criterion $\Delta \chi^2 = T^2$ ensures that each data set is described within the desired confidence level.
The correlated statistical error on any given quantity $q$  is then obtained from the standard error propagation:
	\begin{equation}
	(\sigma_{q})^2 = \Delta \chi^2 \left( \sum_{\alpha,\beta} \frac{\partial q}{\partial p_{\alpha}}C_{\alpha,\beta} \frac{\partial q}{\partial p_{\beta}} \right)~. 
	\label{eqn:error_propagation}
	\end{equation}
The Hessian matrix is defined as ${ H_{\alpha,\beta} = \frac{1}{2} \partial^2 \chi^2 / \partial p_{\alpha} \partial p_{\beta}}$, and thus the covariance matrix $ C = H^{-1}$ is the inverse of the Hessian matrix evaluated at the $\chi^2$ minimum. In order to be able to calculate the fully correlated  1$\sigma$ error bands for the valence PDFs, one can choose $T=1$ in the $\mathsf{xFitter}$ package.

Indeed, we have 13 unknown PDFs parameters in addition to free parameters $\alpha_s$ and ${P}_{c{\bar c/p}}$ which are obtained from the fit.

\section{\label{fit results}Fit results}

\begingroup
\squeezetable
\begin{table*}

\caption{\label{tdata}The list of observable and experimental data with detailed information which we used in our QCD analysis. For each data sets, we indicate the number of data points and their $x$ and $Q^2$ kinematic ranges.}
\begin{ruledtabular}
\begin{tabular}{llllcc}

      Observable
  &  ~~~~~~~~Experiment   & Ref.  & \# Data  & $x$ & $Q^2$ [ GeV$^2$] \\
\cline{1-6}

 DIS $\sigma$  
  & ~~~~~~~~HERA1+2 CC $e^+p$ &\cite{Abramowicz:2015mha}&39 &[$8.0\times 10^{-3}$-0.4]&[300-30000] \\ 
  & ~~~~~~~~HERA1+2 CC $e^-p$ &\cite{Abramowicz:2015mha}&42 &[$8.0\times 10^{-3}$-0.65]&[300-30000]\\ 
  & ~~~~~~~~HERA1+2 NC $e^-p$ &\cite{Abramowicz:2015mha}&159 &[$8.0\times 10^{-4}$-0.65]&[60-50000]\\ 
  & ~~~~~~~~HERA1+2 NC $e^-p$ 460 &\cite{Abramowicz:2015mha} &200 &[$3.48\times 10^{-5}$-0.65]&[1.5-800]\\ 
  & ~~~~~~~~HERA1+2 NC $e^-p$ 575 &\cite{Abramowicz:2015mha}&249&[$3.48\times 10^{-5}$-0.65]&[1.5-800]\\ 
  & ~~~~~~~~HERA1+2 NC $e^+p$ 820 &\cite{Abramowicz:2015mha}&68 &[$6.21\times 10^{-7}$-0.4]&[0.045-30000]\\ 
  & ~~~~~~~~HERA1+2 NC $e^+p$ 920 &\cite{Abramowicz:2015mha}&363 &[$5.02\times 10^{-6}$-0.65]&[1.5-30000]\\ 
  DIS $\mathrm{F_2^p}$  
  & ~~~~~~~~BCDMS $\mathrm{F_2^p}$ 100 GeV &\cite{Benvenuti:1989rh} &83&[0.07-0.75]&[7.5-75]\\ 
  & ~~~~~~~~BCDMS $\mathrm{F_2^p}$ 120 GeV &\cite{Benvenuti:1989rh}  &91 &[0.07-0.75]&[8.75-99]\\ 
  & ~~~~~~~~BCDMS $\mathrm{F_2^p}$ 200 GeV &\cite{Benvenuti:1989rh} &79  &[0.07-0.75]&[17-137.5]\\ 
  & ~~~~~~~~BCDMS $\mathrm{F_2^p}$ 280 GeV &\cite{Benvenuti:1989rh}  &75 &[0.1-0.75]&[32.5-230]\\
  & ~~~~~~~~SLAC $\mathrm{F_2^p}$ & \cite{Whitlow:1991uw} &24 &[0.07-0.85]&[0.59-29.2]\\ 
  DIS $\sigma^{c\bar{c}}$
  & ~~~~~~~~Charm cross section H1-ZEUS combined & \cite{Abramowicz:1900rp} &47 &[1.8$\times10^4$-0.025]&[5.0-120]\\   
  DIS $\mathrm{F_2^c}$
  & ~~~~~~~~EMC $\mathrm{F_2^c}$ &\cite{Aubert:1982tt}&16 &[0.0075-0.421]&[2.47-78.1]\\ 

\end{tabular}
\end{ruledtabular}
\end{table*}
\endgroup

\begingroup
\squeezetable
\begin{table*}

\caption{\label{tIC}
The results for the $\chi^2 $/number of points, correlated $\chi^2$, and the total $\chi^2$/ degree of freedom (dof)  values of each data sets for different Base, and Fits A, B, C. The nuclear effects are considered for Fits.~A, B and C and the intrinsic charm contribution is included for Fits~B and C.} 

\begin{ruledtabular}
\begin{tabular}{lllll}
Experiment & \multicolumn{4}{c}{$\chi^2$/number of points}    \\ 
\cline{2-5}
      
& Base &  Fit~A & Fit~B & Fit~C\\
\hline
  
  HERA1+2 CC $e^+p$ & 50 / 39& 69 / 39& 56 / 39& 55 / 39
 \\
  HERA1+2 CC $e^-p$ & 55 / 42& 54 / 42& 55 / 42& 56 / 42 \\ 
  HERA1+2 NC $e^-p$ & 236 / 159& 242 / 159& 239 / 159& 239 / 159  \\ 
  HERA1+2 NC $e^-p$ 460 &  213 / 200& 217 / 200& 213 / 200& 212 / 200  \\ 
  HERA1+2 NC $e^-p$ 575 & 219 / 249& 225 / 249& 221 / 249& 220 / 249  \\ 
  HERA1+2 NC $e^+p$ 820 &72 / 68& 72 / 68& 72 / 68& 72 / 68 \\ 
  HERA1+2 NC $e^+p$ 920 &461 / 363& 482 / 363& 474 / 363& 476 / 363  \\ 
  BCDMS $\mathrm{F_2^p}$ 100~GeV & 83 / 83& 113 / 83& 95 / 83& 95 / 83   \\ 
  BCDMS $\mathrm{F_2^p}$ 120~GeV & 70 / 91& 79 / 91& 73 / 91& 72 / 91   \\ 
  BCDMS $\mathrm{F_2^p}$ 200~GeV & 90 / 79& 105 / 79& 96 / 79& 95 / 79 \\ 
  BCDMS $\mathrm{F_2^p}$ 280~GeV & 68 / 75& 74 / 75& 71 / 75& 72 / 75  \\ 
  SLAC $\mathrm{F_2^p}$ & 93 / 24& 41 / 24& 59 / 24& 59 / 24 \\ 
  Charm cross section H1-ZEUS combined & 43 / 47& 44 / 47& 44 / 47& 44 / 47   \\ 
  EMC $\mathrm{F_2^c}$ & - & 102 / 16& 73 / 16& 70 / 16  \\ 
 \hline
  Correlated $\chi^2$  & 120& 128& 120& 122  \\
Total $\chi^2$   & 1872 & 2047 & 1959 & 1957   \\ 
  Total $\chi^2$ / dof  & 1872 / 1506& 2047 / 1522& 1959 / 1521& 1957 / 1520  \\

\end{tabular}
\end{ruledtabular}
\end{table*}
\endgroup

The main goal of this paper is the simultaneous determination of the intrinsic charm probability ${P}_{c{\bar c/p}}$ and strong coupling $\alpha_s$. This allows us to study the interaction of these two quantities as well as the influence on the PDFs in the presence of IC contributions. 

For fit analysis, we will include the DIS reduced cross section and differential cross sections data from HERA I+II, the charm combined cross sections data from H1 and ZEUS, and the proton structure function data from BCDMS, SLAC, and EMC. The detailed information and references, number of data points, and  their kinematic range of $x$ and $Q^2$ for each data set are summarized in Table~\ref{tdata}.

To investigate the effect of the EMC data in the present analysis with and without IC contributions, we will divide our QCD analysis into four fits: 

\begin{itemize}
\item \textbf{Base}: We include all the data sets of Table~\ref{tdata} with \emph{exception} of the EMC experimental data, this totals 1519 data points. In the Base fit we consider zero IC contribution, $i.~e.$, $P_{c\bar{c}/p}\equiv0$. We fixed $\alpha_s(M^2_Z) = 0.1182 $ considering to the world average data as a first step. 
\item
\textbf{Fit~A}: We include all the data from the Base fit in addition to the EMC $F_2^c$ data and this gives us 1535 data point. This fit also assumes zero IC contribution and we use a fixed $\alpha_s(M^2_Z) = 0.1182$.

\item
\textbf{Fit~B}: We use all the data sets of the Table~\ref{tdata} and we use a fix $\alpha_s(M^2_Z) = 0.1182$ parameter and include an intrinsic charm probability $P_{c\bar{c}/p}$ value as a free parameter. 

\item
\textbf{Fit~C}: This is all same as Fit~B, except now use a free $\alpha_s(M^2_Z)$.
\end{itemize}

The $\chi^2$/number for each data set after cuts, the total  $\chi^2$, and $\chi^2 / dof$ for all fits are summarized in Table ~\ref{tIC}. According to Table~\ref{tIC} and without taking into account IC contribution in Base fit, as a first step,  the total $\chi^2/dof$ value is $1872/1506=1.243$. There are 13 unknown parameters for PDFs only.

As a second step and the same as Base fit, we have 13 unknown parameters for PDFs  in Fit. A.
In this case and according to Table~\ref{tIC}, we obtain $2047/1522=1.345$ for the total $\chi^2/dof$ for Fit~A.

For a more complete discussion, the investigation of necessity to include a non-zero $P_{c\bar{c}/p}$ in the present QCD analysis and also the effect of including the EMC data would be important. In Fit~B, as a third step, we obtain $1959/1521=1.289$ for the total of $\chi^2/dof$ for Fit~B.  There are 13 unknown parameters for PDFs and an unknown parameter for intrinsic charm probability, therefore the total number of unknown parameters is 14.

Our motivation to present Base, Fit~A and Fit~B  results is to show that an IC component is unnecessary as long as the EMC data remains excluded. 

Finally,  in  Fit~C we  consider the intrinsic charm probability  $P_{c\bar{c}/p}$ and  $\alpha_s(M^2_Z)$ as  free parameters in our QCD analysis, so there are 13 unknown parameters for PDFs and two unknown parameters for strong coupling constant and intrinsic charm probability. In Fit~C, the total number of unknown parameters is 15, which should be obtained by the QCD fit on experimental data. In this fit, we obtain $1957/1520=1.287$  for the total $\chi^2/dof$ with considering intrinsic charm probability  $P_{c\bar{c}/p}$ and 
$\alpha_s(M^2_Z)$, as the free parameters. 

It should be noted that, in the present analysis and for Fit~A, B, and C, the nuclear effects are considered. Although the Fit~C contains the complete analysis,  the comparison of this fit with other above cases with each other would be interested.

Obviously, in comparison Fit~A with Fits~B and C, there are  almost 4\% improvements for $\chi^2$ values and the fit quality.

In Ref.~\cite{Ball:2016neh,Rottoli:2016lsg}, the NNPDF collaboration studied the influence of the EMC data in their analyses with and without the IC contribution. They found $\chi^2$ per point $ = 7.3$  for the EMC data if the charm PDF was generated perturbatively, and  $\chi^2$ per point $ = 4.8$ when the IC contribution was included and the EMC data was rescaled. They improved the fit quality of the EMC data by imposing an additional cut of $x>0.1$ when the charm PDF was fitted.

According to Table.~\ref{tIC}, we note that the  $\chi^2$ per point for the EMC data are 6.37, 4.56 and 4.37 in  Fits~A, B and C, respectively. When the IC  includes in Fits~B and C, we find an improvement in the $\chi^2$ per point for the EMC data of 28\% and 31\%, respectively in comparison with Fit~A. 

\begin{figure*}
\begin{center}
\includegraphics[scale=.58]{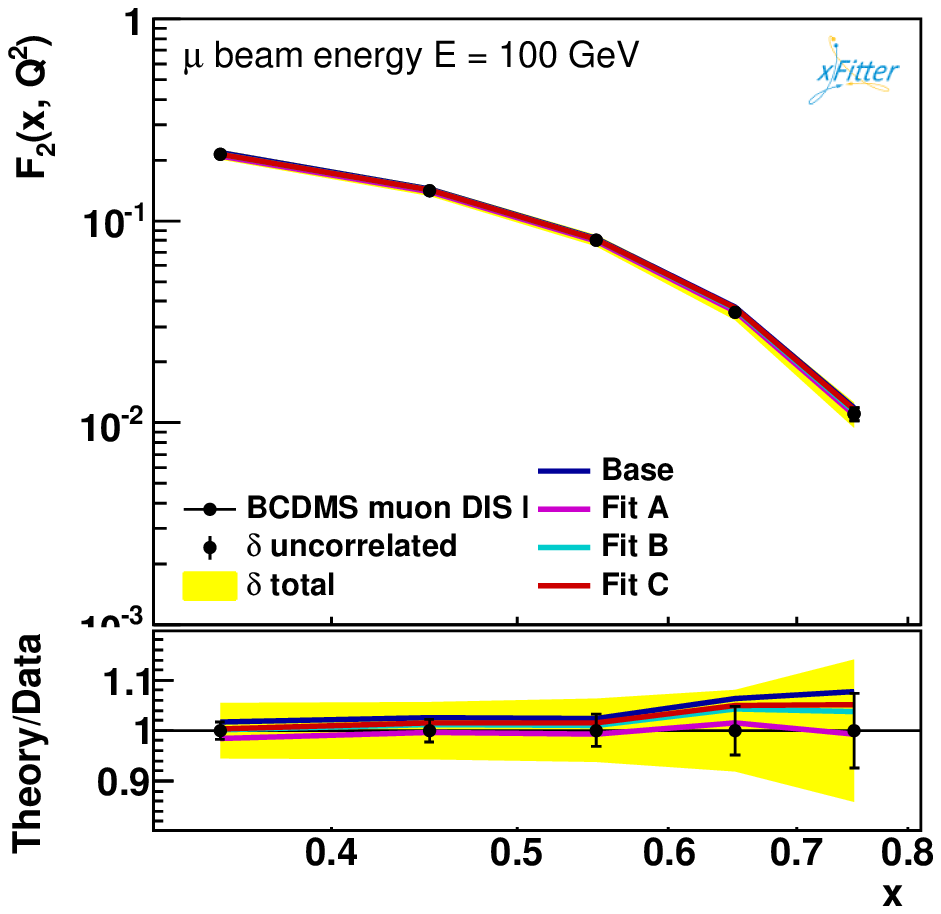}
\includegraphics[scale=.58]{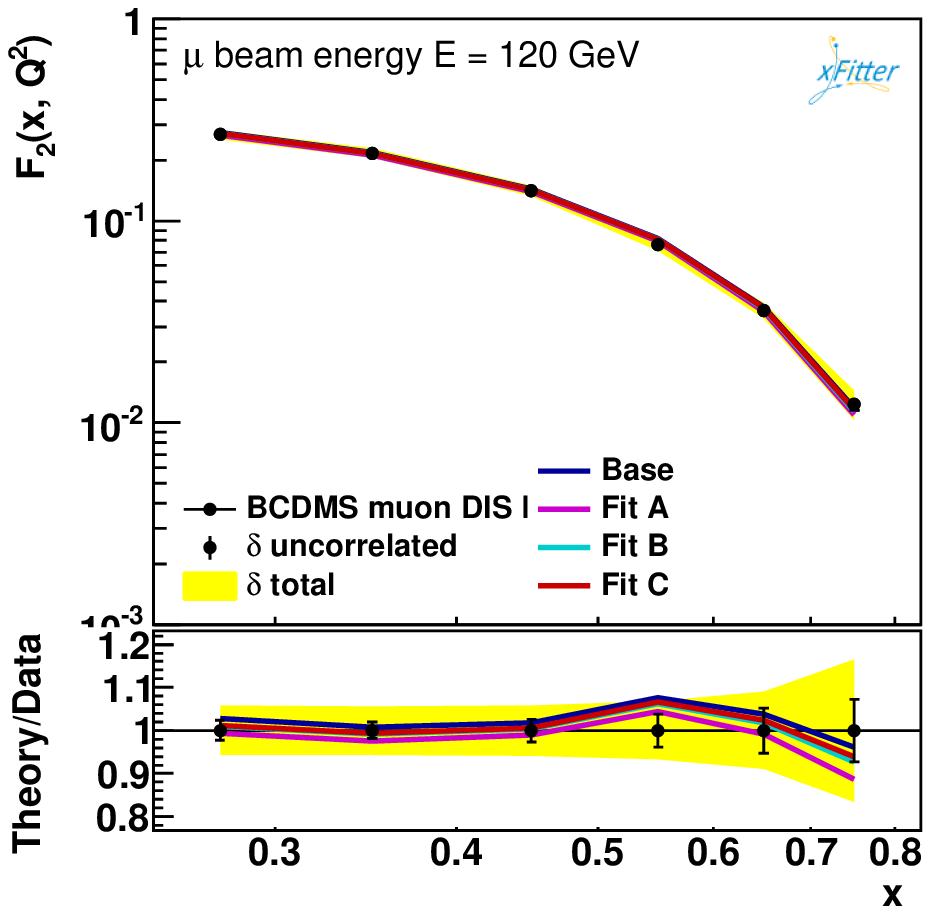}
\includegraphics[scale=.58]{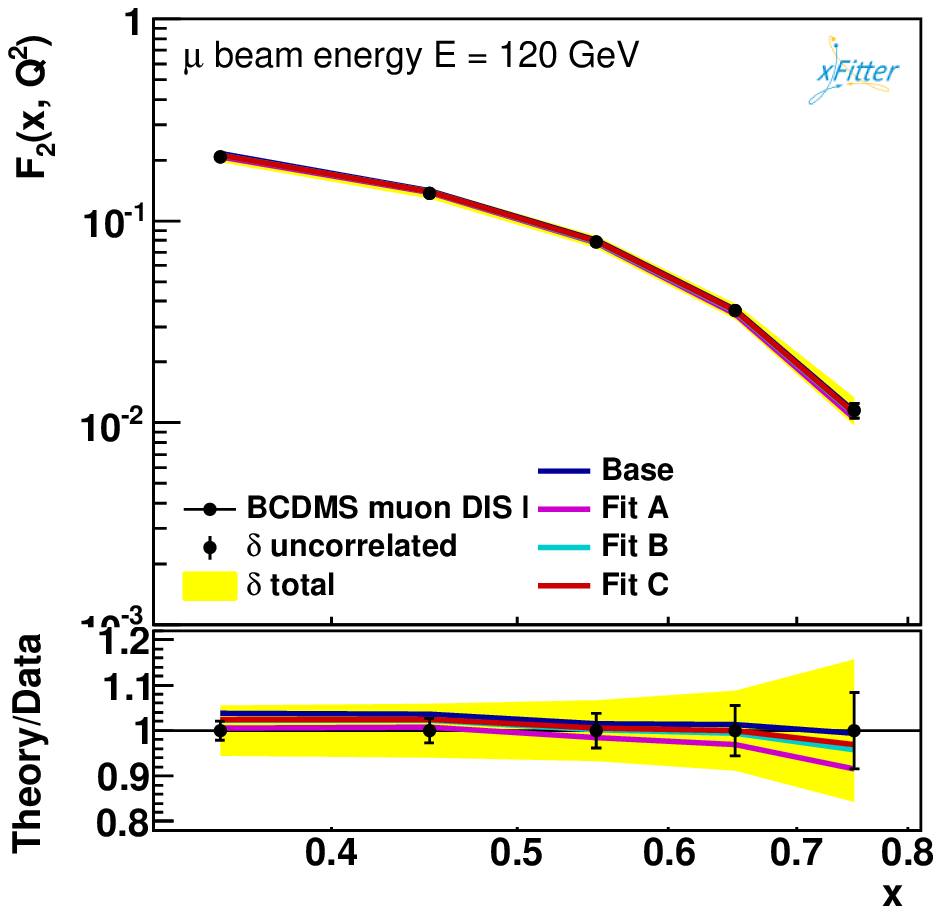}\\

\includegraphics[scale=.57]{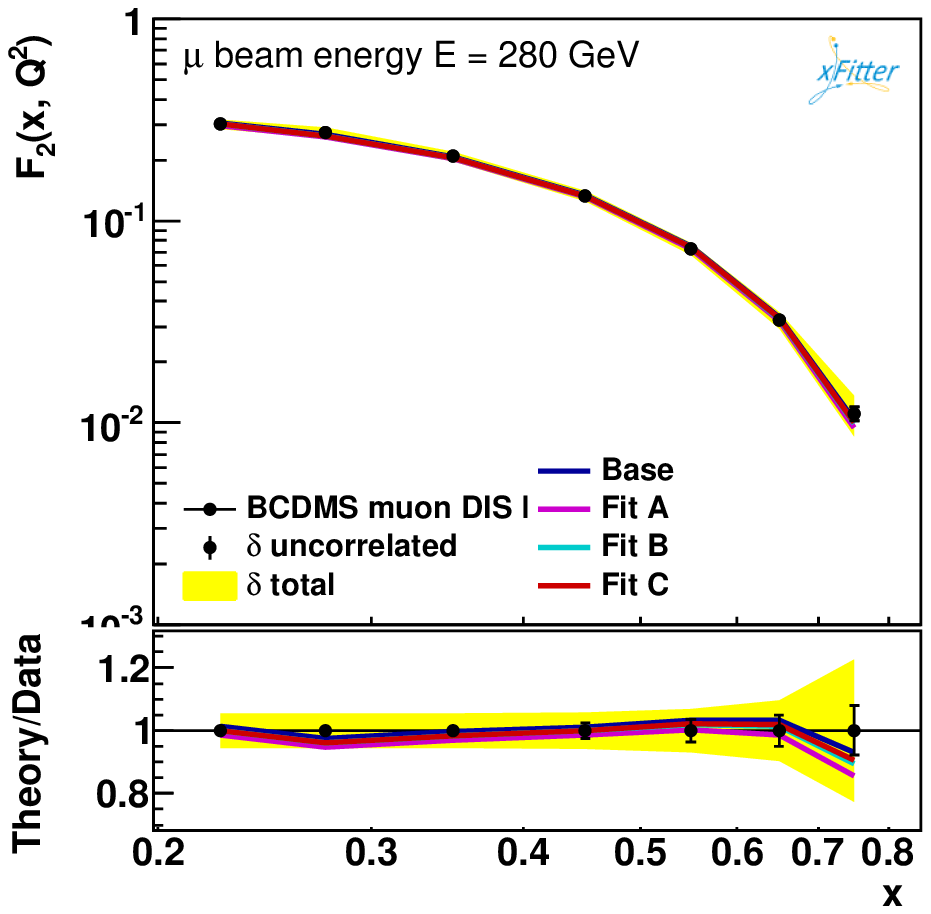}
\includegraphics[scale=.57]{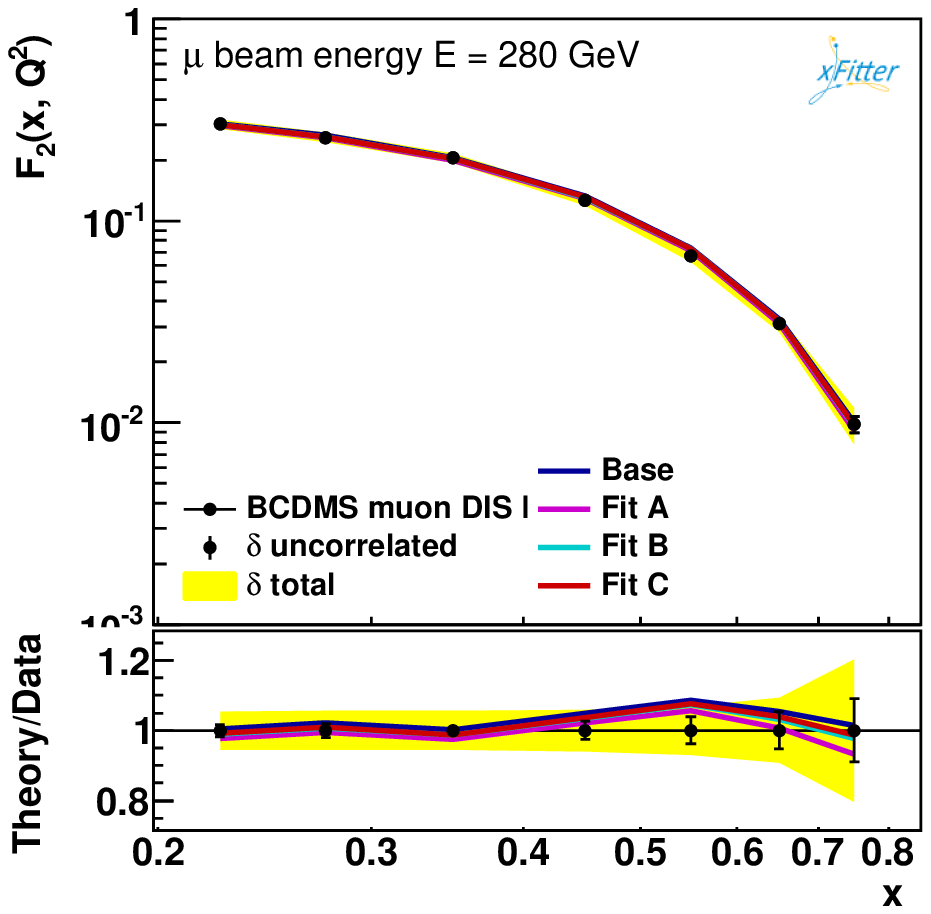}
\includegraphics[scale=.57]{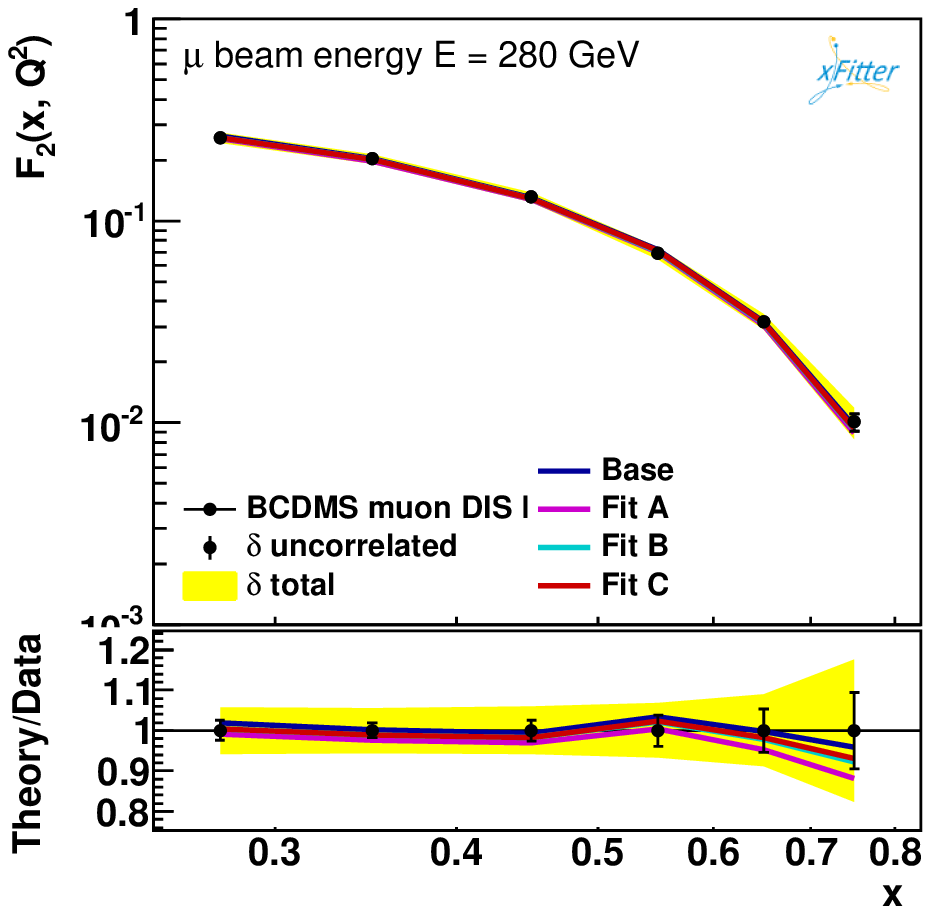} \\

\includegraphics[scale=.57]{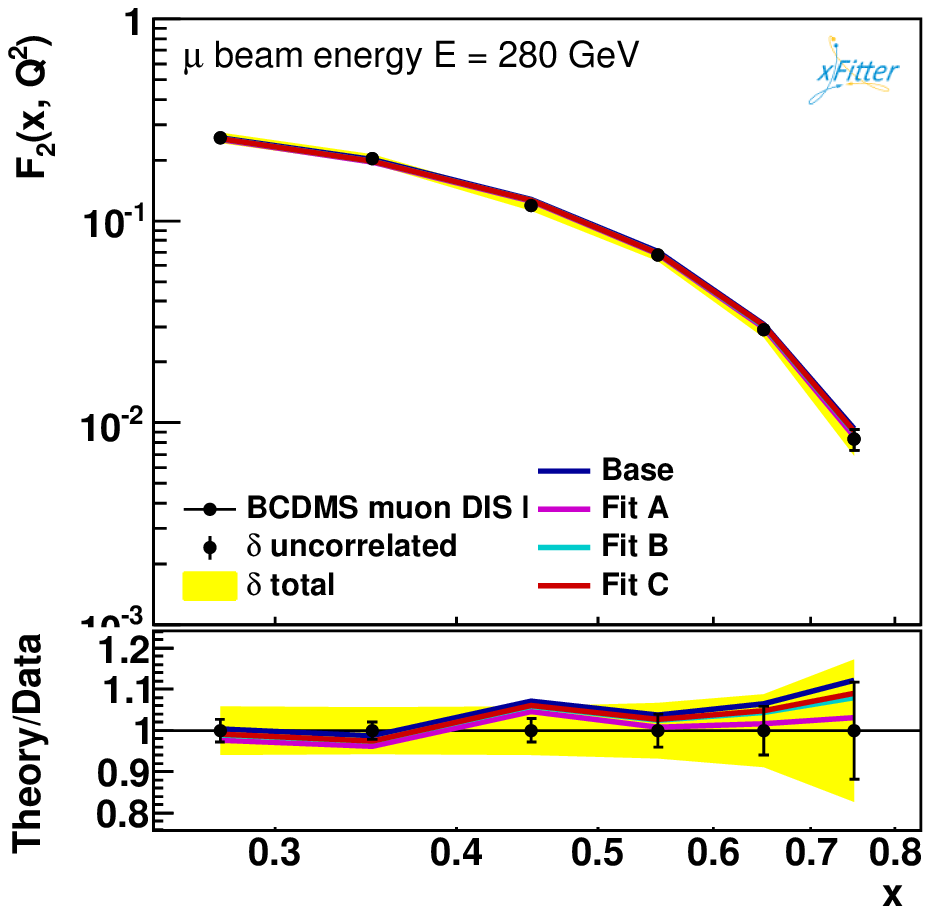}
\includegraphics[scale=.57]{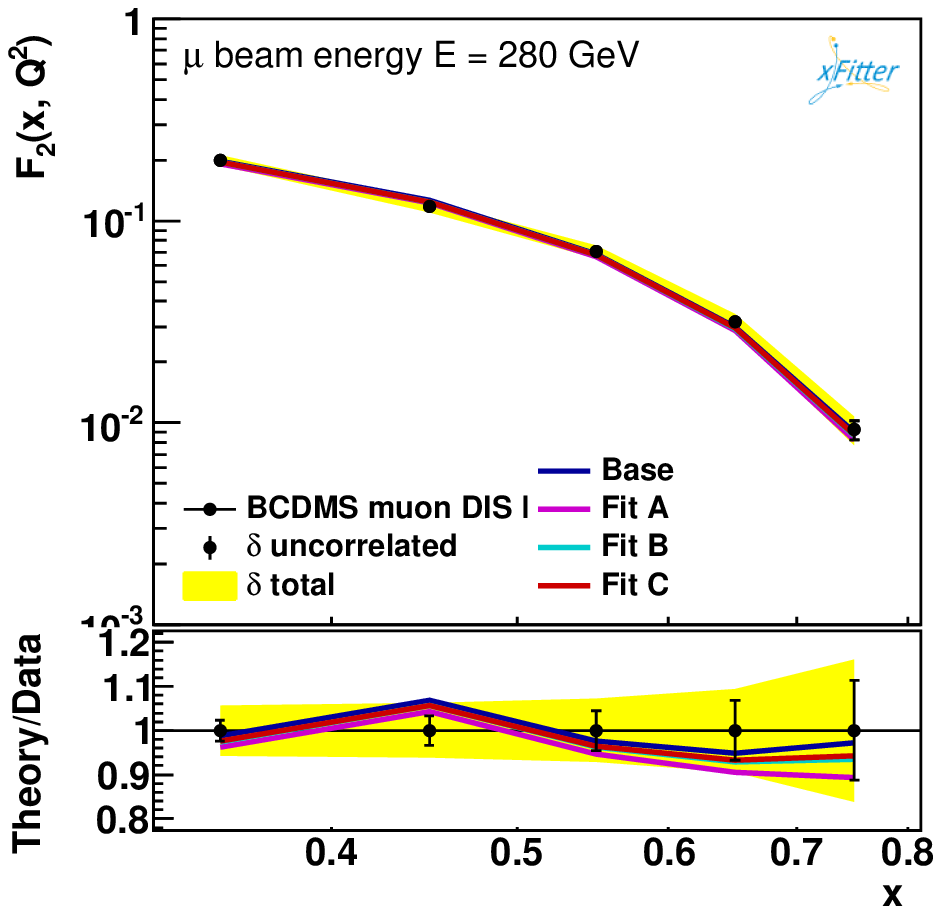}
\includegraphics[scale=.57]{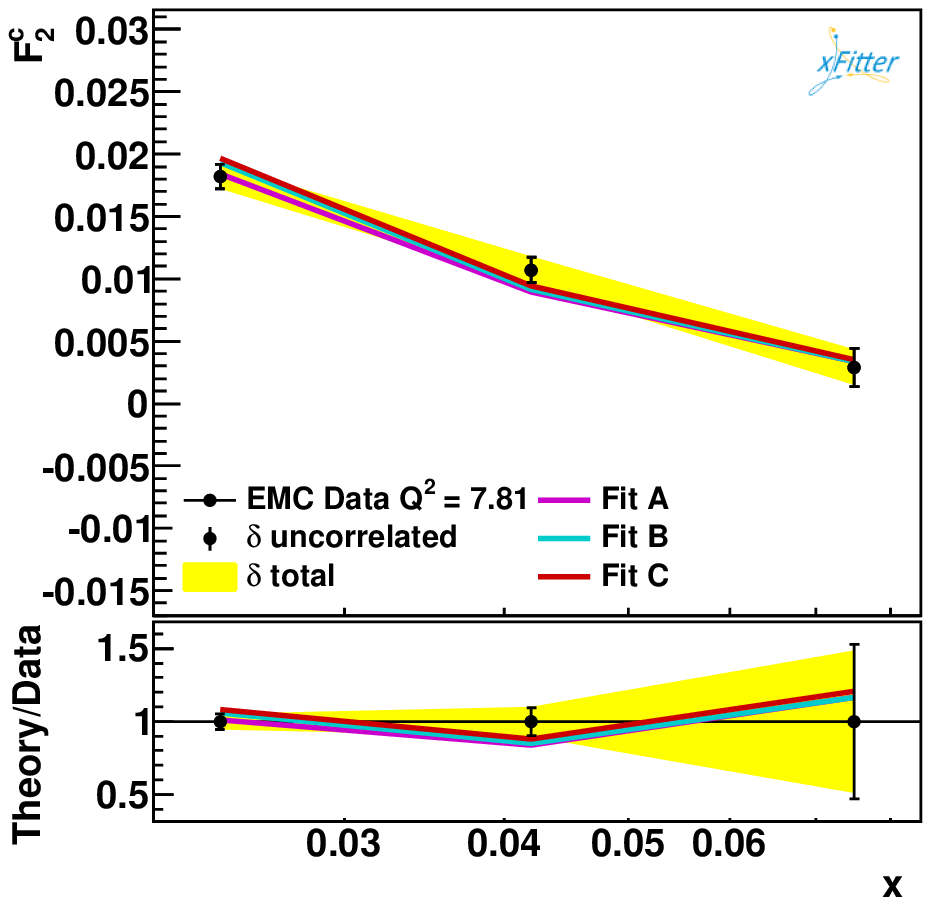} \\
    \subfigure[BCDMS and EMC ]{\label{pic:QCD-Fit-1-b}
\includegraphics[scale=.57]{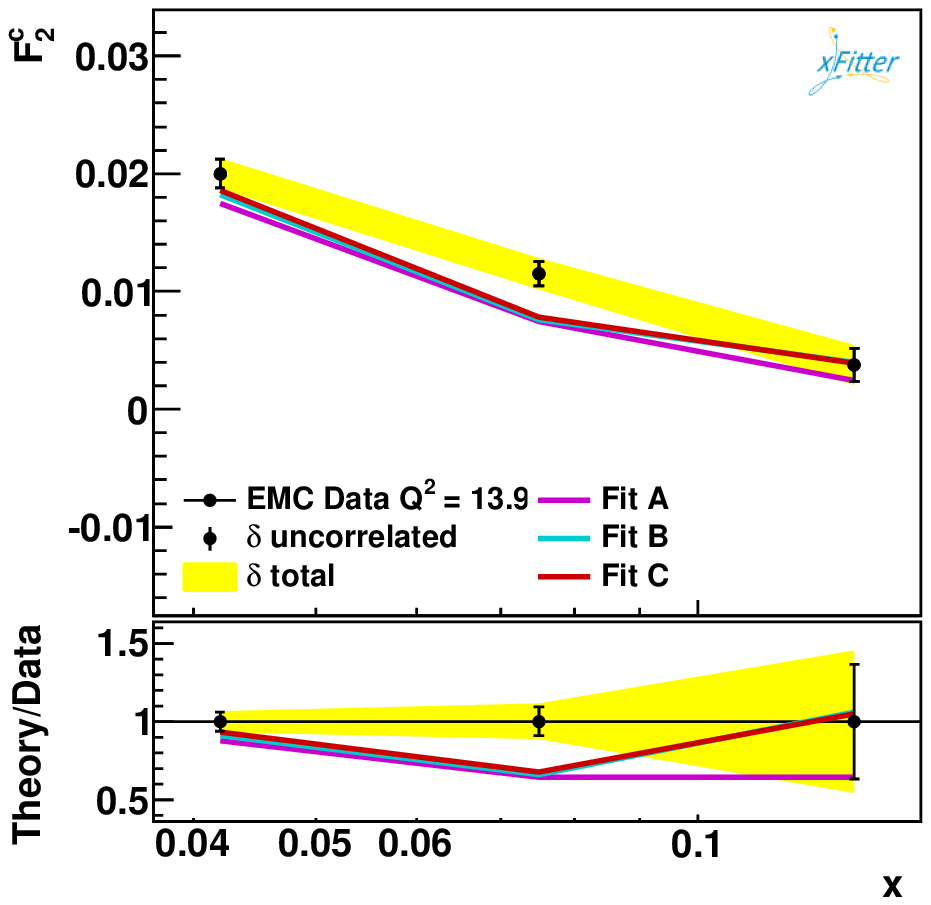}
\includegraphics[scale=.57]{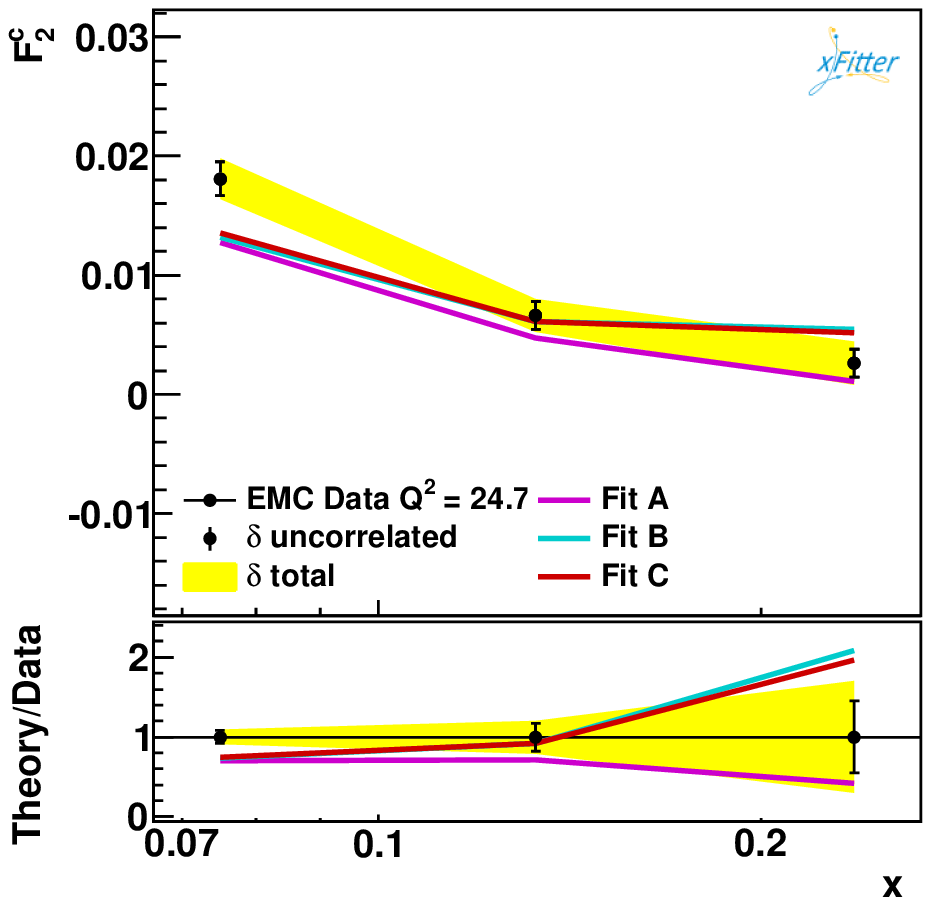}
\includegraphics[scale=.57]{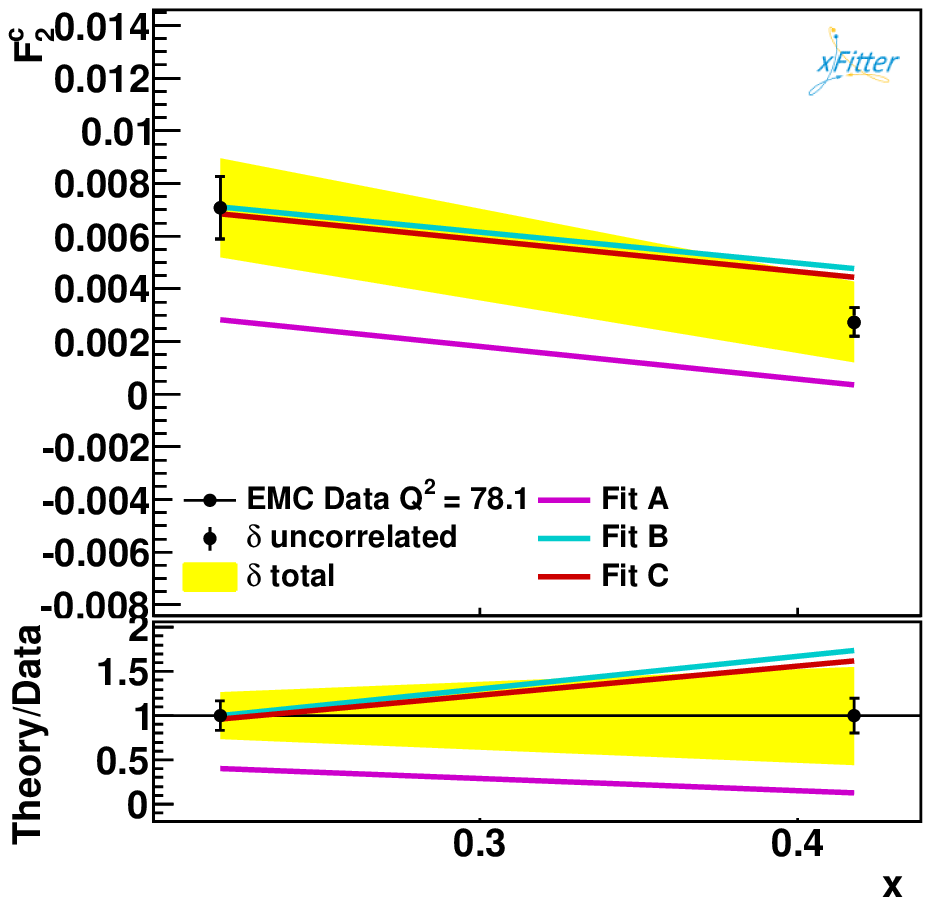}}
  \end{center}
\caption{Our theoretical predictions of the reduced cross section and proton structure function which are obtained from the QCD fits as a function of the momentum fraction $x$ at different values of $Q^2$ for our Base, Fits~A, B and C. We compare to the experimental data from HERA I+II \cite{Abramowicz:2015mha}, the charm cross section from the combined the H1 and ZEUS \cite{Abramowicz:1900rp} (Fig.\ref{pic:QCD-Fit-1}-a), the proton structure function from SLAC \cite{Whitlow:1991uw} and  BCDMS \cite{Benvenuti:1989rh}, and the charm structure function from EMC data  \cite{Aubert:1982tt} (Fig.\ref{pic:QCD-Fit-1}-b).}\label{pic:QCD-Fit-1}

\end{figure*}

To investigate the specific impact of the EMC data with and without IC contribution at NLO, we need to compare our results for our individual fits: Base, Fits~A, B, and C.  In Fig.~\ref{pic:QCD-Fit-1}, we display some samples of our theoretical predictions for all fits in comparison to fixed target DIS data from HERAI+II \cite{Abramowicz:2015mha}, H1-ZEUS combined charm cross section \cite{Abramowicz:1900rp} (Fig.\ref{pic:QCD-Fit-1}-a), BCDMS \cite{Benvenuti:1989rh}, SLAC \cite{Whitlow:1991uw} and EMC \cite{Aubert:1982tt} measurements (Fig.\ref{pic:QCD-Fit-1}-b) and their uncertainties at NLO as a function of $x$ for different values of $Q^2$. According to Fig.~\ref{pic:QCD-Fit-1-b}, the data description turns out to be better including the IC component in comparison  with the Fit.~A results. In fact, we find that in general, it is necessary to include a non-zero $P_{c\bar{c}/p}$ to have a better description of the EMC data.

In Table~\ref{PARIC},  the numerical QCD fit results for the PDFs parametrization according to Eq.~(\ref{eq:xpar-general}),  $\alpha_s$ value  and the intrinsic charm probability ${P}_{c{\bar c/p}}$ are summarized in four separate cases.

It is obvious that the ${P}_{c{\bar c/p}}$ value has significant sensitivity to the $\alpha_s$ value.  
The simultaneous determination of ${P}_{c{\bar c/p}}$ and $\alpha_s$, as shown in Table \ref{PARIC}, gives a somewhat lower
IC probability and an $\alpha_s$ which is more or less in agreement with the world average value of $\alpha_{s}(M_{Z}^{2})=0.1182\pm0.0011$~\cite{Patrignani:2016xqp}. This
is a very reasonable result based on the fact that IC is a non-perturbative phenomenon.

According to Tables \ref{tIC} and \ref{PARIC}, the increase of 8\% of $\chi^2/dof$ in Fit~A with respect to our Base fit, and $\sim4$\% improvement  of $\chi^2/dof$ in Fit~B and Fit~C with respect to Fit~A, is obtained.  Conversely, we observed the significant changes in both, the PDFs parameters and their uncertainties. 
Using the definition of $\Delta{{P}_{c{\bar c/p}}} = {P}_{c{\bar c/p}}^{Fit.C}-{P}_{c{\bar c/p}}^{Fit.B}$, we find a change of $\sim$7.6\% on $\Delta{{P}_{c{\bar c/p}}}/{P}_{c{\bar c/p}}^{Fit.B}$.
It is clear that a simultaneous determination of ${P}_{c{\bar c/p}}$ and $\alpha_s$ in the present analysis can impact the central value of the intrinsic charm probability.

\begin{figure*}
 \begin{center}
\includegraphics[width=0.32\textwidth]{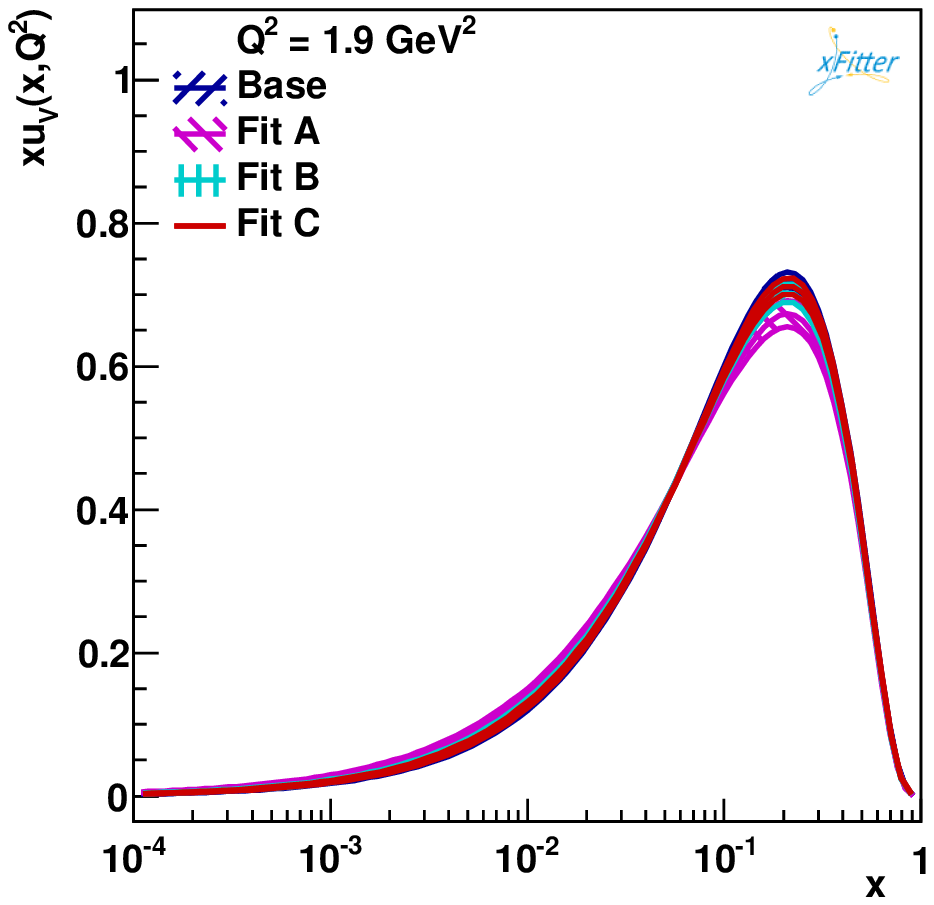}
\includegraphics[width=0.32\textwidth]{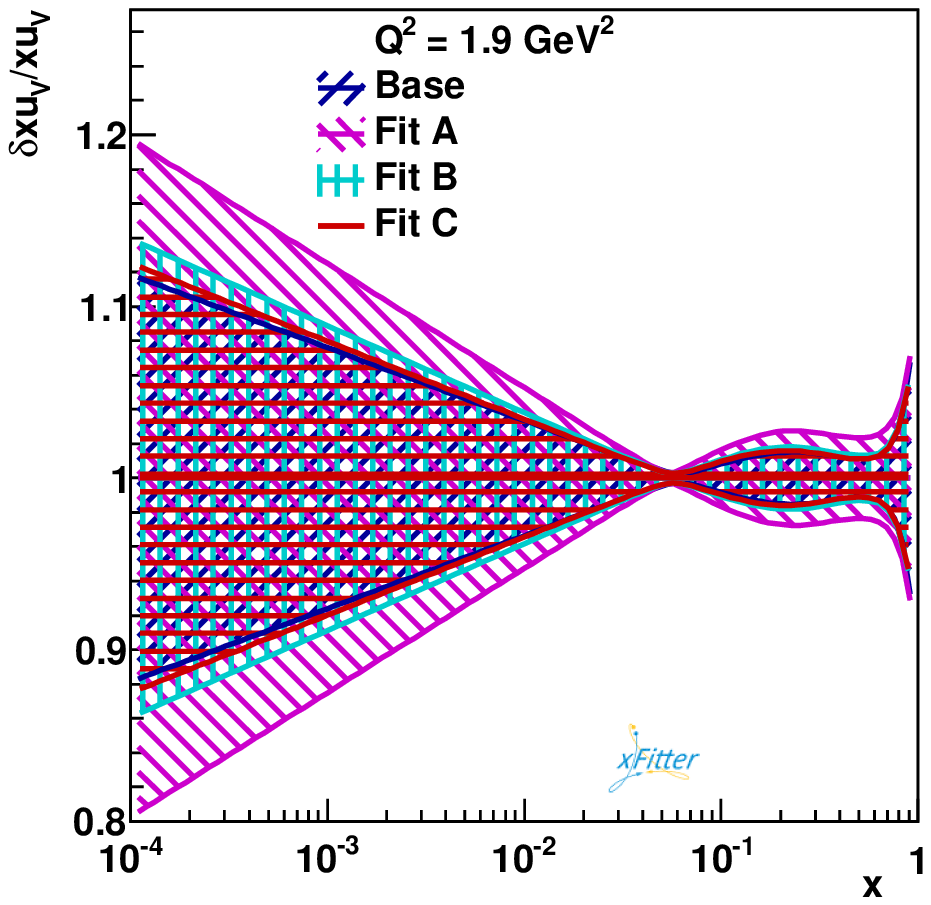}

\includegraphics[width=0.32\textwidth]{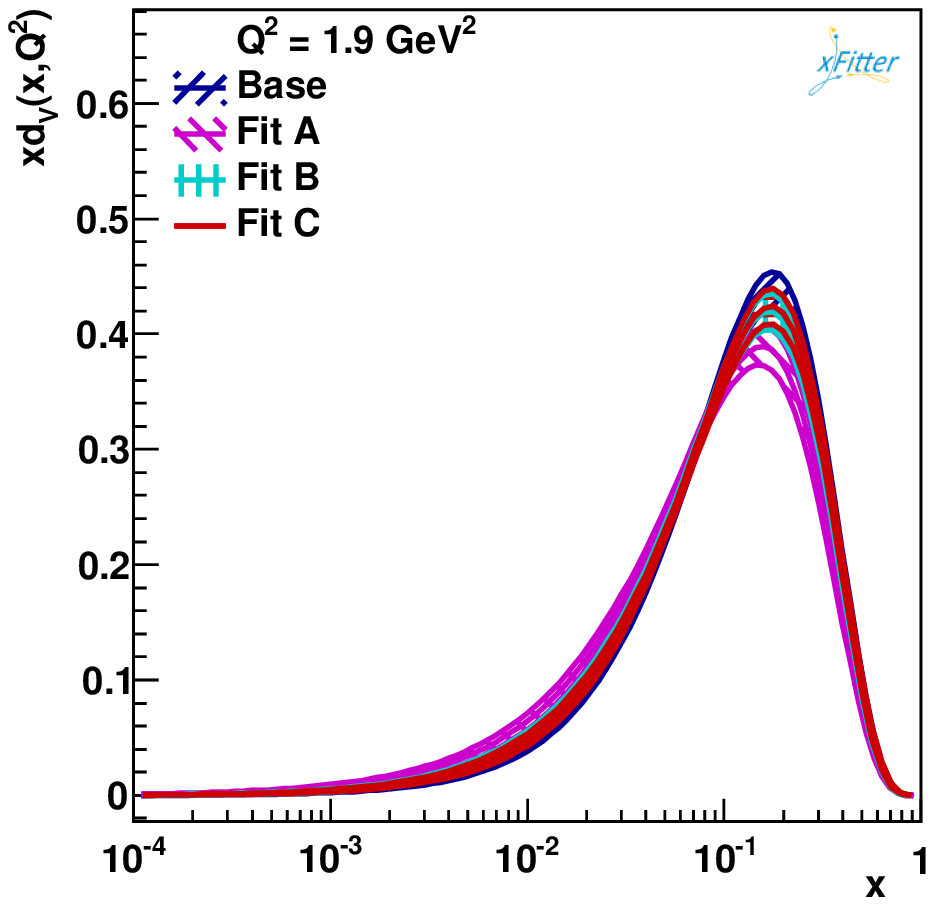}
\includegraphics[width=0.32\textwidth]{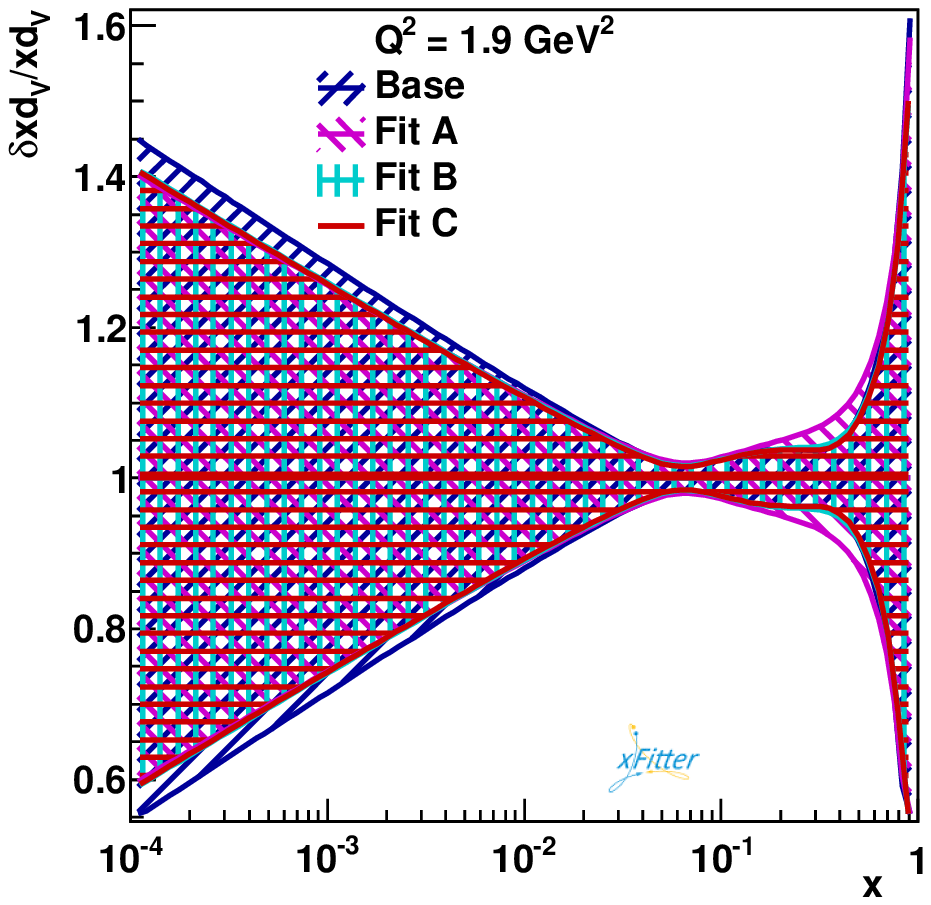}

\includegraphics[width=0.32\textwidth]{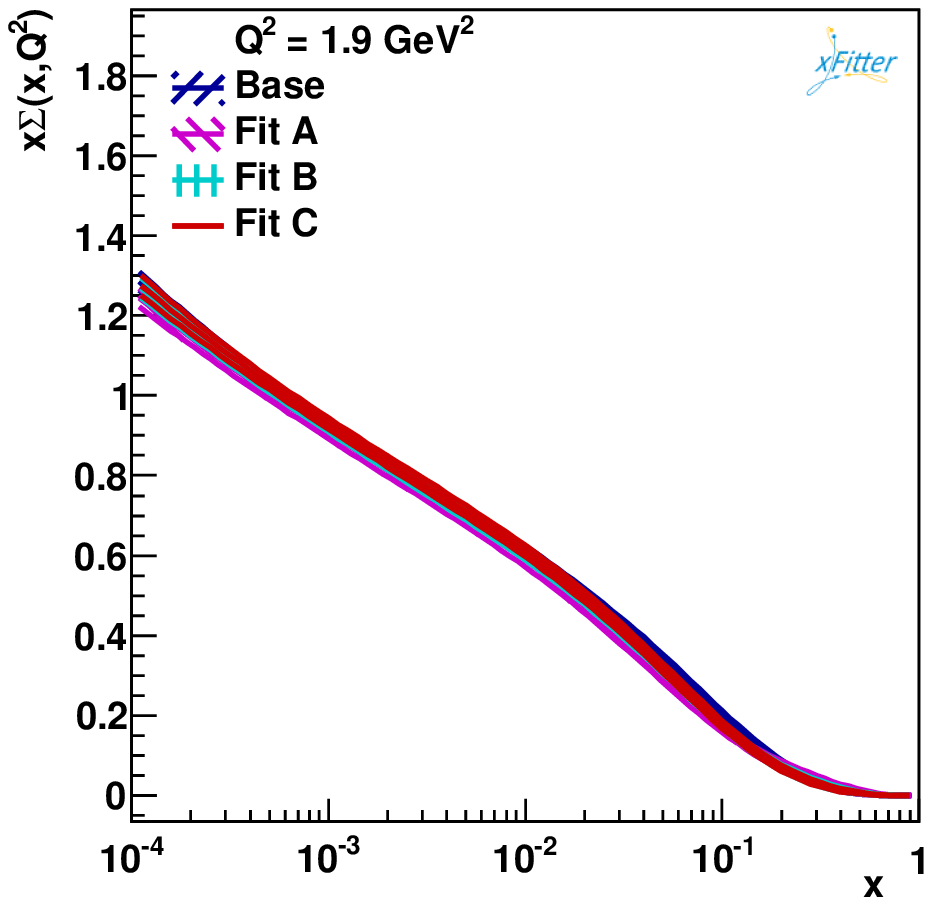}
\includegraphics[width=0.32\textwidth]{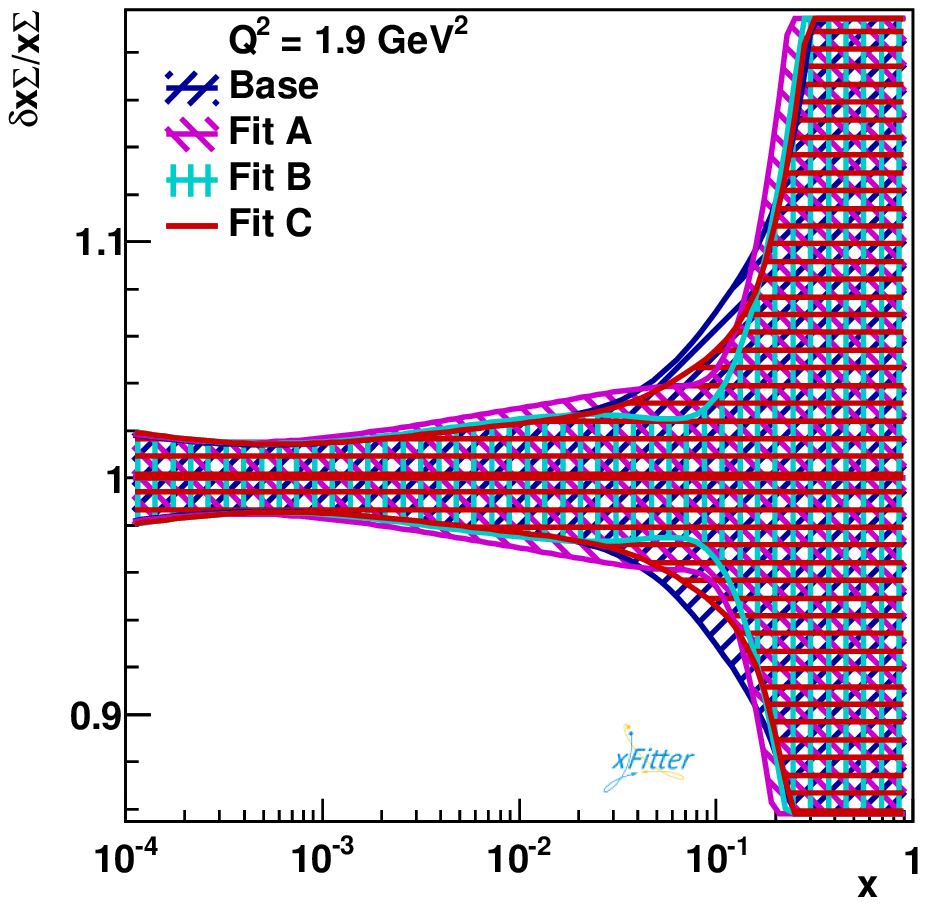}

\includegraphics[width=0.32\textwidth]{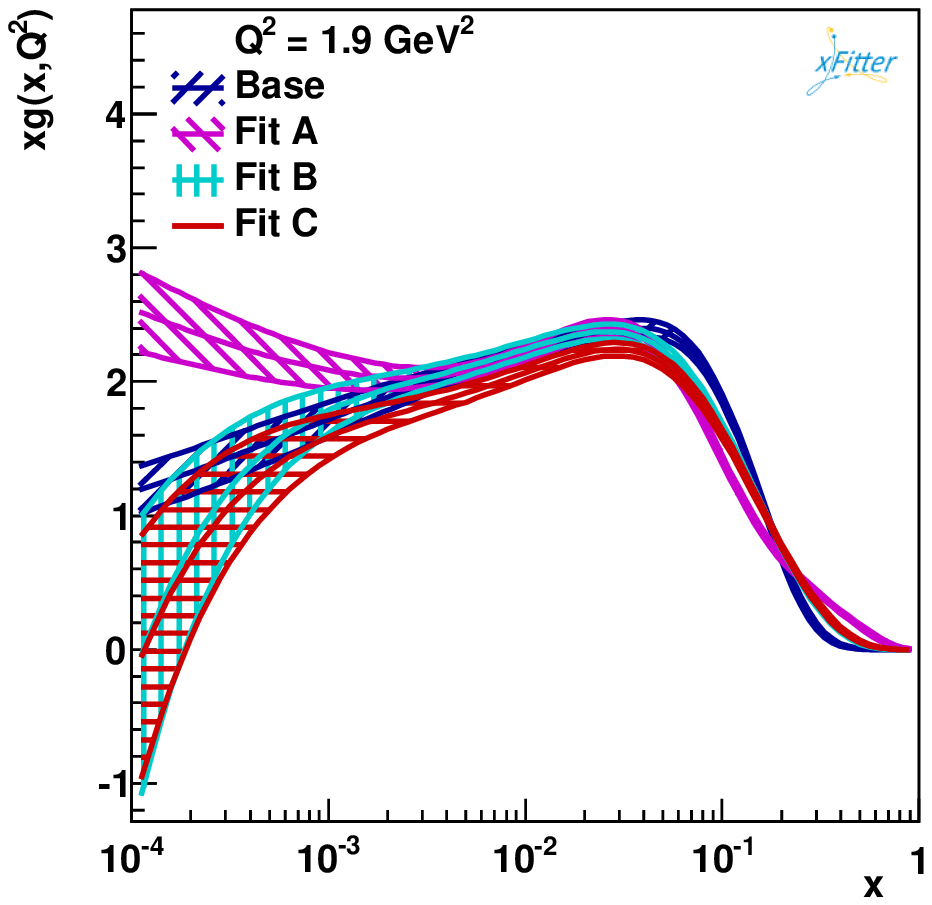}
\includegraphics[width=0.32\textwidth]{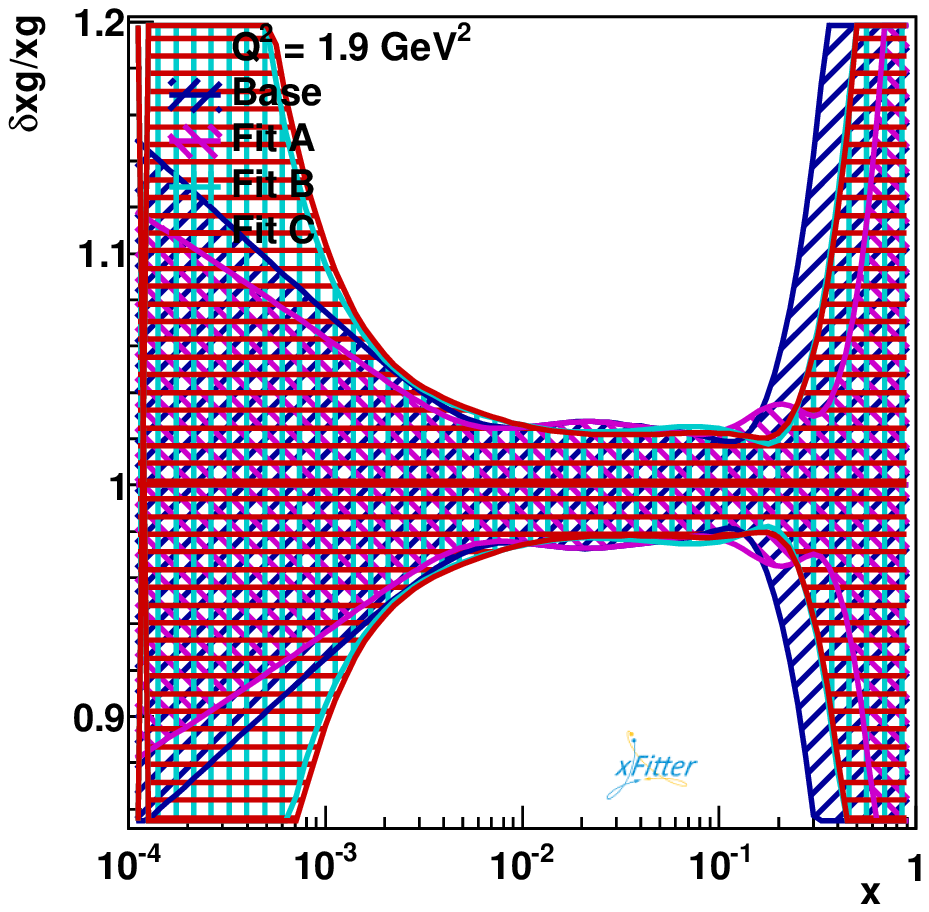}
 \end{center}

\caption{The NLO PDFs of our Base, Fits~A, B and C  results for  $xu_{v}$, $xd_{v}$, $x\Sigma$ and $xg$ at the initial scale of  $Q_{0}^{2}$= 1.9 GeV$^{2}$ as a function of $x$, with $m_c = 1.50$ GeV (left panels). The  relative uncertainties $\delta xq(x,Q^2)/xq(x,Q^2)$ are displayed (right panels) and the Fits~B and C have IC contribution.}

 \label{pic:PDFs}
\end{figure*}

Fig.~\ref{pic:PDFs} illustrates the NLO QCD fit results for valence, sea and gluon PDFs as a function of $x$ at the initial scale of $Q^2_0 = 1.9$ GeV$^2$.  Although there are no significant changes in the valence and sea PDFs in all fits, significant changes in the gluon PDFs are observed (left panels). To clarify the difference between our  four fits, we present the relative uncertainties $\delta xq(x,Q^2)/xq(x,Q^2)$ with respect to the Base fit. According to this figure, the changes of the central values of the PDFs, their uncertainties, or both are observed (right panels).

\begin{figure*}
 \begin{center}

\includegraphics[width=0.32\textwidth]{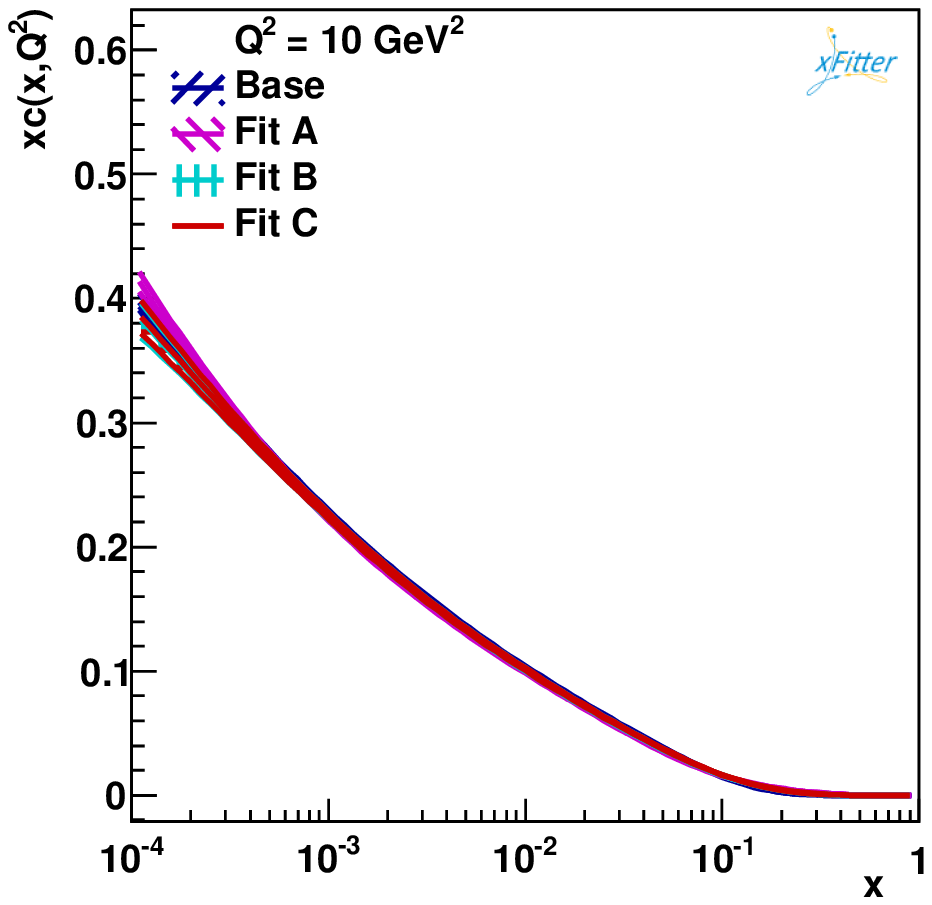}
\includegraphics[width=0.32\textwidth]{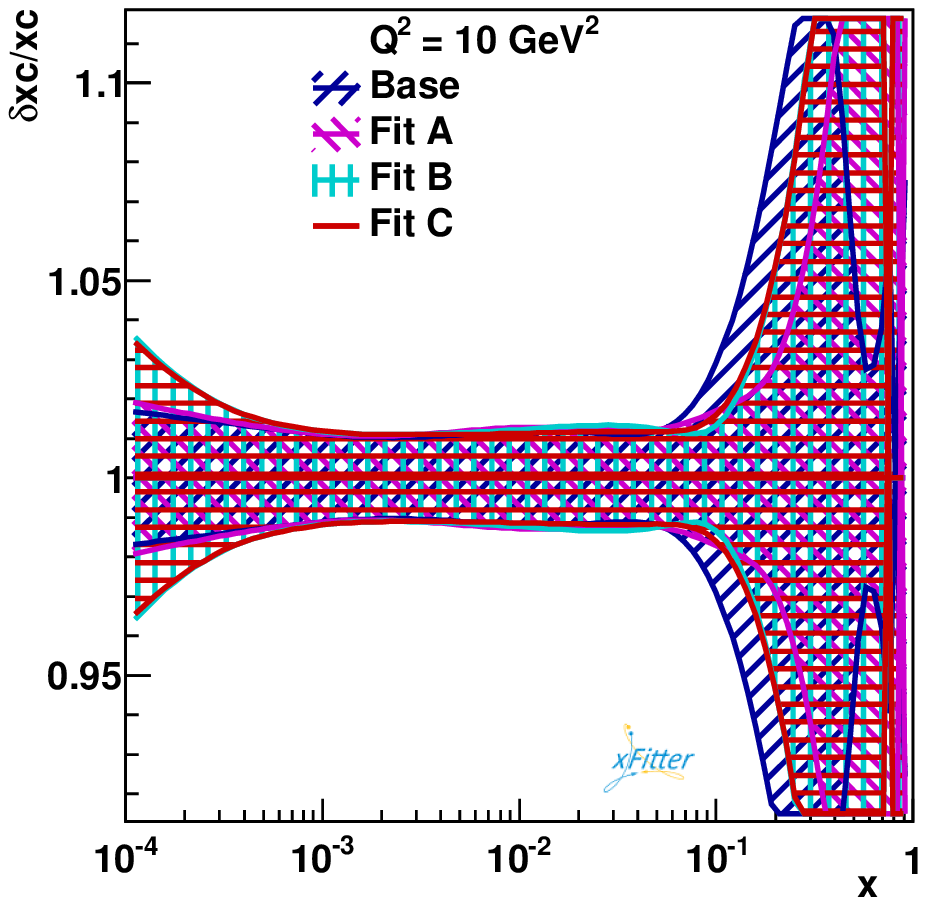}

\includegraphics[width=0.32\textwidth]{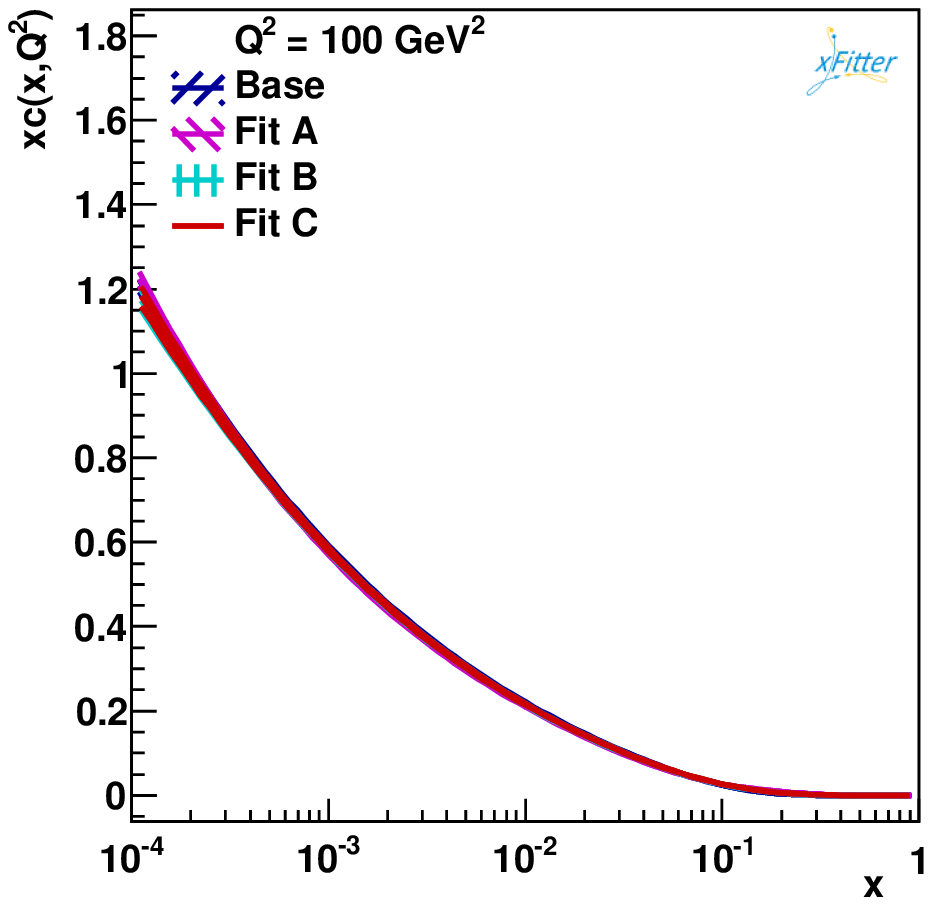}
\includegraphics[width=0.32\textwidth]{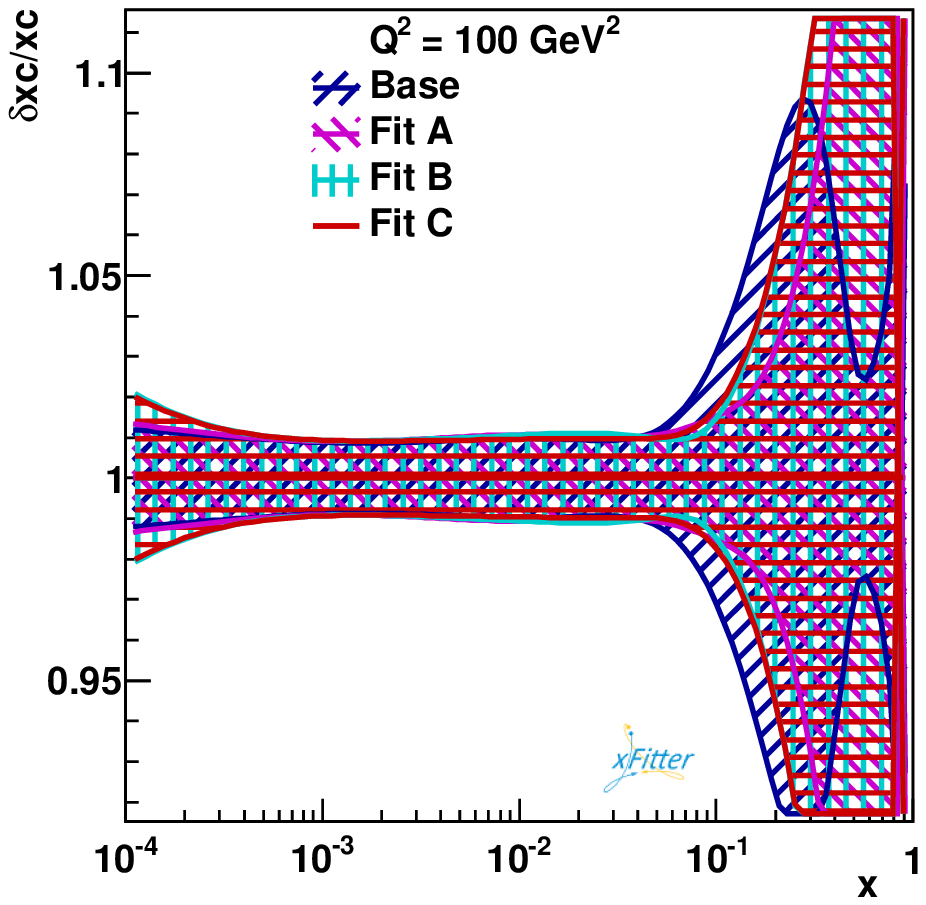}

\end{center}

\caption{The NLO extrinsic charm PDF from the Base and Fits~A, B and C as a function of $x$ for $Q^{2}$= 10 GeV$^{2}$ and $Q^{2}$= 100 GeV$^{2}$ (left panels) and the relative uncertainties $\delta xc(x,Q^2)/xc(x,Q^2)$  (right panels) are shown.}
 \label{pic:c-Sce1-3}
\end{figure*}

In Fig.~\ref{pic:c-Sce1-3}, we present our results for the extrinsic charm PDF based on our four fits as a function of $x$ for $Q^{2}$= 10 GeV$^{2}$ and $Q^{2}$= 100 GeV$^{2}$ at NLO (left panels). The relative uncertainties $\delta xc(x,Q^2)/xc(x,Q^2)$  are also shown (right panels).

Since our analysis fixed the charm mass to 1.5 GeV, we are interested in study the impact of the charm mass value on  ${P}_{c{\bar c/p}}$. Basically, the charm mass impact in our analysis has led us to recalculate the simultaneous fit of ${P}_{c{\bar c/p}}$ and $\alpha_s$ for different values of charm mass. Here, we choose  Fit~C, as a reference point to study the impact of charm mass on PDFs behavior and the extraction ${P}_{c{\bar c/p}}$ and $\alpha_s$ values.

\begingroup
\squeezetable
\begin{table*}
\caption{\label{PARIC} The fit results for parameter values in Eq.~(\ref{eq:xpar-general}), and the intrinsic charm probability ${P}_{c{\bar c/p}}$(\%) and their uncertainties at the initial scale $Q_0^2$=1.9 GeV$^2$ at NLO,  for Base, Fit~A, Fit~B and Fit~C. }
\begin{ruledtabular}
\begin{tabular}{lllll}	 
     Parameter   
&Base& ~~~~~~~Fit~A~~~~~~~& ~~~~~~~Fit~B~~~~~~~&Fit~C\\ 
     \hline
  $B_{u_v}$ & $0.834 \pm 0.018$& $0.753 \pm 0.028$& $0.804 \pm 0.020$& $0.819 \pm 0.021$  \\ 
  $C_{u_v}$ &$3.964 \pm 0.056$& $4.142 \pm 0.042$& $4.054 \pm 0.047$& $4.022 \pm 0.049$  \\ 
  $E_{u_v}$ &$2.84 \pm 0.40$& $4.84 \pm 0.57$& $3.63 \pm 0.44$& $3.40 \pm 0.46$  \\ 
  $B_{d_v}$ &$1.152 \pm 0.070$& $0.958 \pm 0.064$& $1.075 \pm 0.062$& $1.092 \pm 0.064$  \\ 
  $C_{d_v}$ & $5.43 \pm 0.33$& $5.21 \pm 0.28$ & $5.27 \pm 0.29$& $5.23 \pm 0.29$  \\ 
  $C_{\bar{U}}$ &$4.44 \pm 0.63$& $3.56 \pm 0.83$& $4.4 \pm 1.2$& $4.6 \pm 1.2$  \\ 
  $D_{\bar{U}}$ &-0.34 (Fixed)&  -0.65 (Fixed)& -0.4 (Fixed)& -0.4 (Fixed)  \\ 
  $A_{\bar{D}}$ &$0.201 \pm 0.011$& $0.210 \pm 0.014$& $0.209 \pm 0.012$& $0.212 \pm 0.013$  \\ 
  $B_{\bar{D}}$ &$-0.1460 \pm 0.0074$& $-0.1376 \pm 0.0081$ & $-0.1406 \pm 0.0075$& $-0.1398 \pm 0.0078$  \\ 
  $C_{\bar{D}}$ &$11.9 \pm 2.3$& $19.3 \pm 2.0$& $16.1 \pm 1.9$& $15.3 \pm 1.8$  \\ 
    $Bg$ & $0.235 \pm 0.073$& $-0.107 \pm 0.026$& $-0.359 \pm 0.066$& $-0.377 \pm 0.061$  \\ 
  $Cg$ & $11.1 \pm 1.4$& $2.55 \pm 0.26$& $5.10 \pm 0.45$& $4.72 \pm 0.44$  \\ 
  $A'_g$ & $15 \pm 12$& $-54 \pm 13$& $0.90 \pm 0.11$& $0.86 \pm 0.11$  \\ 
  $B'_g$ &$0.43 \pm 0.15$& $0.888 \pm 0.072$& $-0.405 \pm 0.054$& $-0.416 \pm 0.051$  \\ 
  $C'_g$ & 25.0 (Fixed)& 25.0 (Fixed)& 25.0 (Fixed)& 25.0 (Fixed)  \\ 
    $f_s$ & 0.31 (Fixed) & 0.31 (Fixed)&   0.31 (Fixed) & 0.31 (Fixed)\\ 
 \hline
    $\alpha_s(M^2_Z)$ & 0.1182 (Fixed)& 0.1182 (Fixed)& 0.1182 (Fixed)& $0.1191 \pm 0.0008$  \\ 
  ${P}_{c{\bar c/p}}(\%)$ & 0 & 0 & $1.017\pm0.20 $& $0.94 \pm 0.20$  \\ 
\end{tabular}
\end{ruledtabular}
\end{table*}
\endgroup

As mentioned above, to compute the perturbative part of structure-function, we utilize the GM-VFN scheme, which assumes that the charm distribution is generated perturbatively by gluon and light quark splittings. The detail of this is sensitive to the charm quark mass, which is not precisely known. We are going to use the RT-opt heavy quark scheme and we are going to vary the charm mass 1.2 GeV to 1.8 GeV and  we will scan the $\chi^2$ to determine the optimal value \cite{Abramowicz:1900rp}. In the present paper, we consider the initial value of $Q^2_0$ for the evolution to be below the charm threshold $\mu_c$.

\begin{figure*}
\begin{centering}
 \resizebox{0.6\textwidth}{!}{
\includegraphics{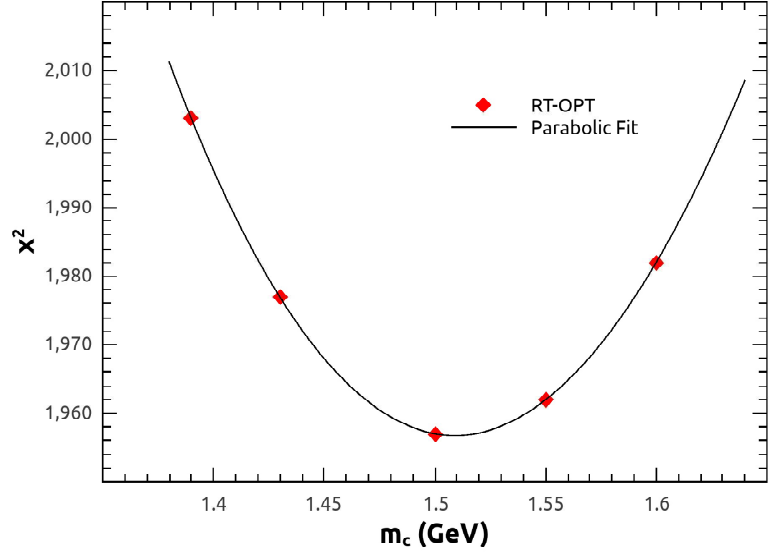}}
\par\end{centering}
\caption{The parabolic fit to the QCD $\chi^2$ values as a function of charm mass in the RT-opt scheme.}
\label{fig:chi2-p5} 
 \end{figure*}

 The $\chi^2(m_c)$ value is calculated for each fits and the optimal values of $m_c^{opt}$ parameter is determined in a given scheme from a parabolic fit  to the $\chi^2(m_c)$ values as
\begin{equation}
\chi^2(m_c) = \chi^2_{min} + \left(\frac{m_c - m_c^{ opt}}{\sigma(m_c^{opt})}\right)^2~,
\label{kappafit}
\end{equation}
where $\chi^2_{min}$ is the $\chi^2$ value at the minimum and $\sigma(m_c^{opt})$ is the fitted experimental uncertainty on $m_c^{opt}$.
The procedure of the $\chi^2$-scan is illustrated in Fig.~\ref{fig:chi2-p5} for the optimal RT (RT-opt) scheme by fitting the experimental data. 
According to this figure, we find the minimum of $\chi^2$ value for the charm mass of 1.50 GeV. This was our motivation to choose $m_c = $ 1.50 GeV in the presence of IC.

\begingroup
\squeezetable
\begin{table*}
\caption{\label{t3} The extracted values of the intrinsic charm probability ${P}_{c{\bar c/p}}$, from Fit~B and Fit~C, and strong coupling constant $\alpha_s(M^2_Z)$, for different values of charm mass.}
\begin{ruledtabular}
\begin{tabular}{ll|ccccc}	 
     &$m_c$ (GeV) & 1.39 & 1.43& 1.50& 1.55& 1.60\\
\hline

\multirow{ 2}{*}{Fit~B  }&${P}_{c{\bar c/p}}$ (\%)& $0.78 \pm 0.18$& $0.86 \pm 0.19$& $1.02 \pm 0.20$& $1.24 \pm 0.22$& $1.40 \pm 0.24$\\
&$\alpha_s (M^2_Z)$  & $~0.1182 ~$ (Fixed)& $~0.1182 ~$ (Fixed)& $~0.1182 ~$ (Fixed)& $~0.1182 ~$ (Fixed) & $~0.1182 ~$ (Fixed)\\ \hline
\multirow{ 2}{*}{Fit~C  }&${P}_{c{\bar c/p}}$ (\%)& $0.97 \pm 0.20$& $0.96 \pm 0.20$& $0.94 \pm 0.20$& $0.92 \pm 0.21$& $0.88 \pm 0.20$\\
&$\alpha_s (M^2_Z)$  & $~~~~0.1152 \pm 0.0008~~~~$& $ 0.1166 \pm 0.0008~~~~$& $0.1191 \pm 0.0008~~~~$& $0.1208 \pm 0.0008~~~~$ & $0.1225 \pm 0.0008~~~~$\\
\end{tabular}
\end{ruledtabular}
\end{table*}
\endgroup

In Table \ref{t3}, we summarize the influence of the charm mass for the determination of ${P}_{c{\bar c/p}}$ and the strong coupling constant using Fit~B and Fit~C. According to this table and for Fit B, ${P}_{c{\bar c/p}}$ goes from 0.78\% to 1.4\% as increases $m_c$ from 1.39 to 1.6 GeV. For Fit~C, we found ${P}_{c{\bar c/p}}$ changes from 0.97\% to 0.88\% as $m_c$ changes and we also find the $\alpha_s(M_Z^2)$ increase from 0.1152 to 0.1225 with the similar uncertainties.

 As expected, we show the intrinsic charm probability  and the strong coupling constant values depend on the charm mass value. The stronger dependence of ${P}_{c{\bar c/p}}$ and charm mass is due to the fact that the heavier $m_c$ makes it harder to create $c\bar{c}$ pairs to obtain 5-particle Fock states. 

\begin{figure*}
 \begin{center}
\includegraphics[width=0.32\textwidth]{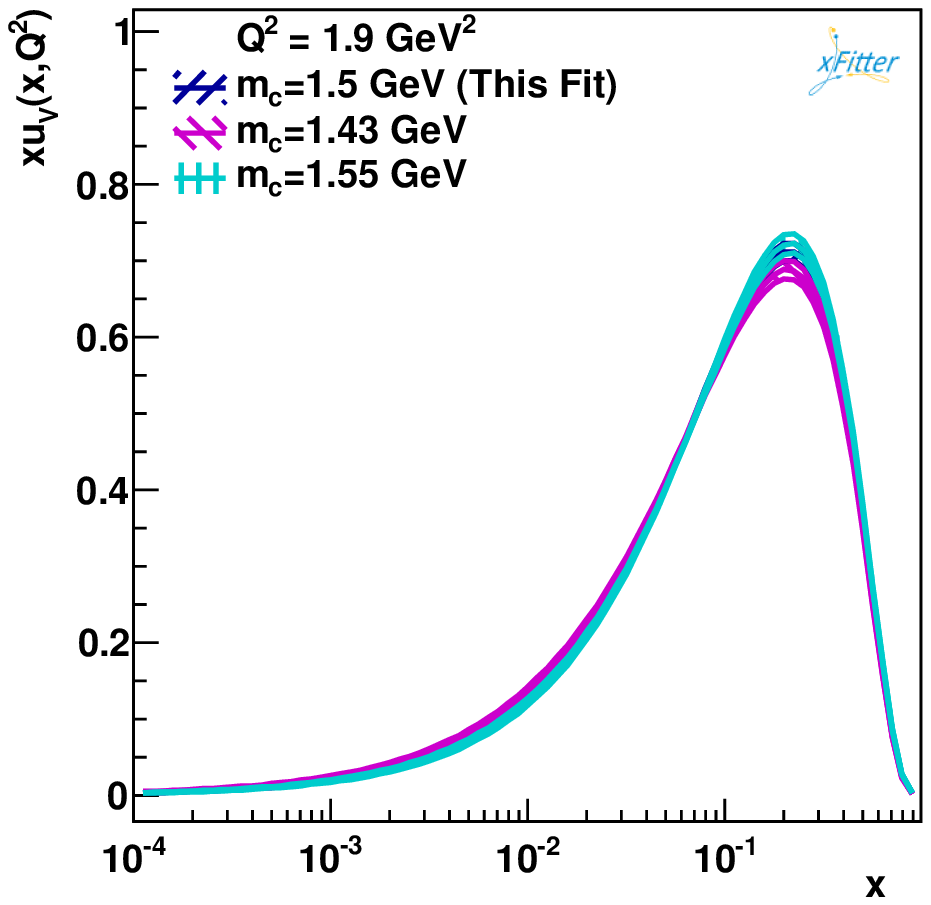}
\includegraphics[width=0.32\textwidth]{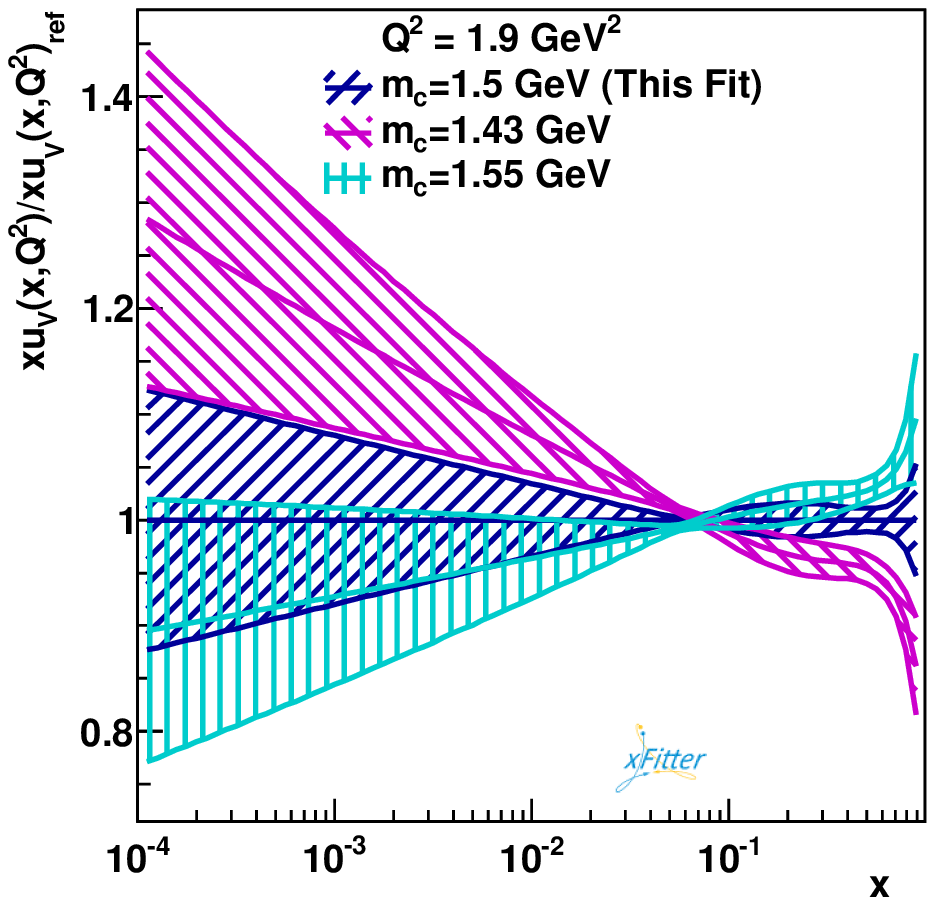}

\includegraphics[width=0.32\textwidth]{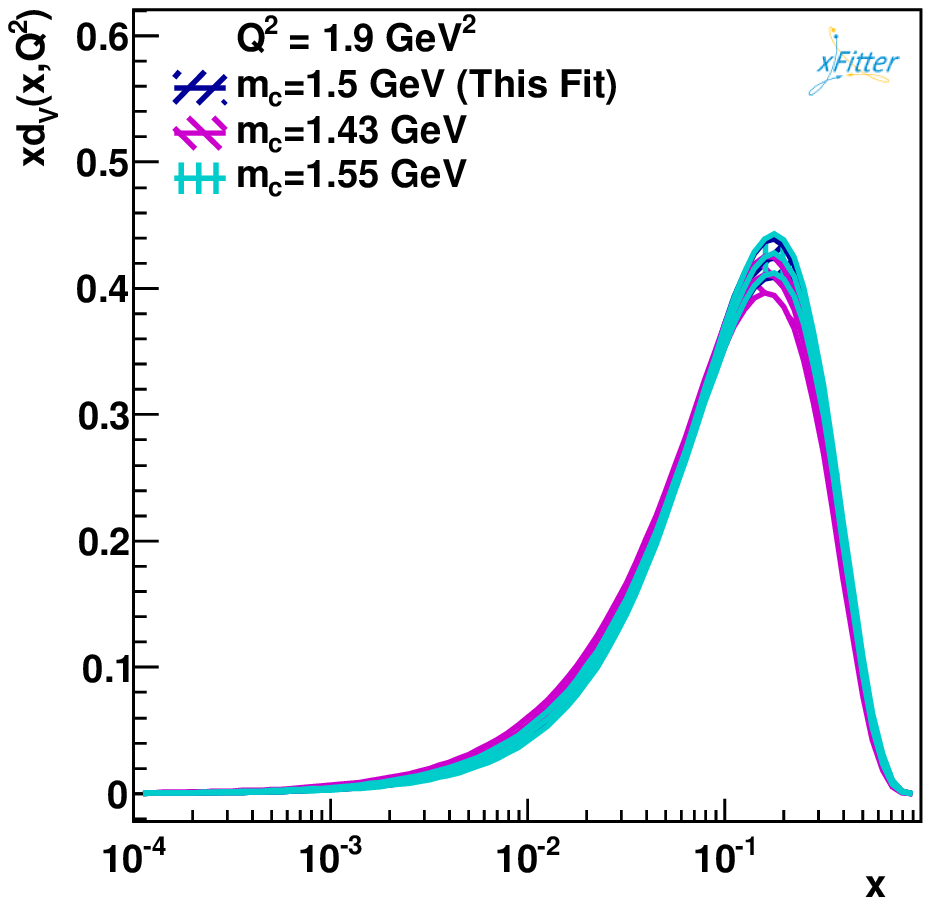}
\includegraphics[width=0.32\textwidth]{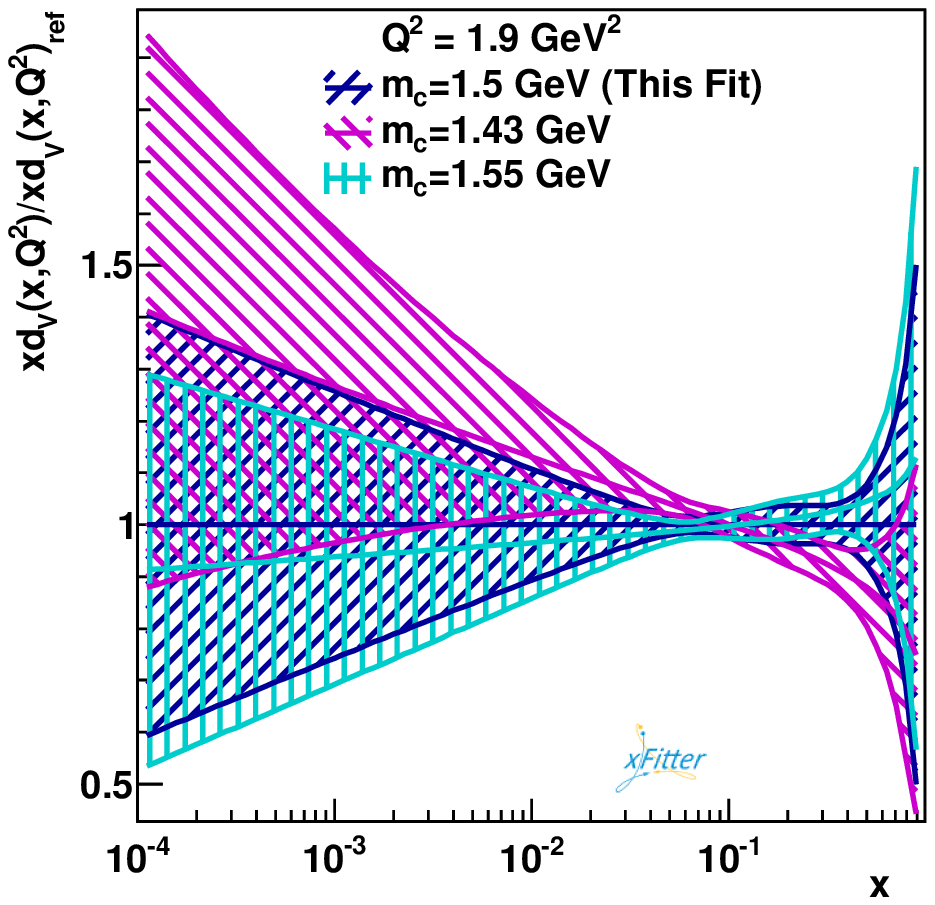}

\includegraphics[width=0.32\textwidth]{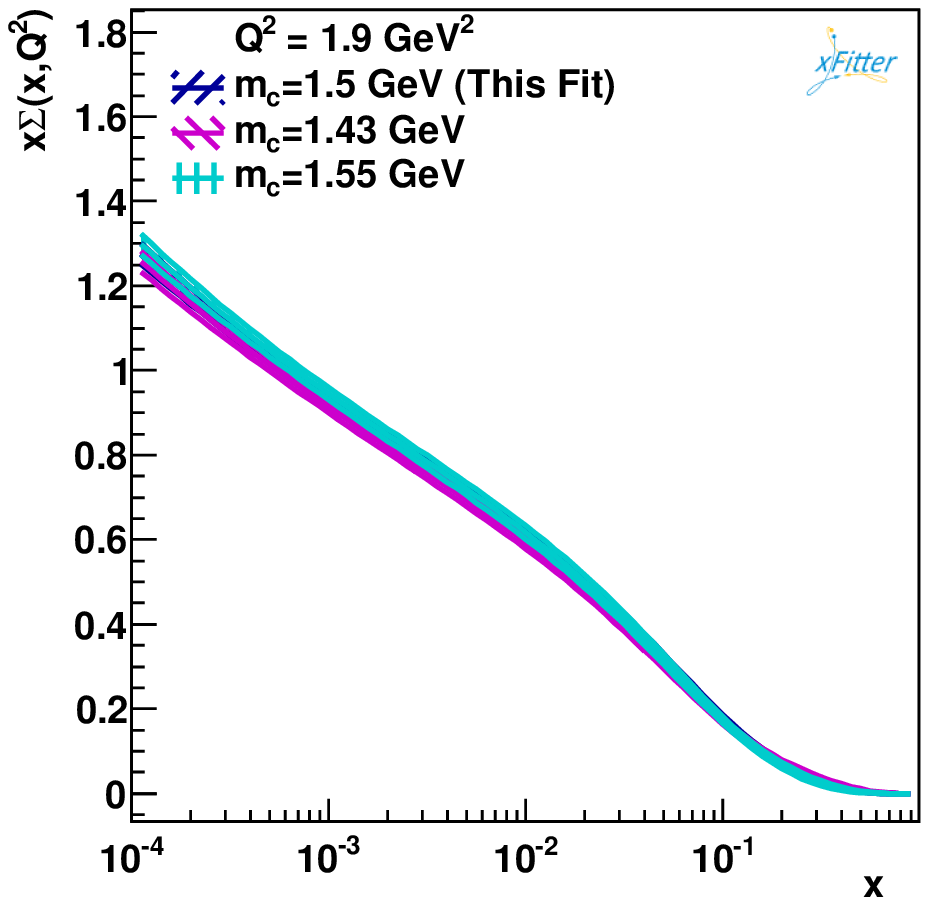}
\includegraphics[width=0.32\textwidth]{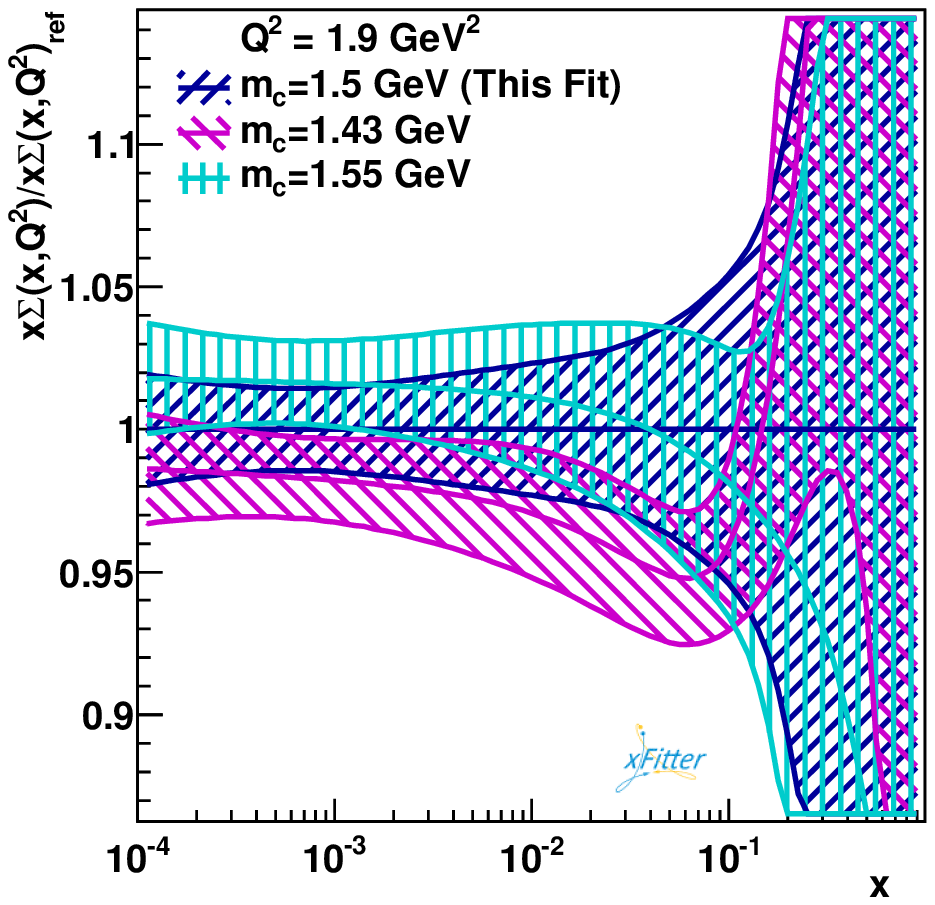}

\includegraphics[width=0.32\textwidth]{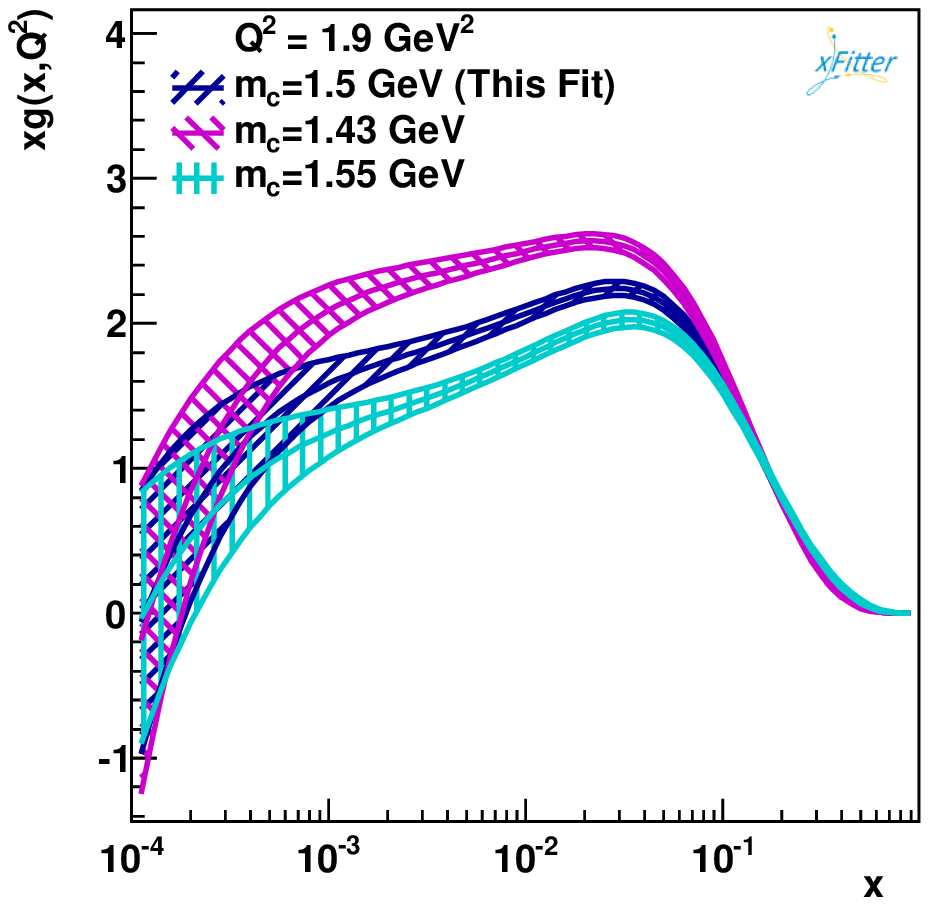}
\includegraphics[width=0.32\textwidth]{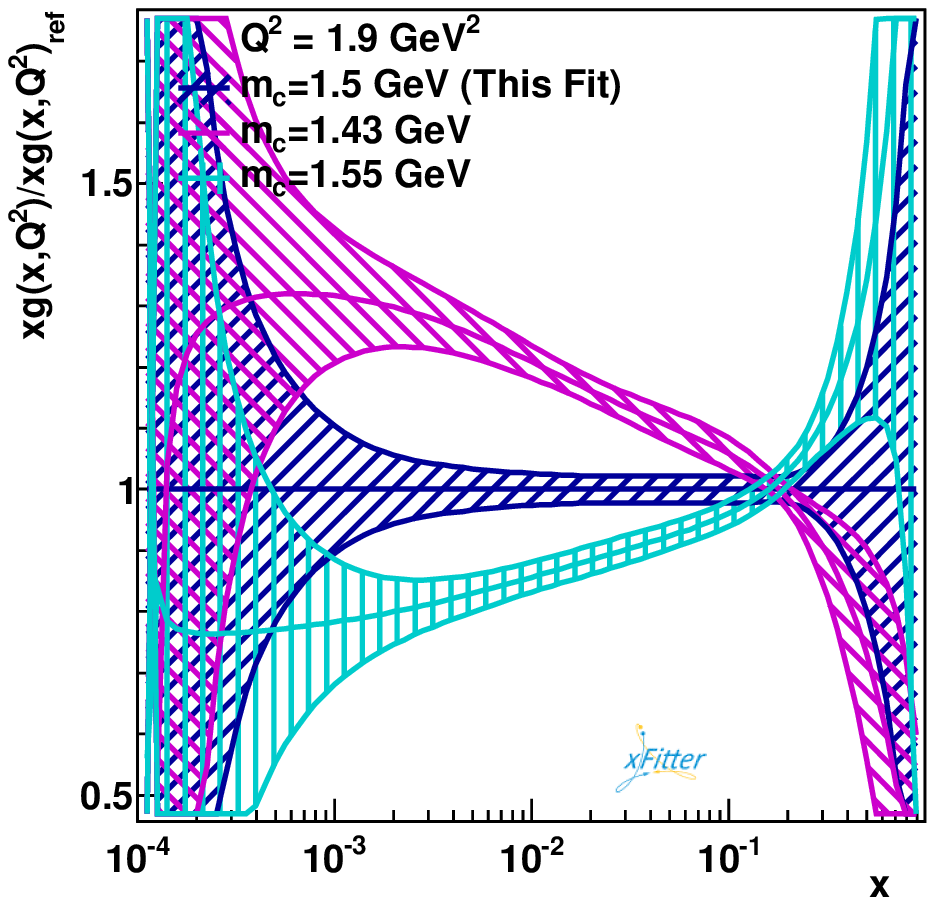}
\end{center}

\caption{The NLO PDFs from Fit~C (This Fit) for different charm mass values $m_c=$1.43, 1.5 and 1.55 GeV at the initial scale of  $Q_{0}^{2}$= 1.9 GeV$^{2}$ as a function of $x$ (left panels). This fit uses the IC contribution. The relative uncertainty ratios  $xq(x,Q^2)/xq(x,Q^2)_{ref}$ with respect to $m_c=$1.5 GeV (right panels) are shown.} \label{pic:PDFs-mc}
\end{figure*}

In Fig.~\ref{pic:PDFs-mc}, we present the valence, sea and gluon PDFs with their uncertainties for different values of $m_c$ are presented as a function of $x$ at the input scale of $Q_0^2=1.9$ GeV$^2$ based on our results for Fit.~C. Although  there are no significant changes in the valence and sea PDFs when comparing different charm mass values,  significant changes in the gluon PDFs are observed (left panels). Also we present the relative uncertainty ratios  $xq(x,Q^2)/xq(x,Q^2)_{ref}$ with respect to $m_c=1.5$ GeV. According to this figure, the changes of the central values of the PDFs and  their uncertainties  for different values of the charm mass are  especially  observed for the gluon PDF (right panels).

Since the gluon PDF plays an important role in the total cross section of the top quark production, we used the results of Fit~C with different charm quark masses and Fit~B with fix $m_c=1.5$ to calculate the top quark cross section at the LHC using the HATHOR package \cite{Aliev:2010zk}. 

According to Fig.~\ref{pic:PDFs-mc}, the gluon densities show significant variation for the different cases with varying charm mass values for Fit C, as a complete fit to study the impact of charm mass. We obtain the total top quark cross section as 804.7$\pm$6.6, 828.7$\pm$8.0 and 855.7$\pm$7.8, using  different values of  charm mass $m_c=1.43, 1.5$ and $1.55$, respectively. Also, we find the total top quark cross section to be  817.8$\pm$6.4 using  Fit~B when $m_c=1.5$. These results are in agreement with the recent CMS top quark cross section measurements \cite{CMS:2016rtp}, 834 $\pm$ 25 (stat.) $^{+118}_{-104}$(syst.) $\pm$ 23(lumi.) at the LHC for 13 TeV.

\begin{figure*}
 \begin{center}

\includegraphics[width=0.32\textwidth]{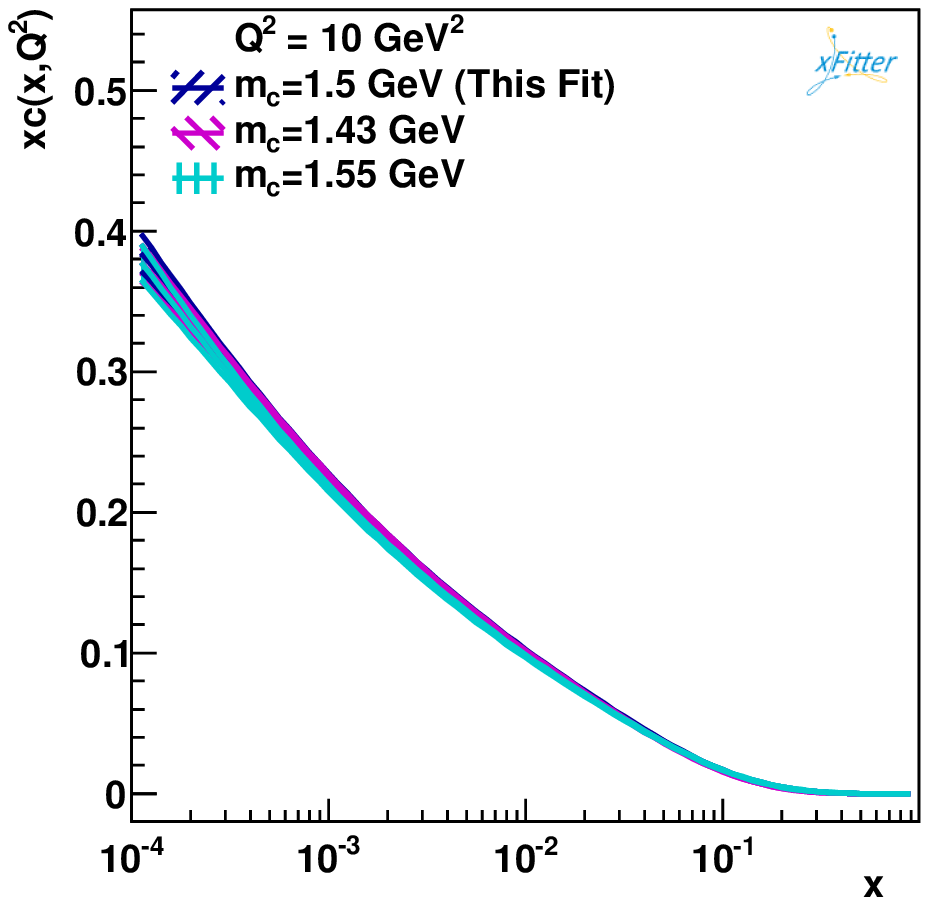}
\includegraphics[width=0.32\textwidth]{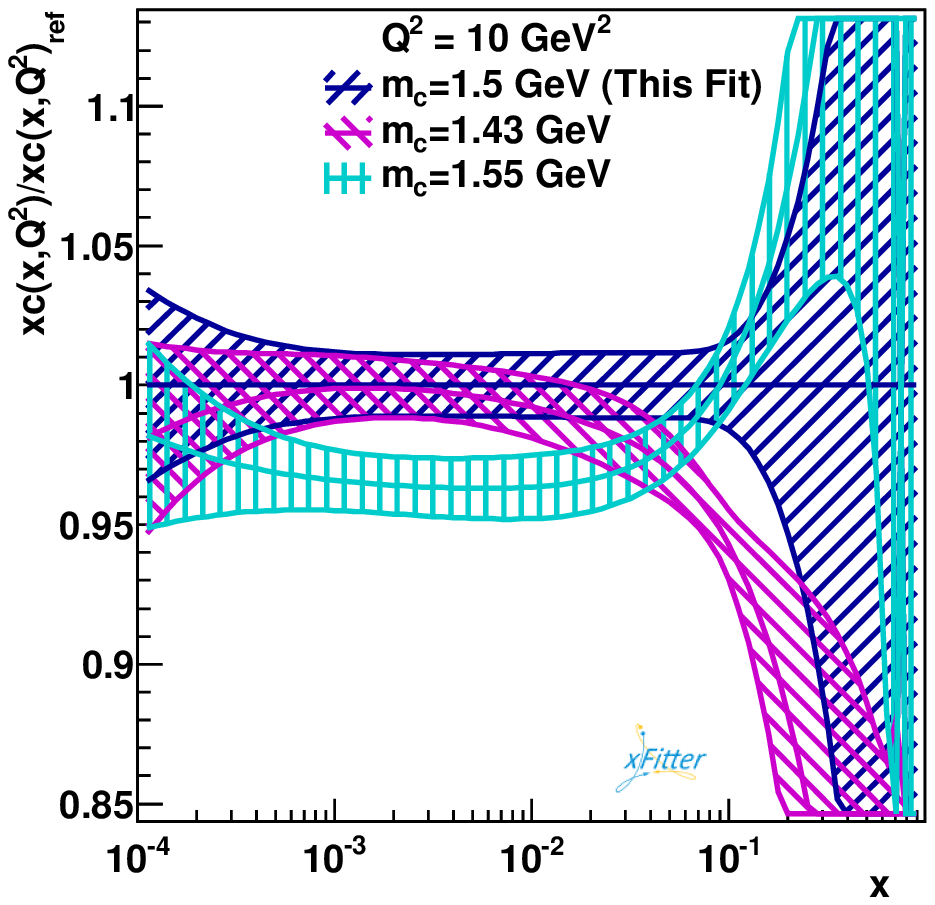}

\includegraphics[width=0.32\textwidth]{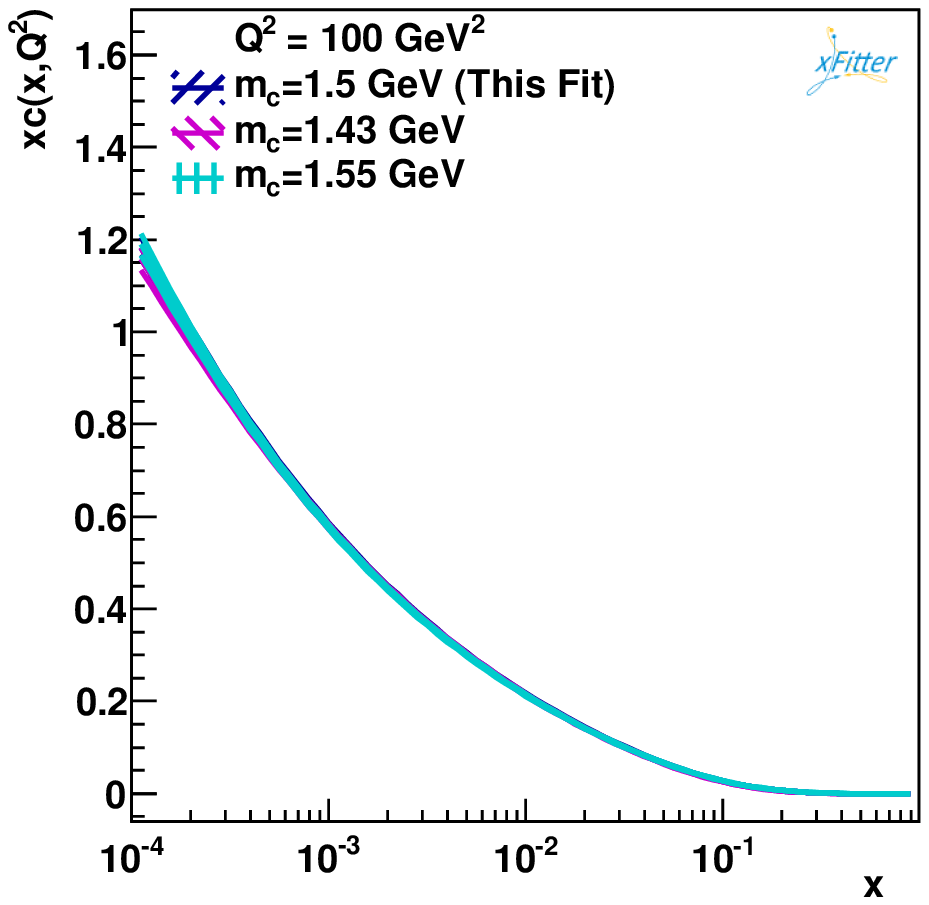}
\includegraphics[width=0.32\textwidth]{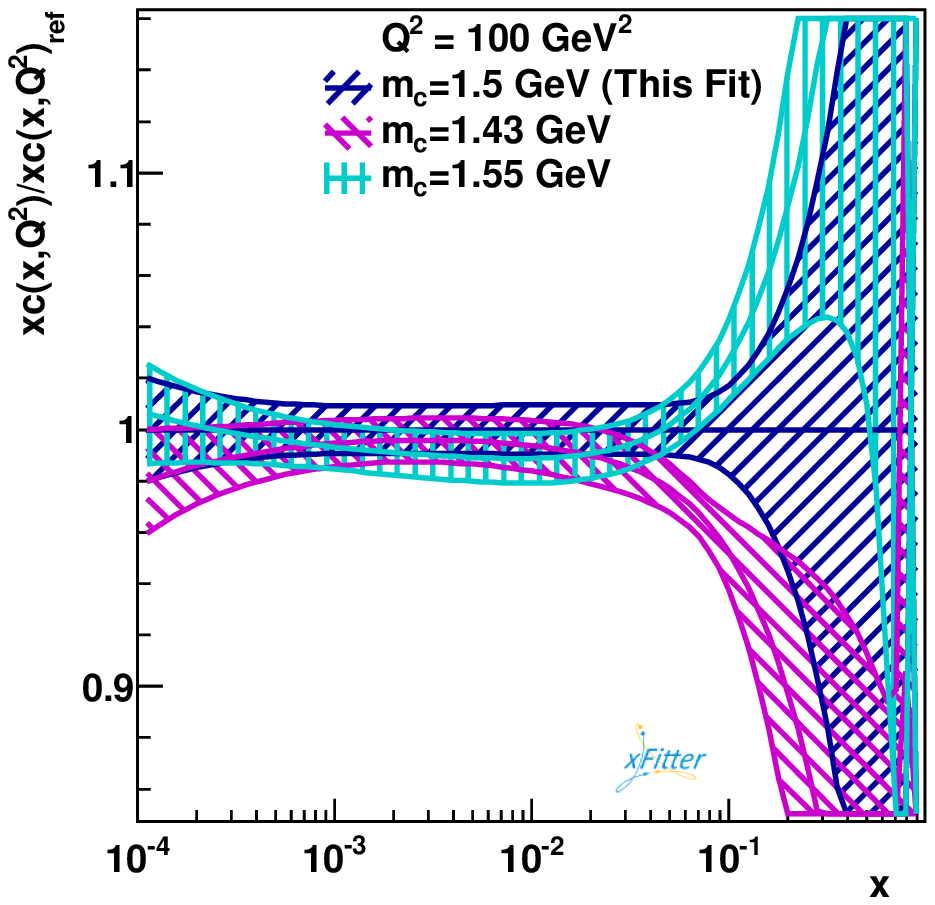}

\end{center}

\caption{The NLO extrinsic charm PDF extracted from Fit~C (This Fit) with different charm mass values $m_c=$1.43, 1.5 and 1.55 GeV  for $Q^{2}$= 10 GeV$^{2}$ and $Q^{2}$= 100 GeV$^{2}$ as a function of $x$ (left panels). This fit include the IC contribution. The relative uncertainty ratios  $xc(x,Q^2)/xc(x,Q^2)_{ref}$ with respect to $m_c=$1.5 GeV (right panels) are shown.} \label{pic:c-mc}
\end{figure*}

Fig.~\ref{pic:c-mc} shows the NLO extrinsic charm PDF extracted from Fit C with different charm quark mass values for range of $Q^{2}$ as a function of $x$,  for different values of $Q^{2}$= 10, 100 GeV$^{2}$ (left panels). The relative uncertainty ratios  $xc(x,Q^2)/xc(x,Q^2)_{ref}$ in respect to $m_c=$1.5 GeV (right panels) are shown.

In Table \ref{t4}, we present the intrinsic charm probability values extracted from Fit~C with the present analysis for different values of charm mass and compare with other predictions from the literature. The extracted value of intrinsic charm probability is in good agreement with the other predictions of the models.

In Table \ref{t5}, we present the momentum fraction of the IC distribution $<x>_{c + \bar{c}}^{int}$ according to Eq. (\ref{eq:m fraction}) and total charm momentum fraction, $<x>_{c+\bar{c}}^{tot} = <x>_{c + \bar{c}}^{int} + <x>_{c + \bar{c}}^{ext}$, extracted in  Fit~C, and compared to the NNPDF3IC and CT14-IC at $Q^2$ = 4 GeV$^2$. Our results for 
$<x>_{c + \bar{c}}^{int}$ and $<x>_{c+\bar{c}}^{tot}$ are obtained using ${P}_{c{\bar c/p}}$(\%)= $0.94 \pm 0.20$.  The momentum fraction of the IC distribution and the total momentum fraction of the charm PDF are in good agreement with other models. Note that in the CT analysis several models for IC are studied whilst NNPDF3IC assumed a specific
parametrisation of the charm PDF which is fitted simultaneously to light quarks and gluons.

\begin{figure*}
\vspace{3cm}
\begin{centering}

  \resizebox{0.60\textwidth}{!}{ 
\includegraphics{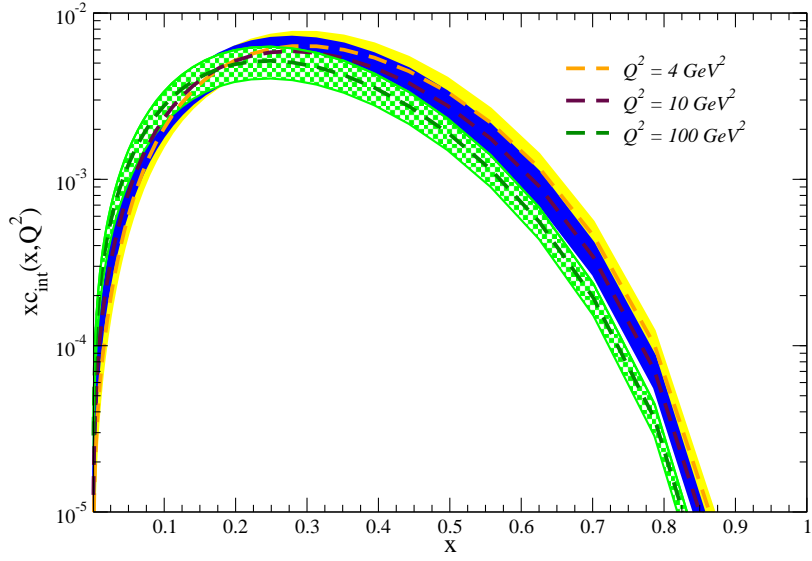}}

\par\end{centering}  
 \caption{The intrinsic charm distribution as a function of $x$ and its uncertainties at $Q^2=$4, 10 and 100 ~GeV$^2$ with ${P}_{c{\bar c/p}} = 0.94 \pm 0.2$\%.
}\label{pic:intrinsic}
\end{figure*}

\begin{figure*}
\vspace{2cm}
\begin{centering}
\resizebox{0.6\textwidth}{!}{ 
\includegraphics{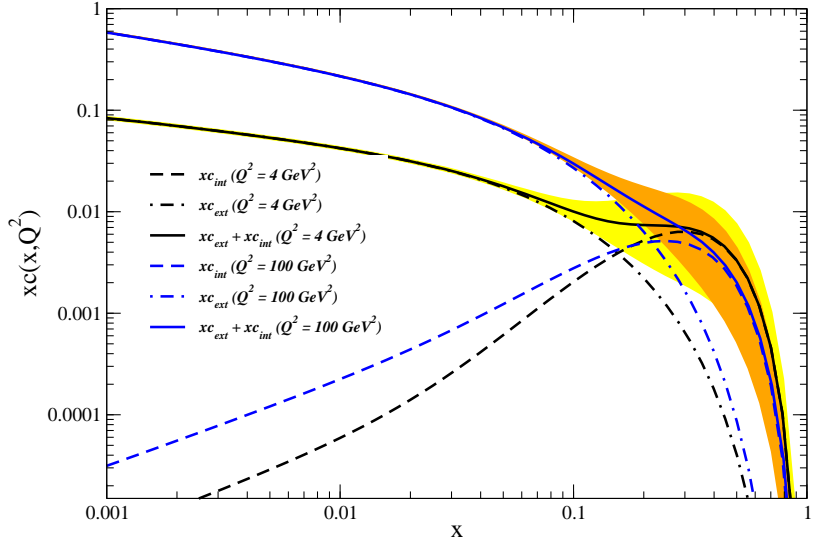}}
\par\end{centering} 
 \caption{The comparison of the intrinsic charm $xc_{int}(x,Q^2)$, extrinsic charm $xc_{ext}(x,Q^2)$, and total charm distribution 
 $xc(x,Q^2) = (xc_{int} + xc_{ext})(x,Q^2)$, extracted from Fit~C as a function of $x$ for $Q^2=$ 4 and 100 ~GeV$^2$ with ${P}_{c{\bar c/p}} = 0.94 \pm 0.2$\%. 
}\label{pic:xctotal}
\end{figure*}

In Fig.~\ref{pic:intrinsic}, we show the intrinsic charm distribution $xc_{int}(x,Q^2)$ with its uncertainty  as  a function of $x$  at different values of $Q^2$ with ${P}_{c{\bar c/p}} = 0.94 \pm 0.2$\%. In Fig.~\ref{pic:xctotal}, we display the  intrinsic charm  $xc_{int}$, extrinsic charm $xc_{ext}$ and the total charm $xc_{int} + xc_{ext}$  distributions with their uncertainties with ${P}_{c{\bar c/p}} = 0.94 \pm 0.2$\%, as a function of $x$ and $Q^2$.

\begin{figure*}
\vspace{2cm}

 \resizebox{0.32\textwidth}{!}{ 
\includegraphics{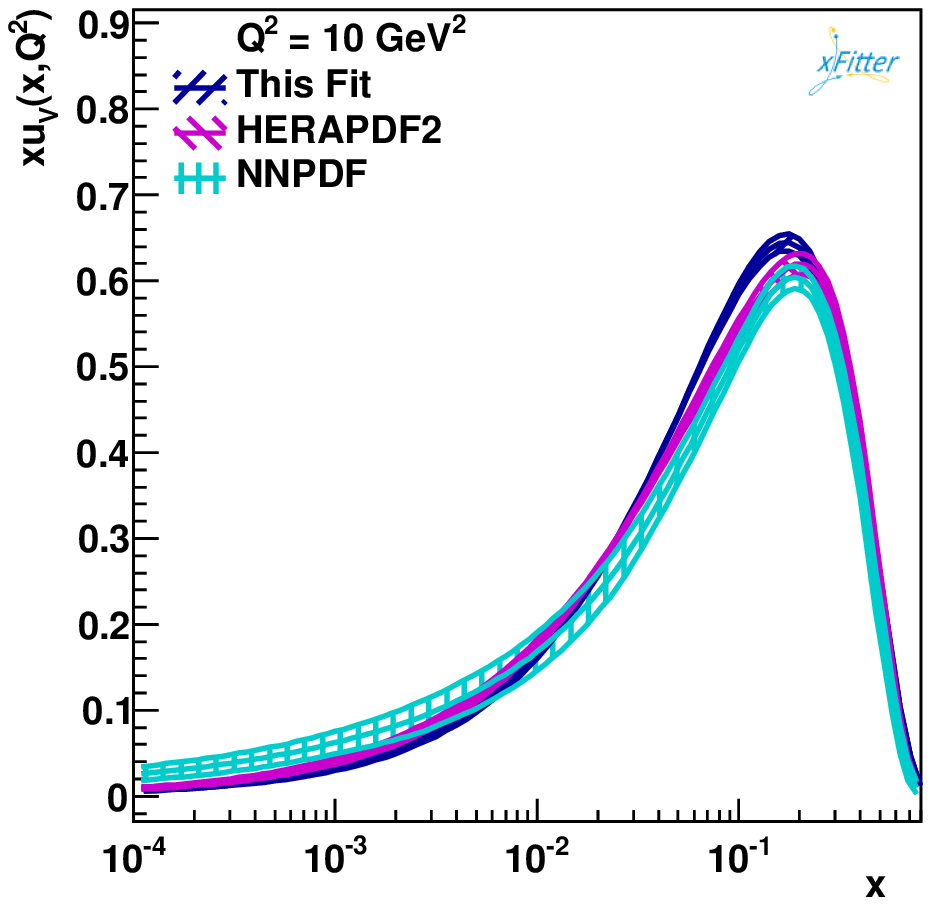}}
 \resizebox{0.32\textwidth}{!}{ 
\includegraphics{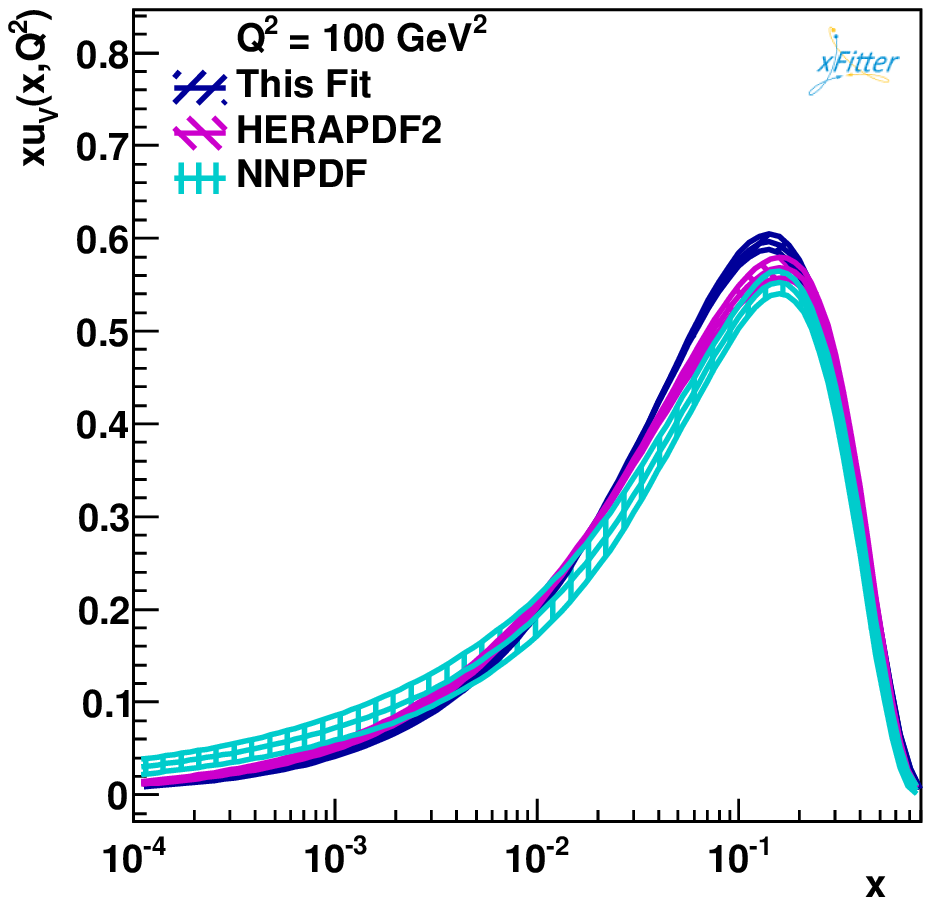}}

 \resizebox{0.32\textwidth}{!}{ 
\includegraphics{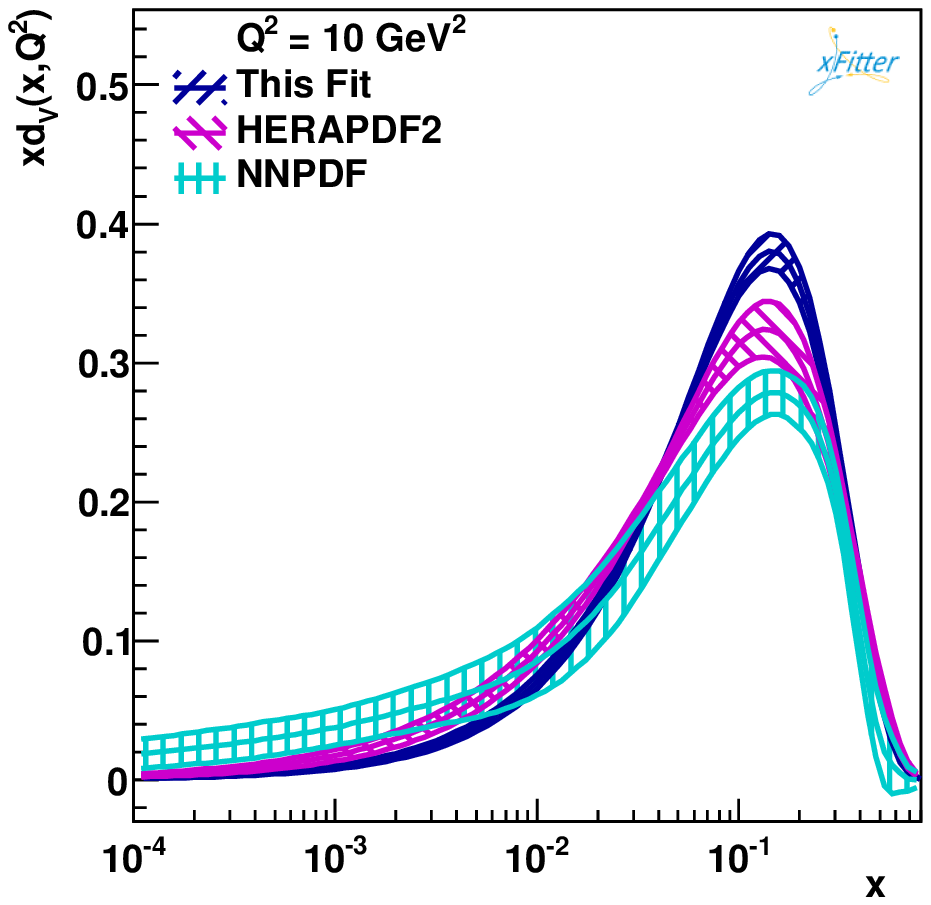}}
 \resizebox{0.32\textwidth}{!}{ 
\includegraphics{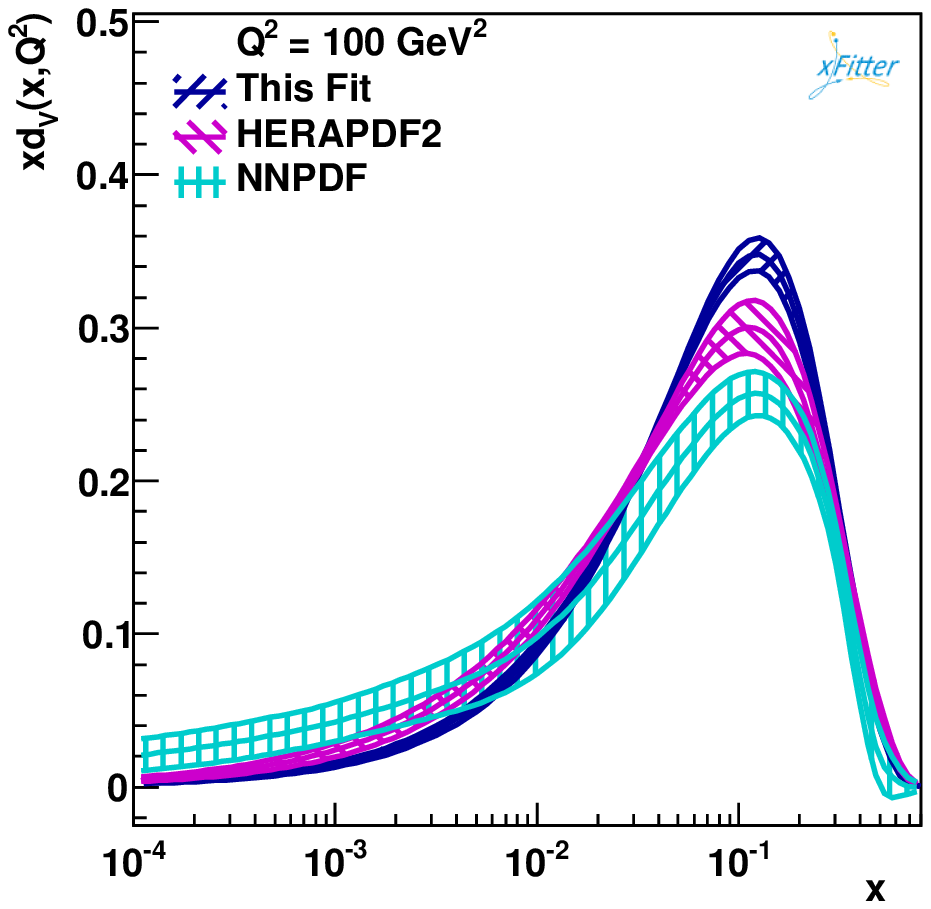}}

 \resizebox{0.32\textwidth}{!}{ 
\includegraphics{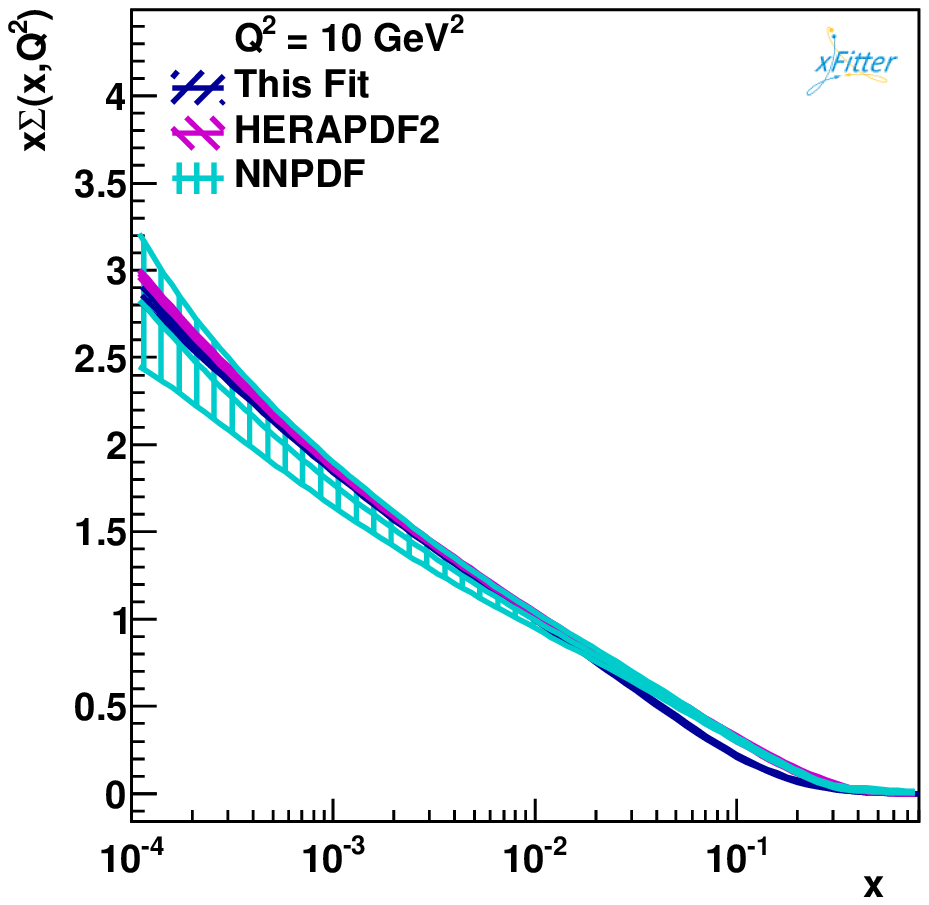}}
 \resizebox{0.32\textwidth}{!}{ 
\includegraphics{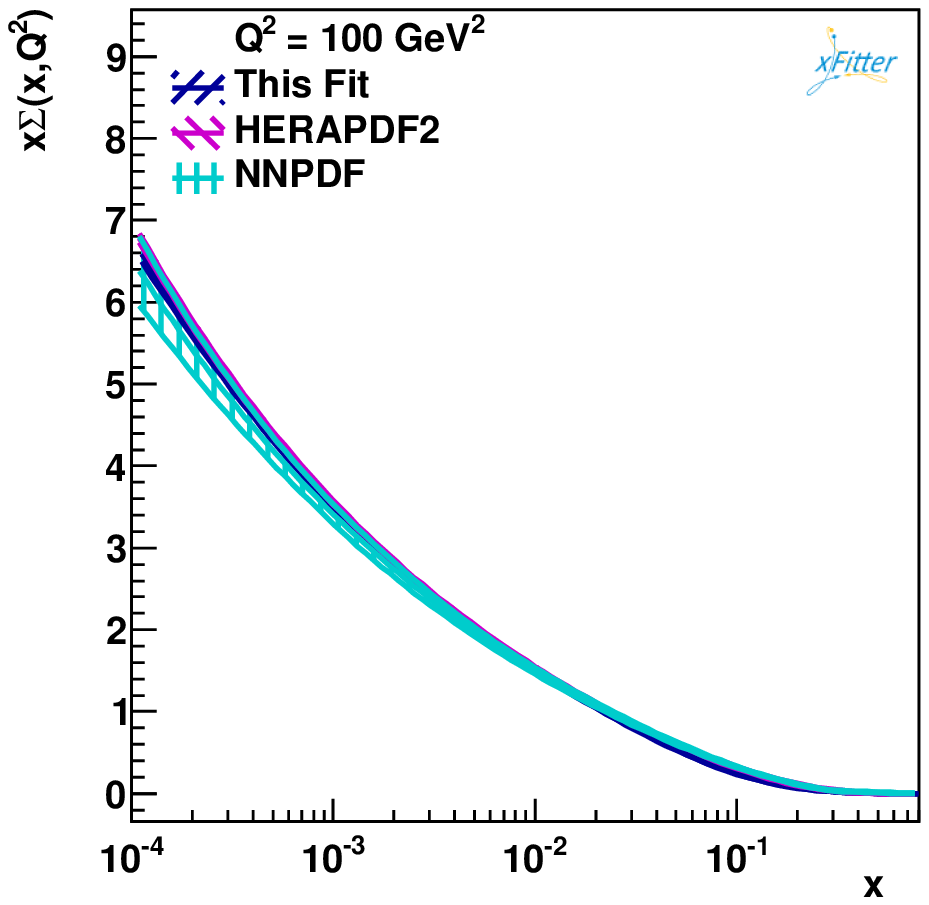}}

 \resizebox{0.32\textwidth}{!}{ 
\includegraphics{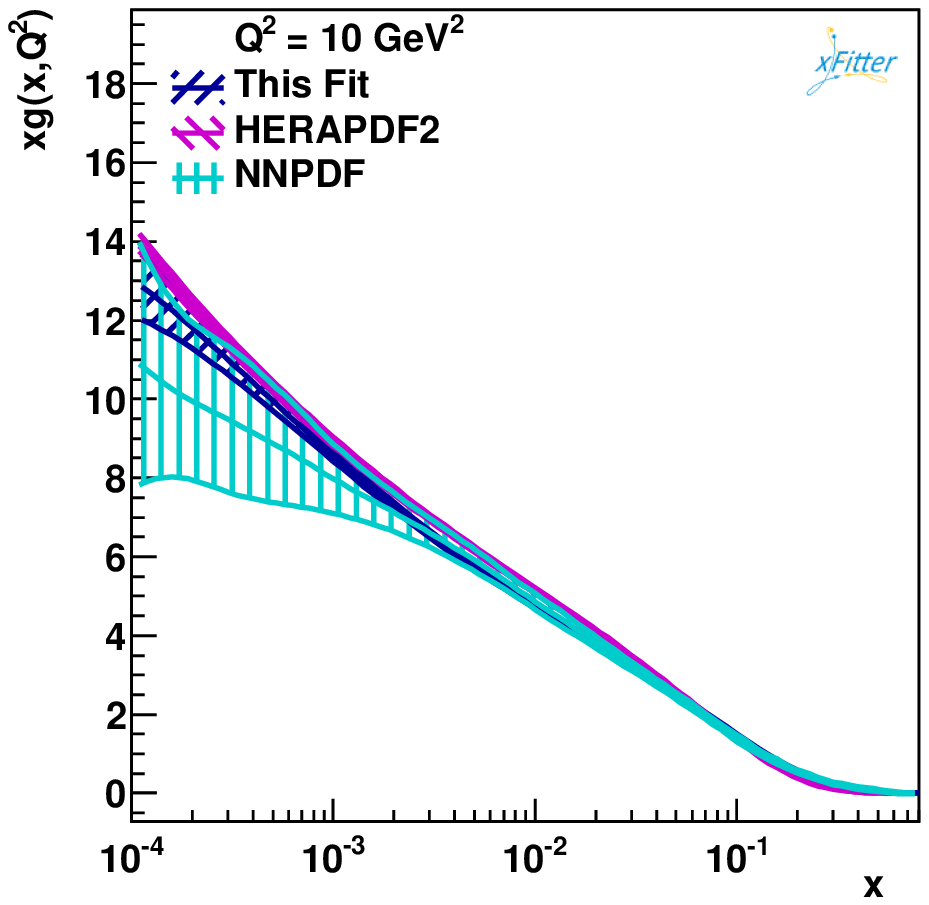}}
 \resizebox{0.32\textwidth}{!}{ 
\includegraphics{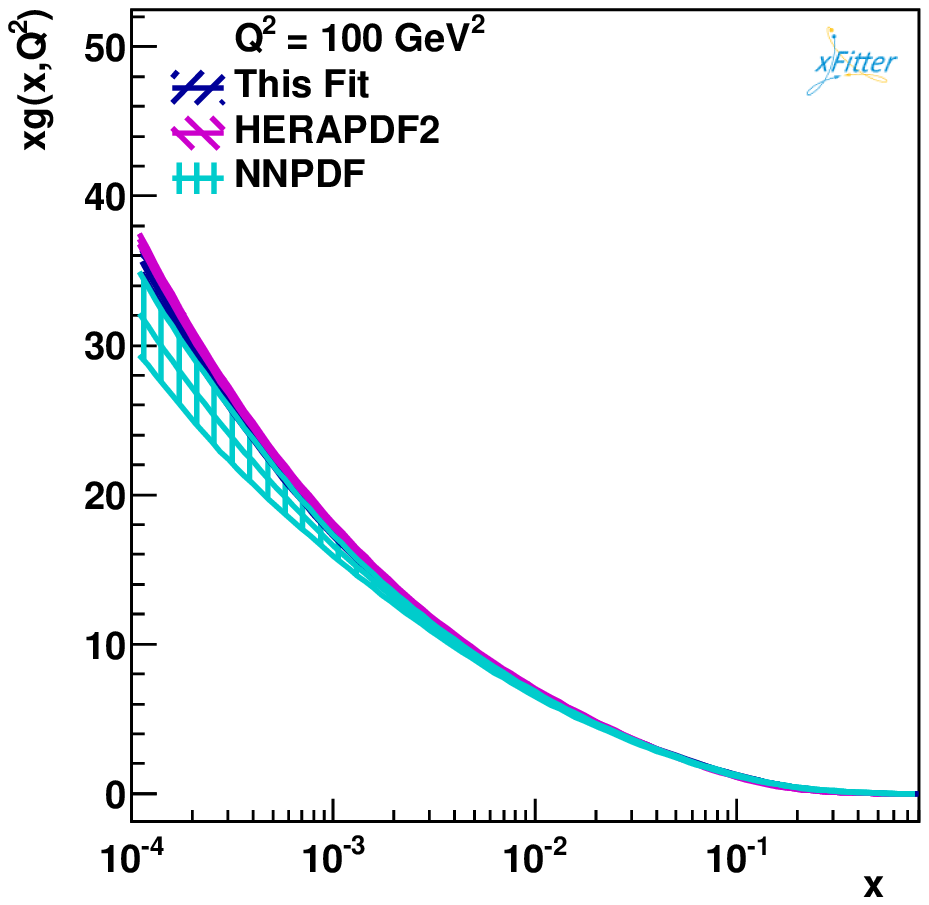}}

\caption{The comparison of our NLO results for $xu_{v}$, $xd_{v}$, $x\Sigma$ and $xg$ PDFs in presence of IC from Fit~C (This Fit). We compare with  HERAPDF2 \cite{Abramowicz:2015mha} and NNPDF3IC \cite{Ball:2017nwa} results, for $Q^{2}$= 10 GeV$^{2}$ and $Q^{2}$= 100 GeV$^{2}$ as a function of $x$.} \label{pic:PDFs-Other}
\end{figure*}

In Fig.~\ref{pic:PDFs-Other}, we display the valence, sea and gluon PDFs extracted in our QCD Fit~C for selected $Q^2$ values  as a function of $x$, and compared them with the results obtained by HERAPDF2 \cite{Abramowicz:2015mha} and NNPDF3IC \cite{Ball:2017nwa}. Note that the NNPDF collaboration used a different methodology to parametrize the PDFs,  and chose different input scales and the cut values. 
 
\begin{figure*}
\vspace{2cm}
 \resizebox{0.32\textwidth}{!}{ 
\includegraphics{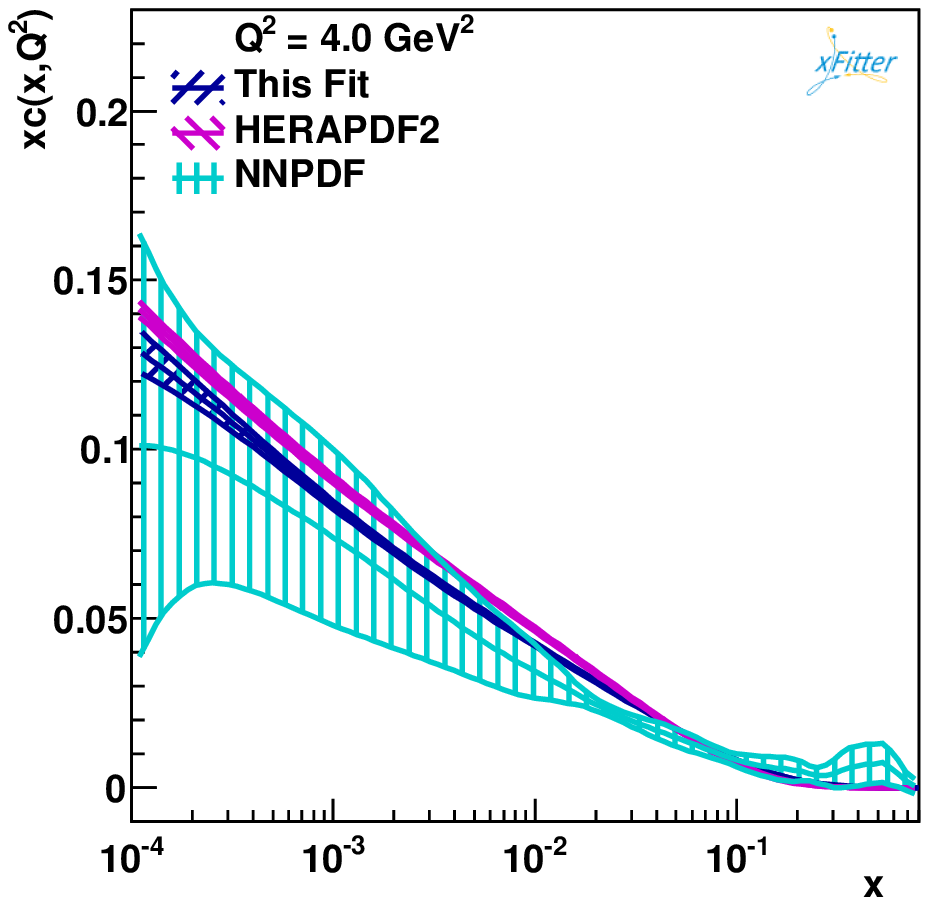}}
 \resizebox{0.32\textwidth}{!}{ 
\includegraphics{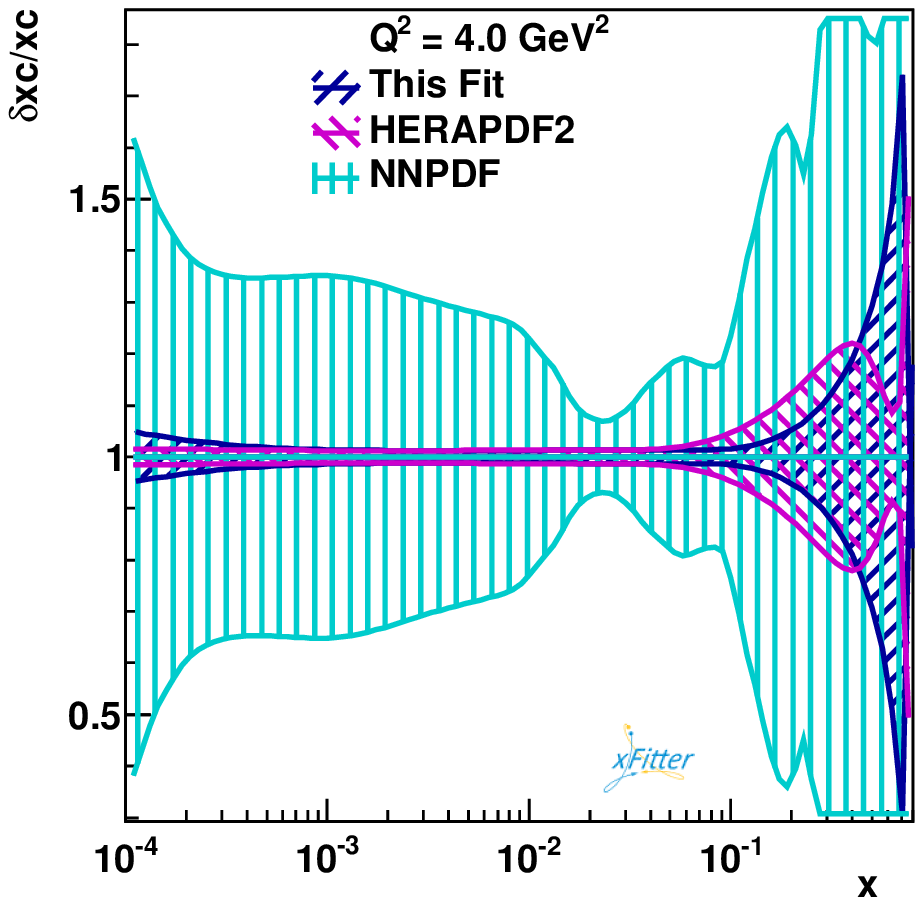}}

 \resizebox{0.32\textwidth}{!}{ 
\includegraphics{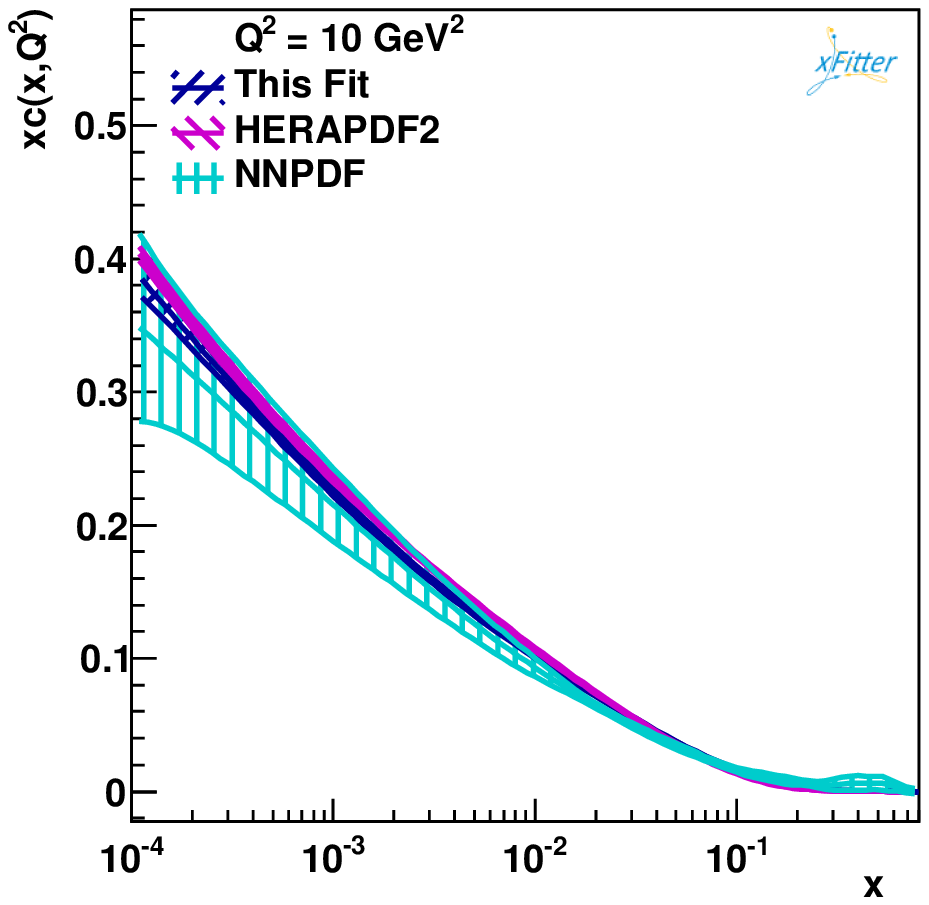}}
 \resizebox{0.32\textwidth}{!}{ 
\includegraphics{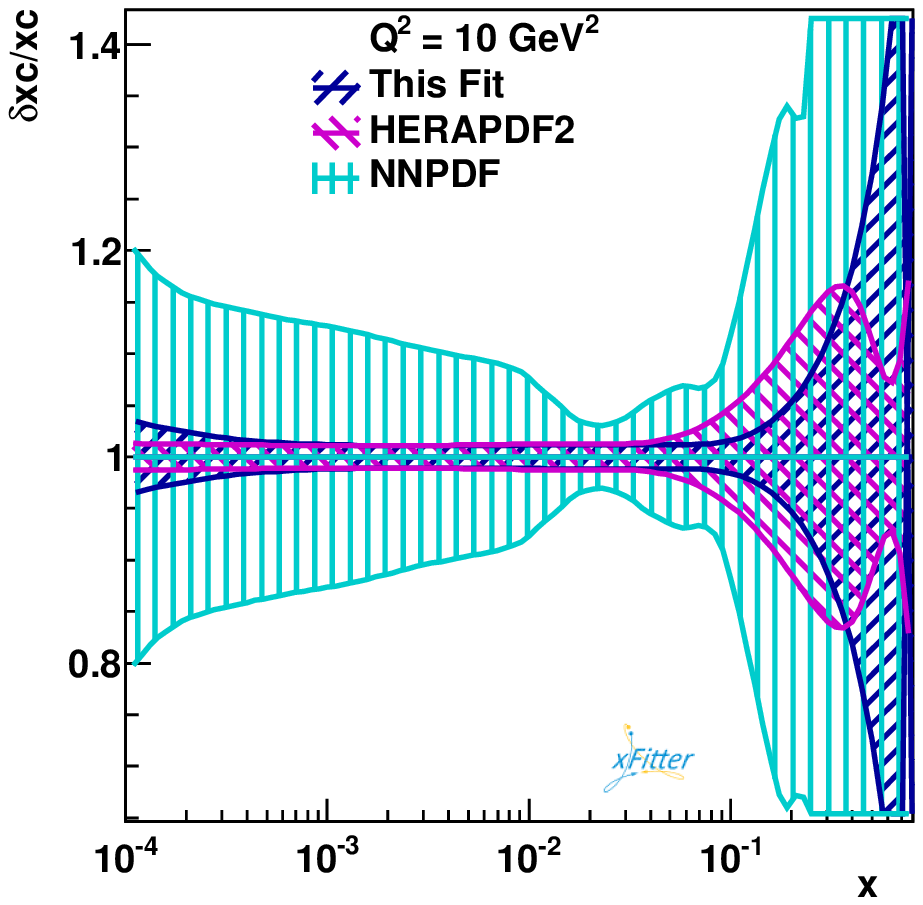}}

 \resizebox{0.32\textwidth}{!}{ 
\includegraphics{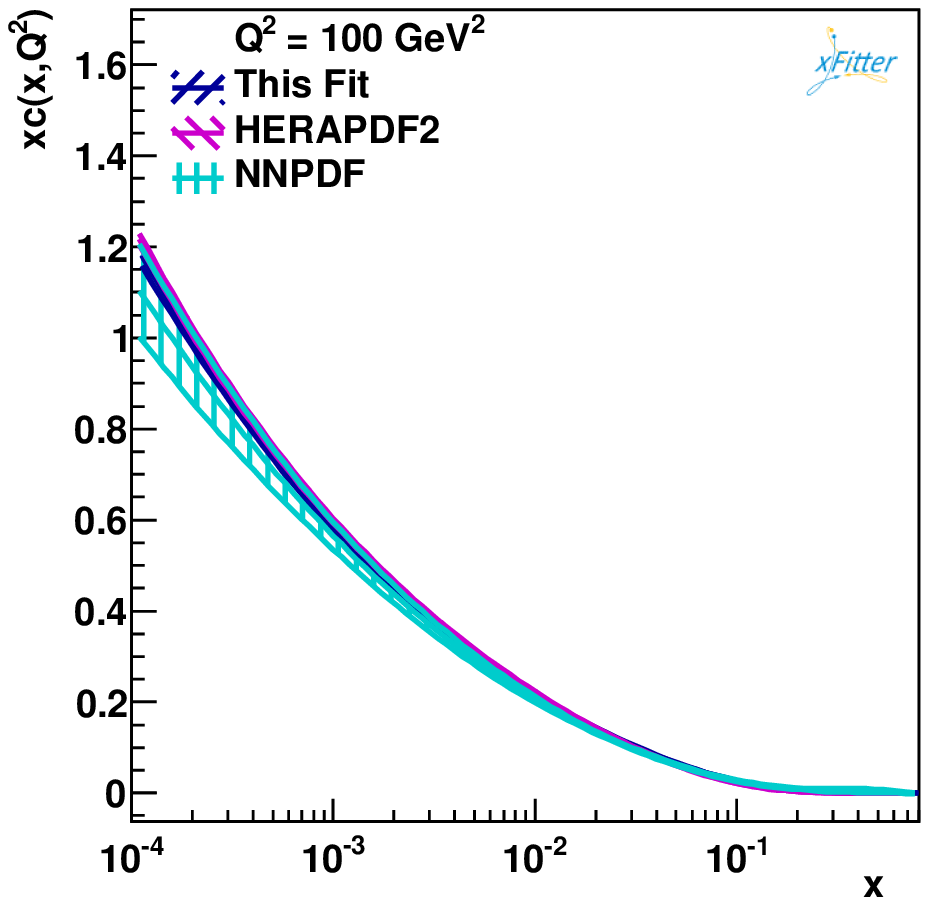}}
 \resizebox{0.32\textwidth}{!}{ 
\includegraphics{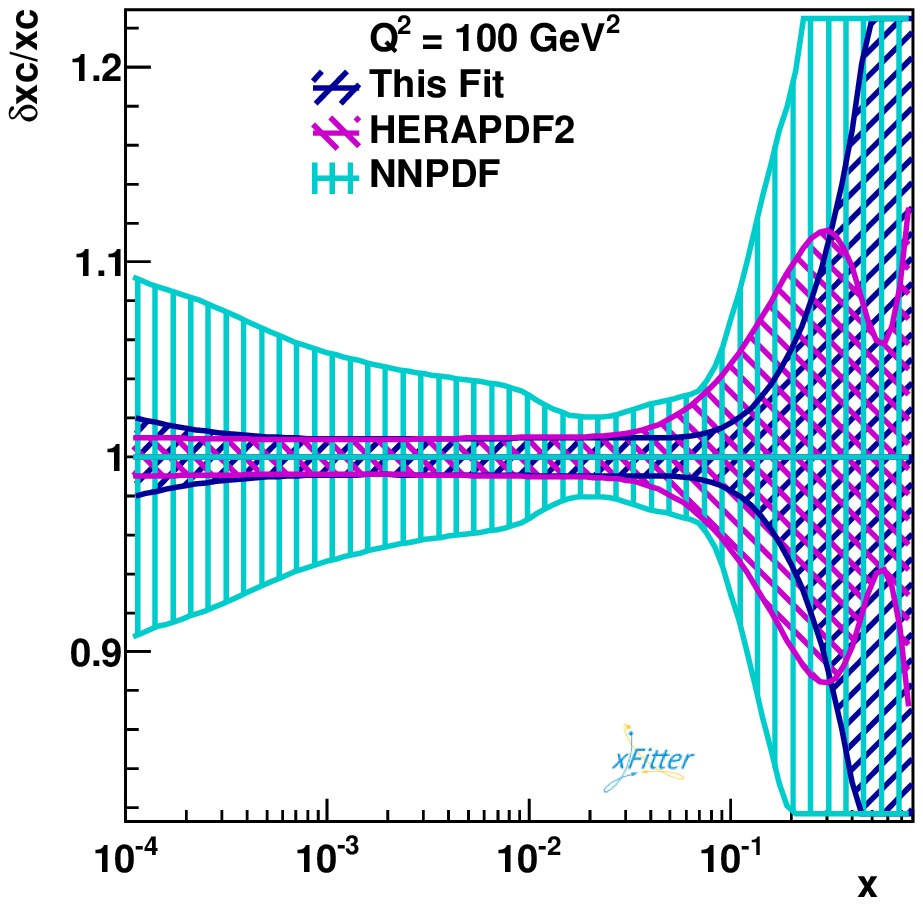}}

 \resizebox{0.32\textwidth}{!}{ 
\includegraphics{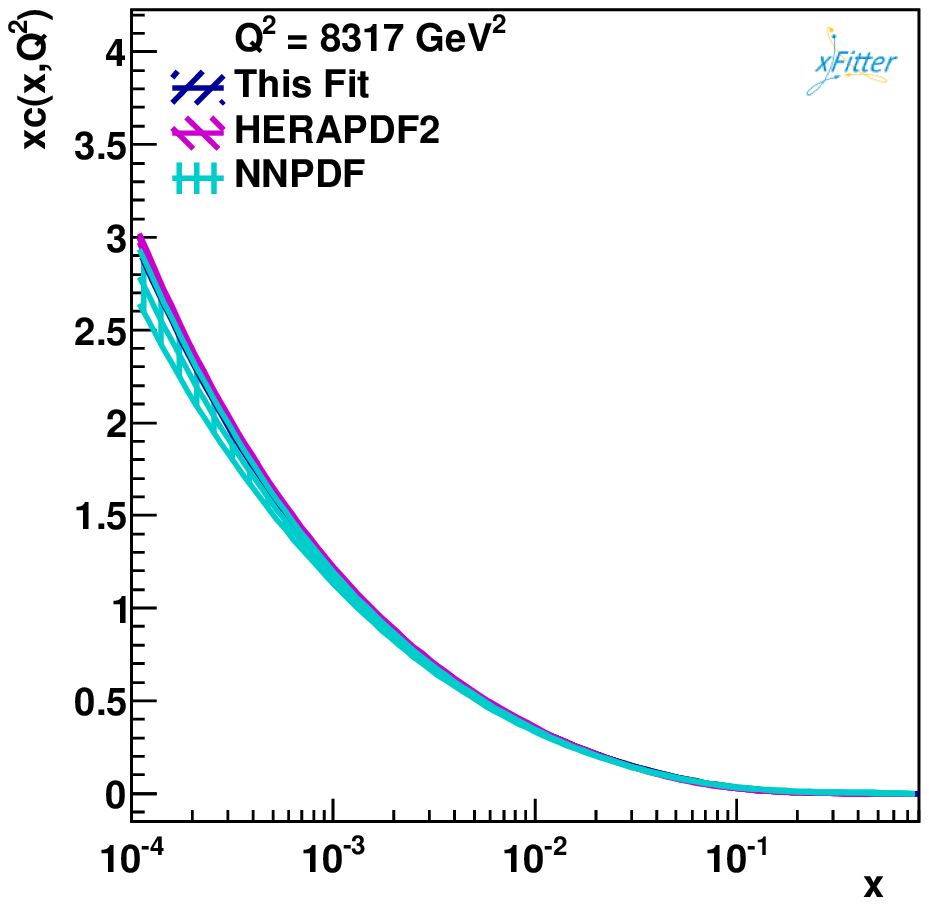}}
 \resizebox{0.32\textwidth}{!}{ 
\includegraphics{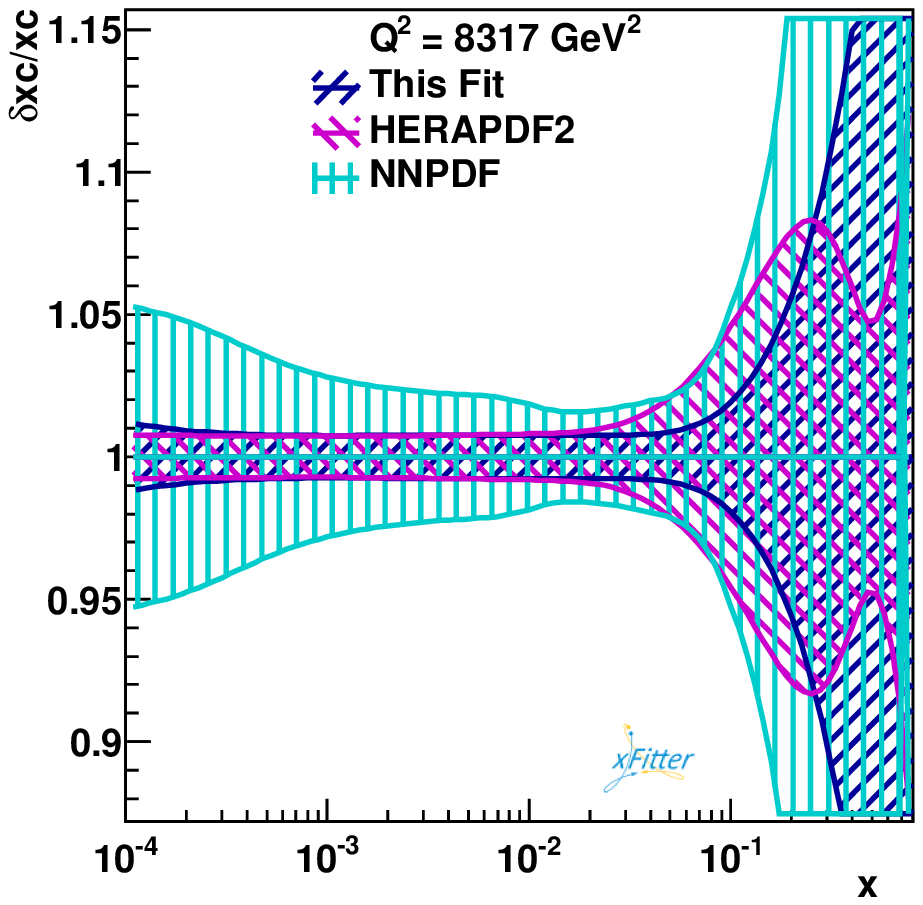}}

\caption{The extrinsic charm PDFs from Fit~C (left panels) and the relative uncertainties $\delta xq(x,Q^2)/xq(x,Q^2)$ (right panels) for the selected scales
$Q^{2}$= 4,  10 , 100, 8317 GeV$^{2}$, as a function of $x$, and compared with HERAPDF2 \cite{Abramowicz:2015mha} and NNPDF3IC \cite{Ball:2017nwa}.} \label{pic:Cratio}
\end{figure*}

 Finally, in Fig.~\ref{pic:Cratio} we present our results for the extrinsic charm PDF (top panels) with its uncertainty and the relative charm PDF uncertainty $\delta xc/xc$ (bottom panels)  as a function of $x$ at varied scale of $Q^2$, and also compared with  HERAPDF2 \cite{Abramowicz:2015mha} and NNPDF3IC \cite{Ball:2017nwa}.  
\begingroup
\squeezetable
\begin{table}
\caption{\label{t4} The predictions of intrinsic charm probability ${P}_{c{\bar c/p}}$ in the different approaches.}
\begin{ruledtabular}
\begin{tabular}{lc}	 

     Reference (Approach) & ${P}_{c{\bar c/p}}$(\%)\\

\hline
BHPS model (Light-cone) \cite{BHPS1} & $\simeq$1 \\
Brodsky \textit{et al} 
  (Light-cone + LHC data) \cite{Bednyakov:2017vck} & $\leq$ 1.93 \\
Harris \textit{et al} 
  (PGF NLO) \cite{Harris:1995jx} & 0.86\\
Hoffmann and Moore 
  (PGF NLO) \cite{Hoffmann:1983ah} & 0.31 \\
Steffens \textit{et al} 
  (Meson cloud) \cite{Steffens:1999hx} & $\simeq$0.4 \\
Dulat \textit{et al}
  (PQCD-NNLO)  \cite{Dulat:2013hea} & $\leq$ 2 \\
Jimenez-Delgado 
  (PDF) \cite{Jimenez-Delgado:2014zga} & 0.3-0.4; $\simeq$1.0 \\\hline
  Our results (PDF + Light-cone ``Fit~C'')\\ 
  $m_c = 1.43$ GeV & $0.96 \pm 0.20$\\
  $m_c = 1.5$ GeV & $0.94 \pm 0.20$\\
  $m_c = 1.55$ GeV & $0.92 \pm 0.20$\\

\end{tabular}
\end{ruledtabular}
\end{table}
\endgroup 

\begingroup
\squeezetable
\begin{table}
\caption{\label{t5} The intrinsic charm moment fraction $<x>_{c + \bar{c}}^{int}$ according to Eq. (\ref{eq:m fraction}) and total charm momentum fraction, $<x>_{c+\bar{c}}^{tot} = <x>_{c + \bar{c}}^{int} + <x>_{c + \bar{c}}^{ext}$ compared to the NNPDF3IC and CT14-IC at $Q^2$ = 4 GeV$^2$.}
\begin{ruledtabular}
\begin{tabular}{llcc}

     Reference & Approach & $<x>_{c + \bar{c}}^{int}$(\%) & $<x>_{c+\bar{c}}^{tot}$(\%) \\

\hline
NNPDF3IC \cite{Ball:2017nwa}& Valence Like & 0.5& 1.2\\
CT14-BHPS1 \cite{Hou:2017khm}& Valence Like& 0.6 &1.64\\
CT14-SEA1 \cite{Hou:2017khm}& Sea Like& 0.6 &1.61\\

\hline
This Fit($m_c$=1.5GeV$^2$ ) &Valence Like & 0.48 $\pm$ 0.1 & 1.06 $\pm$ 0.6\\

\end{tabular}
\end{ruledtabular}
\end{table}
\endgroup

\section{\label{Discussion}Discussion and conclusion}
We performed a QCD analysis using fixed target DIS data from HERAI+II \cite{Abramowicz:2015mha}, H1-ZEUS combined charm cross section \cite{Abramowicz:1900rp}, BCDMS \cite{Benvenuti:1989rh}, SLAC \cite{Whitlow:1991uw} and EMC \cite{Aubert:1982tt} considering IC contribution at NLO. We determine the PDFs with their corresponding uncertainties and also the intrinsic charm probability and strong coupling constant using the $\mathsf{xFitter}$ package \cite{xFitter,Alekhin:2014irh}.

To investigate the effect of the EMC data in the present analysis and to clarify the results, we divided our QCD analysis into four steps. As a first step, we performed a fit with all data sets of Table~\ref{tdata}, with
the \emph{exception} of the EMC  experimental data. We used a fixed $\alpha_s(M^2_Z)$ value of 0.1182 taken from the world average \cite{Patrignani:2016xqp} for the Base fit.

We then include all the data from the Base fit in addition to the EMC $F_2^c$ data with assuming zero IC contribution in Fit A.  

Taking into account the IC contribution and considering ${P}_{c{\bar c/p}}$ as a free parameter, we performed our QCD analysis for two separate Fit B and Fit C. 
In Fit B, we used all data sets which used in Fit A and we fixed strong coupling $\alpha_s(M^2_Z)$. In this case we consider the intrinsic charm probability $P_{c\bar{c}/p}$ as a free parameter. We found that ${P}_{c{\bar c/p}}$  depends on the $\alpha_s$ value, so the simultaneous determination of these parameters in the QCD analysis, taking into account IC is important. In Fit C, we considered $P_{c\bar{c}/p}$ and $\alpha_s(M^2_Z)$ as free parameters. 

The extracted value of the strong coupling constant $\alpha_s(M_Z^2) = 0.1191 \pm 0.0008$ at NLO in Fit~C,  is in good agreement with the world average value of $\alpha_{s}(M_{Z}^{2})=0.1182\pm0.0011$~\cite{Patrignani:2016xqp}. This extracted value is also consistent with our recent reported results of $\alpha_{s}(M_Z^2)$= 0.1199 $\pm$ 0.0031, based on charged current neutrino-nucleon  DIS data at NLO \cite{GHAFARI-PRC}. For ${P}_{c{\bar c/p}}$, we extracted $1.017 \pm 0.2$ for Fit~B and $0.94 \pm 0.2$ for Fit~C. We found 7.6\% difference between Fit~B and Fit~C on ${P}_{c{\bar c/p}}$ values by comparing these fits.

The Q$^2$-evolution of intrinsic charm PDF using ${P}_{c{\bar c/p}} = 0.94 \pm 0.2$\% in the proton  which extracted in the present analysis, are presented. The extracted value of ${P}_{c{\bar c/p}}$ in the present QCD analysis is consistent with the recent reported upper limit of $1.93\%$ of Ref.~\cite{Bednyakov:2017vck},  which is obtained for the first time from LHC measurements \cite{Aaboud:2017skj}.

To find the impact of the charm mass in the present analysis, we study the influence on the extraction of ${P}_{c{\bar c/p}}$. For Fit B, ${P}_{c{\bar c/p}}$ goes from 0.78\% to 1.4\% as increases $m_c$ from 1.39 to 1.6 GeV. For Fit~C, we also find ${P}_{c{\bar c/p}}$ changes from 0.97\% to 0.88\% as $m_c$ changes and the $\alpha_s(M_Z^2)$ increase from 0.1152 to 0.1225.

Since the gluon PDF plays an important role in the total cross section of top quark production, we find a total top quark cross section of 817.8$\pm$6.4 using by Fit~B when $m_c=1.5$. In Fit~C, we find the cross section of 804.7$\pm$6.6, 828.7$\pm$8.0 and 855.7$\pm$7.8 for $m_c=1.43$, $m_c=1.5$ and $m_c=1.55$, respectively. 

The parabolic fit to $\chi^2$ as a function of $m_c$ is used to find the optimal value of $m_c$. In this analysis, we choose the value of $m_c$ equal to 1.50 GeV. We performed our QCD analysis based on this value of $m_c$ in presence of IC.

In the comparison of our PDFs to others in the literature, the valence, sea, gluon and charm PDFs and their uncertainties are in good agreement with other theoretical models. This agreement is despite of different phenomenology and differences in the parametrization and choose of the data sets.

In our previous results, we have shown the differential cross section of $\gamma+c$-jet is sensitive to intrinsic charm contribution \cite{Rostami:2015iva}. In the near future, we will investigate the role of IC contributions in other processes, such as $\gamma+c/b$, and $Z/W +c/b$ production cross section measurements in $pp$ collisions at the LHC.

\begin{acknowledgements}

We thank S. Brodsky for valuable and  constructive suggestions during the planning and development of this research work. We would also like to thank F. Olness, A. Glazov, O. Zenaiev, Z. Karamloo and M. Azizi   for helpful discussions, comments and carefully reading the manuscript. A. K. thanks SITP (Stanford Institute for Theoretical Physics) and the Physics Department of SMU (Southern Methodist University) for their hospitality at the beginning
of this work. A. K. is also grateful to CERN TH-PH division for the hospitality
where a portion of this work was performed. H. A. is also grateful to the DESY FH group for the hospitality and financial support at DESY. 

\end{acknowledgements}

\balance

\begin{figure*}[htp]
 \begin{center}
\includegraphics[scale=.58]{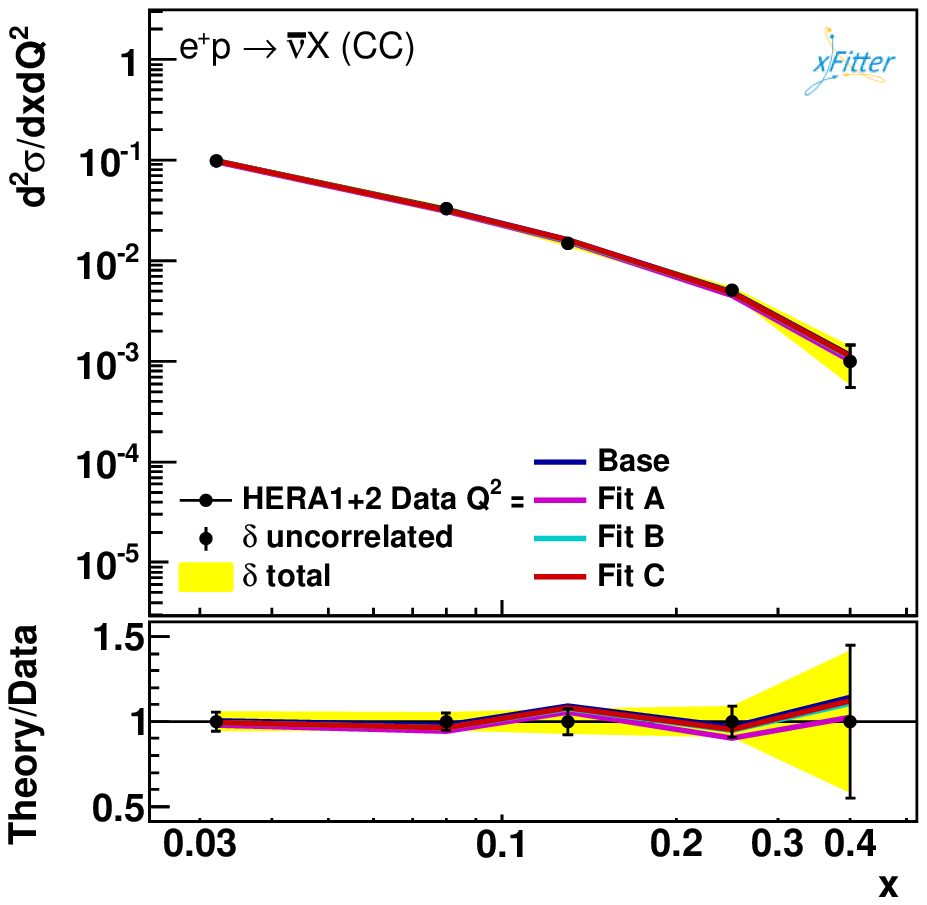}
\includegraphics[scale=.58]{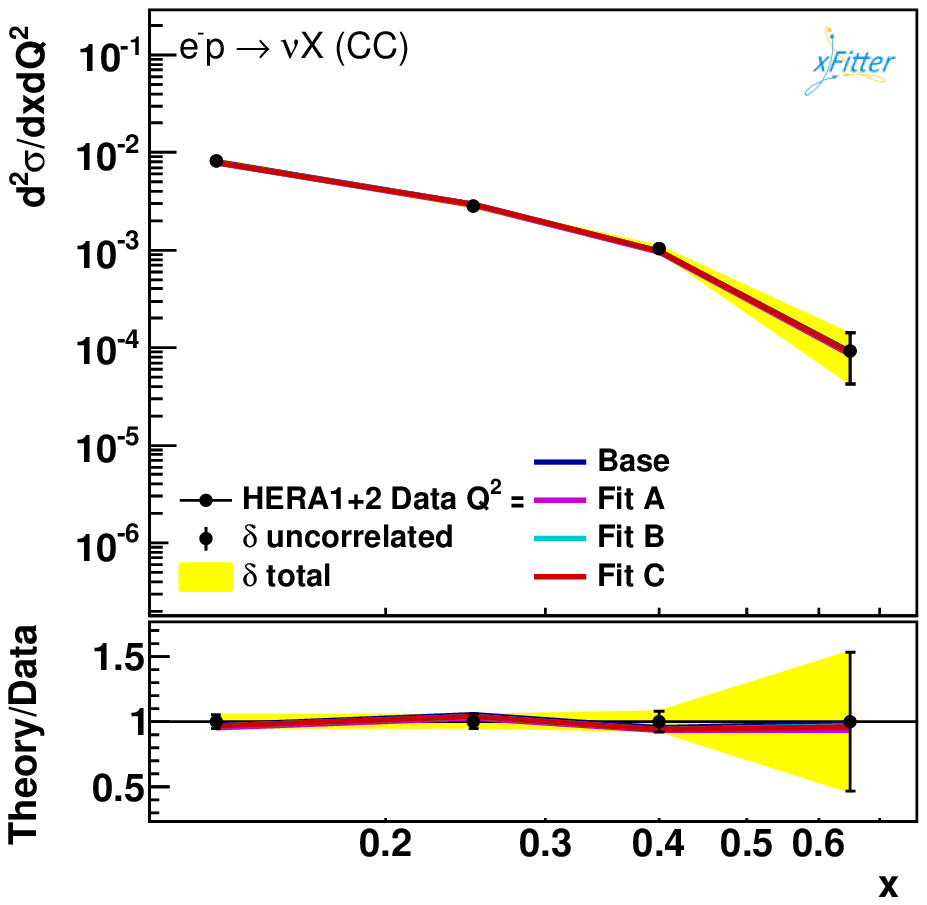}
\includegraphics[scale=.58]{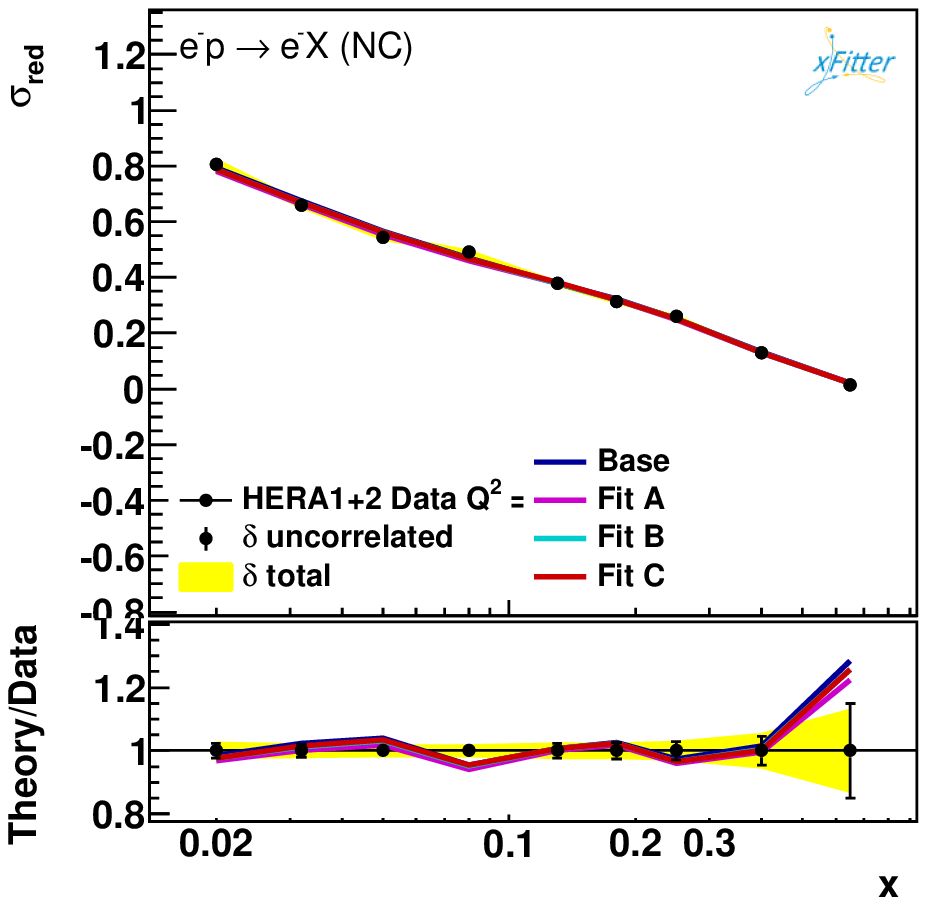}\\

\includegraphics[scale=.58]{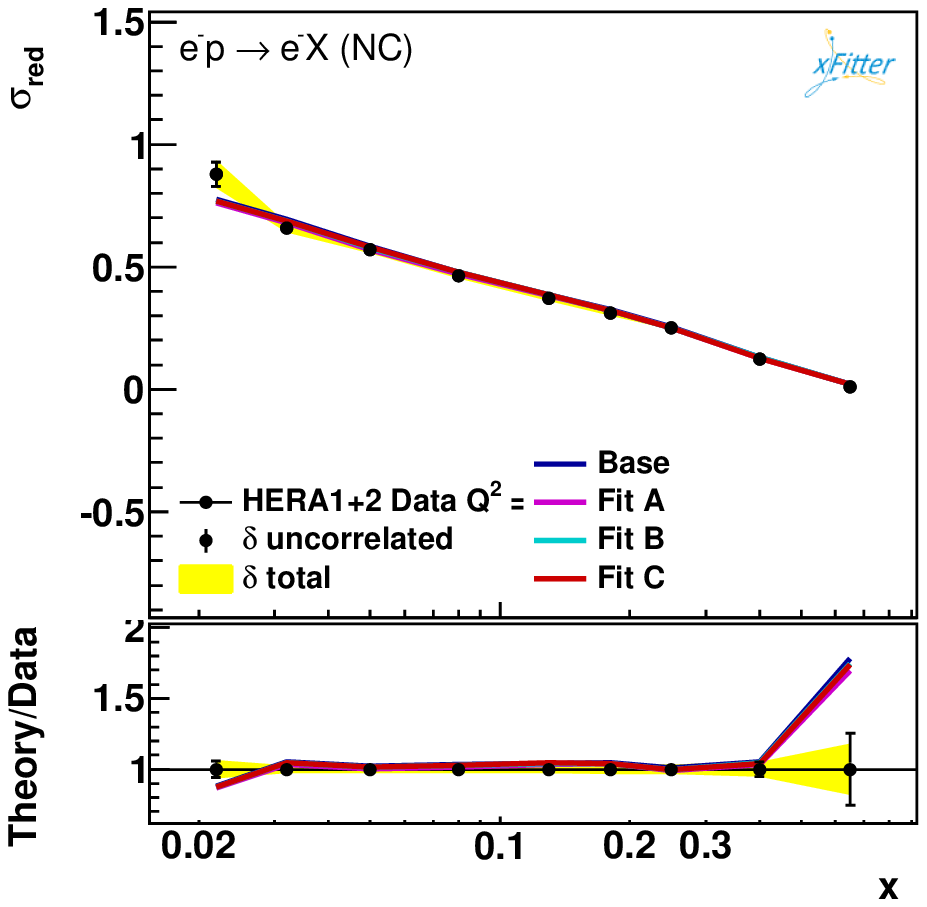}
\includegraphics[scale=.58]{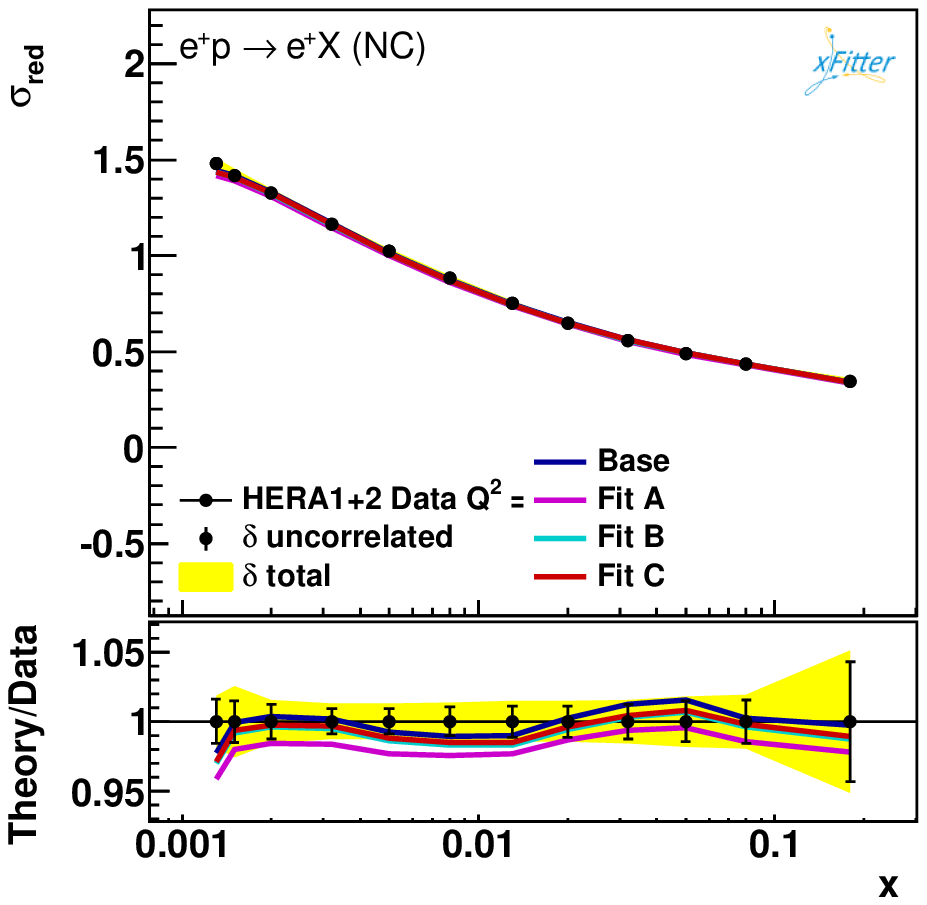}
\includegraphics[scale=.58]{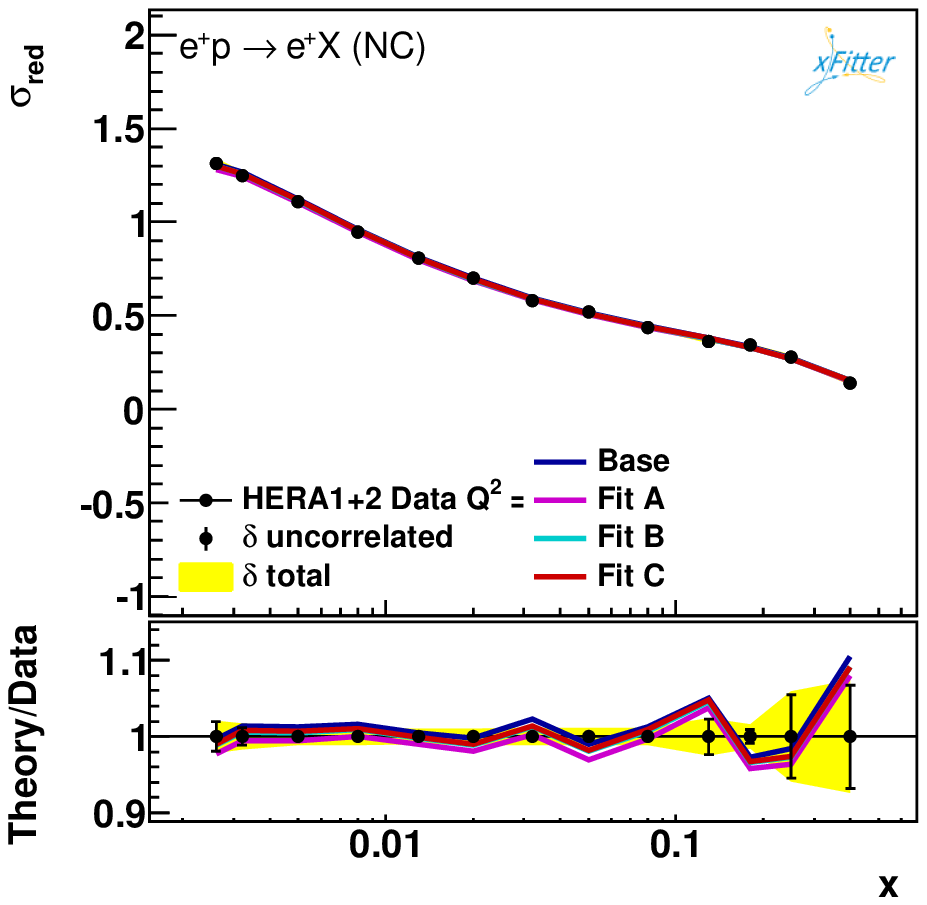}\\

\includegraphics[scale=.58]{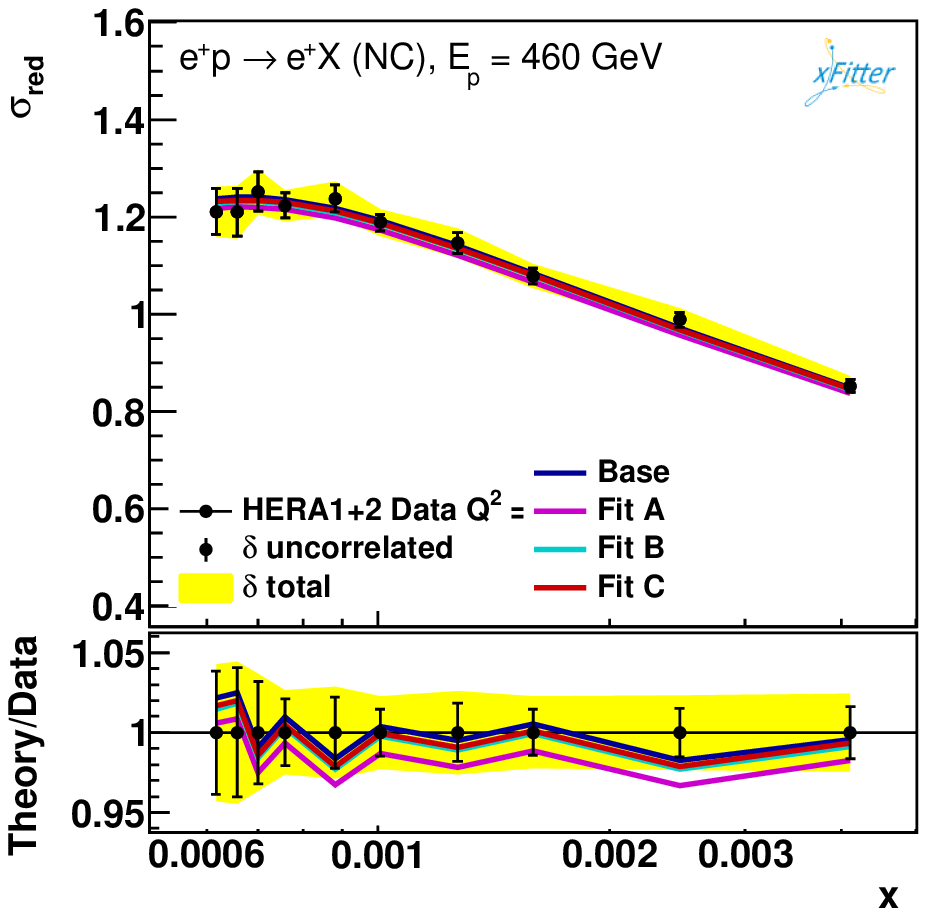}
\includegraphics[scale=.58]{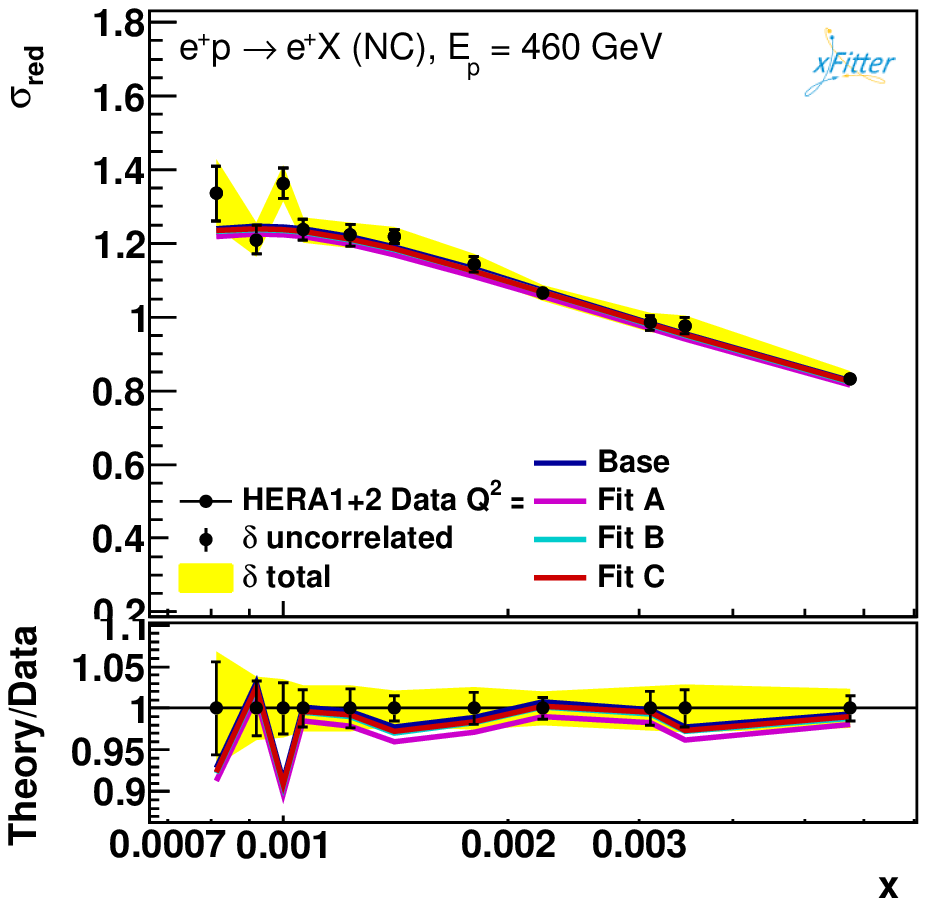}
\includegraphics[scale=.58]{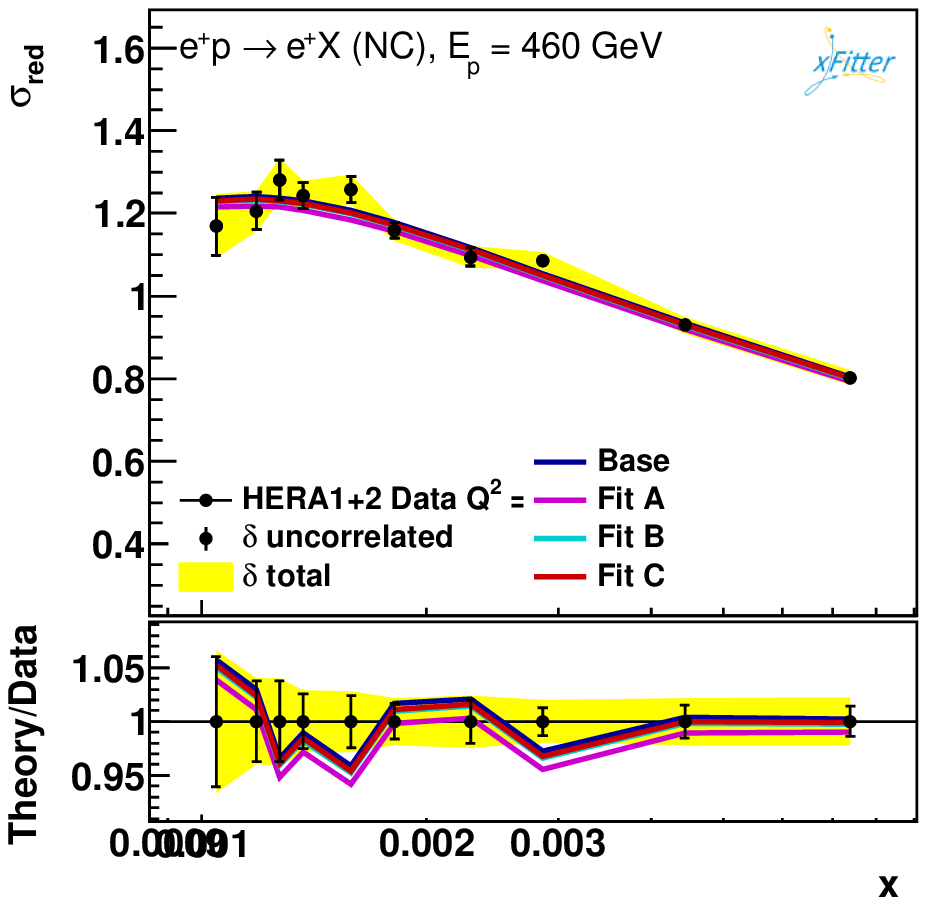}\\
    \subfigure[HERA I+II and combined H1 and ZEUS charm cross sections]{\label{pic:QCD-Fit-1-a}
\includegraphics[scale=.58]{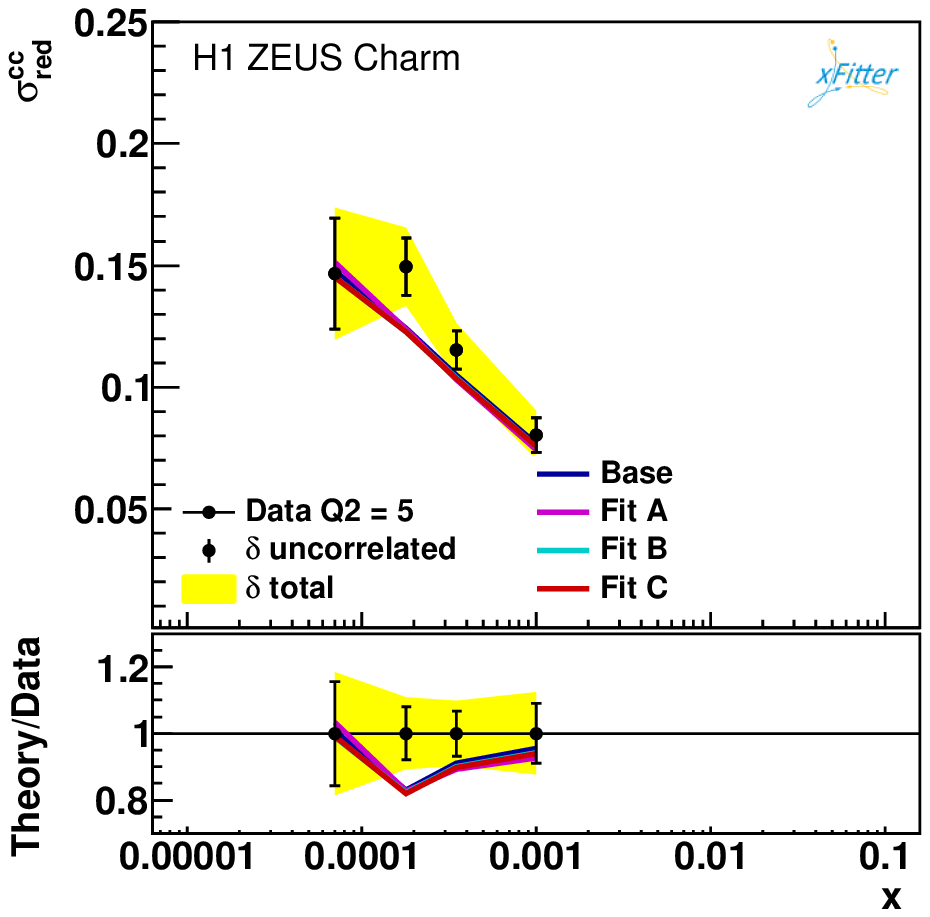}
\includegraphics[scale=.58]{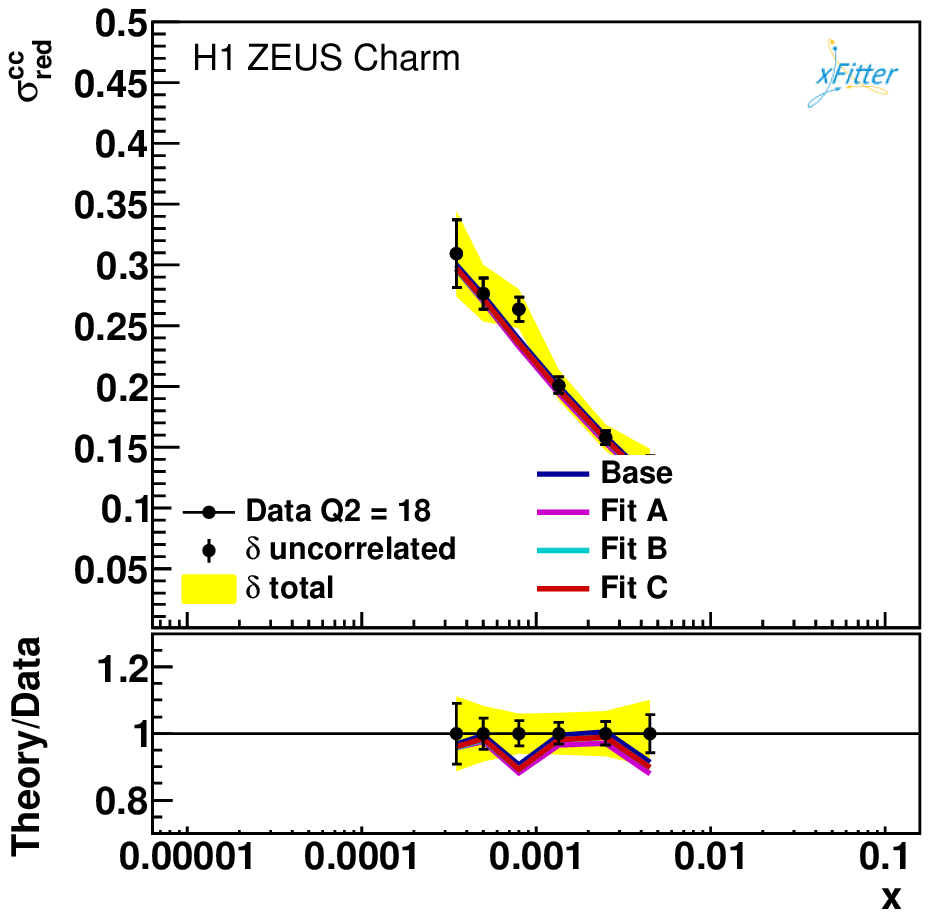}
\includegraphics[scale=.58]{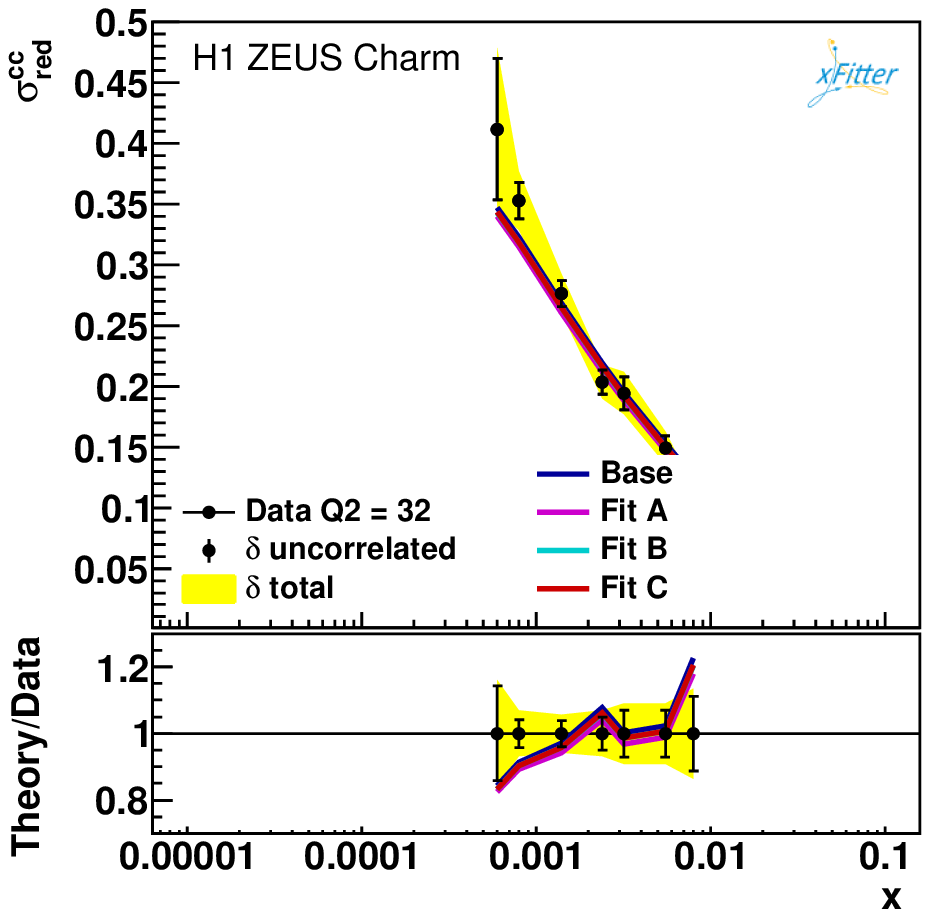}}\\

\end{center}
\end{figure*}

\newpage

\end{sloppypar}

\end{document}